\documentclass[AMA,LATO1COL]{WileyNJD-v2}

\usepackage{amssymb}
\usepackage{subfig}
\usepackage{colortbl}

\articletype{}%

\received{\today}
\revised{Day Month Year}
\accepted{Day Month Year}

\begin{document}

\title{Fog-supported delay-constrained energy-saving live migration of VMs over MultiPath TCP/IP 5G connections}

\author[1]{Enzo Baccarelli}

\author[1]{Michele Scarpiniti}

\author[1]{Alireza Momenzadeh}



\authormark{BACCARELLI \textsc{et al.}}

\address[1]{\orgdiv{Department of Information Engineering, Electronics and Telecommunications (DIET)}, \orgname{Sapienza University of Rome}, \\\orgaddress{\state{via Eudossiana 18, 00184 Rome}, \country{Italy}}}



\corres{Michele Scarpiniti, Sapienza University of Rome. \email{michele.scarpiniti@uniroma1.it}}


\abstract[Summary]{The incoming era of the Fifth Generation (5G) Fog Computing (FC)-supported Radio Access Networks (shortly, 5G FOGRANs) aims at exploiting the virtualization of the computing/networking resources. The goal is to allow battery-powered wireless devices to augment their limited native resources through the seamless (e.g., live) migration of Virtual Machines (VMs) towards nearby Fog data centers. For this purpose, the bandwidths of the multiple  heterogeneous Wireless Network Interface Cards (WNICs) typically equipping the current wireless devices may be used in parallel under the control of the emerging MultiPathTCP (MPTCP) protocol. However, due to the fading and mobility-induced phenomena impairing the underlying RANs, the energy consumptions and migration delays of the current state-of-the-art VM migration techniques may drastically reduce their expected benefits. Motivated by these considerations, in this paper, we analytically characterize, implement in software and numerically test the optimal minimum-energy Settable-Complexity Bandwidth Manager (SCBM) for the live migration of VMs over MPTCP-supported 5G FOGRAN connections. The key features of the proposed SCBM are that: (i) its implementation complexity is settable on-line, in order to attain the targeted energy consumption-vs.-implementation complexity tradeoff; (ii) it minimizes the energy consumed by the migration process under hard QoS constraints on the allowed migration time and downtime; and, (iii) by leveraging a suitably designed adaptive mechanism, it is capable to quickly react to (possibly, unpredicted) fading and/or mobility-induced abrupt changes of the operating conditions without requiring and forecasting support. We numerically test and compare the energy and delay performances of the proposed adaptive SCBM under several 3G/4G/WiFi FOGRAN scenarios by considering a number of synthetic and real-world workloads. The obtained numerical results point out that: (i) the energy savings of the proposed adaptive SCBM over the benchmark one currently implemented by legacy Xen, KWM and VMware hypervisors are typically over 25\% and approach 70\% when the migrated applications are memory-write intensive; and, (ii) the MPTCP may reduce the energy consumption of the proposed SCBM over legacy SinglePathTCP (SPTCP) more than 50\% under strict real-time limits on the allowed migration times.}



\maketitle


\section{Introduction}
\label{sec:sec1}

Smartphones are already our ubiquitous technological assistants. Since 2011, the worldwide smartphone penetration overtook that of wired PCs by reaching 80\% in US and Europe. Cisco envisions that the average number of connected mobile devices per person will reach 6.6 in 2020, mainly fostered by the expected pervasive use of Internet of Everything (IoE) applications \cite{A1}. This will be also due to the fact that future smartphones will increase their capability to use in an efficient way the multiple heterogeneous Wireless Network Interface Cards (WNICs) that already equip them by exploiting an emerging bandwidth-pooling transport technology, generally referred to as MultiPath TCP (MPTCP) \cite{A2}.

However, since smartphones are (and will continue to be) resource and energy limited, Mobile Cloud Computing (MCC) is expected to effectively support them by providing device augmentation \cite{A3}. MCC is, indeed, a quite novel technology that relies on the synergic cooperation of three different paradigms, namely, cloud computing, mobile computing and wireless Internet. Its main goal is to cope with the inherent resource limitation of the mobile devices by allowing them to offload computation and/or memory-intensive applications (such as, for example, image processing, voice recognition, online gaming and social networks) towards virtualized data centers \cite{A3}. However, performing application offloading from mobile devices to remote large-scale cloud-based data centers (e.g., Amazon and Google, to name a few) suffers from the large communication delay and limited bandwidth typically offered by multi-hop cellular Wide Area Networks (WANs) \cite{A3}. The dream is, indeed, to allow mobile devices to: (i) exploit their multiple WNICs to perform bandwidth aggregation through \textit{MPTCP} \cite{A2}; and, then: (ii) leverage the sub-millisecond access latencies promised by the emerging \textit{Fifth Generation} (5G) multi Radio Access Networks (RANs) \cite{A4}, in order to perform \textit{seamless} application offloading towards \textit{proximate} virtualized data centers, generally referred to as \textit{Fog nodes} \cite{A1}. For this purpose, the application to be offloaded from the mobile device is shipped in the form of a \textit{Virtual Machine} (VM) and, then, the offloading process takes place as a \textit{live} (e.g., seamless) VM \textit{migration} from the mobile device to the serving Fog node \cite{A5}.

\subsection{Convergence of Fog, 5G and MPTCP: the reference 5G FOGRAN technological platform}
\label{sec:ssec1.1}

The reason why the convergence of the (aforementioned) three pillar paradigms of Fog Computing, 5G Communication and MPTCP is expected to enable fast seamless migrations of VMs stems out from their native features, that are, indeed, complementary (see Table \ref{tab:tabA} for a synoptic win-to-win overview).

\begin{table*}[htbp]
\centering
\resizebox{0.85\textwidth}{!}{
\begin{tabular}{p{1.6cm}p{3.2cm}c
>{\columncolor[HTML]{E6E6E6}}p{1.6cm}
>{\columncolor[HTML]{E6E6E6}}p{3.3cm}
cp{1.6cm}p{3.2cm}}
\toprule
\multicolumn{2}{c}{\textbf{Fog}}  & &  \multicolumn{2}{c}{\cellcolor[HTML]{E6E6E6}\textbf{5G}}  & &  \multicolumn{2}{c}{\textbf{MPTCP}}  \\
\midrule
Edge location and context awareness & Lasting one-hop from the served devices, Fog nodes may exploit context-awareness for the support of delay-sensitive VM migration & & 
Ultra-low access latency & Sub-millisecond access latencies are obtained by combining massive MIMO, millimeter-wave, micro-cell and interference-suppression technologies & & 
Bandwidth aggregation & By utilizing in parallel the available bandwidth of each path, MPTCP connections may provide larger throughput than SPTCP connections \\[3ex]
\midrule
Pervasive deployment & Fog nodes are a capable to support pervasive services, such as community services & & 
On demand resource provisioning & Wireless bandwidth is dynamically provided an on-demand and per-device basis & & 
Robustness & Path failure is recovered by shifting the traffic from the failed path to the remaining active ones \\[2ex]
\midrule
Virtualized environment & Software clones of the served wireless devices are dynamically bootstrapped onto the Fog servers as VMs & & 
Resource multiplexing and isolation & Processing of the received flows is performed by a centralized Network Processor, that performs resource multiplexing/isolation on a per-flow basis through NFV & & 
Backward compatibility & MPTCP connections utilize the same sockets of the standard SPTCP ones. Hence, SPTCP-based applications may continue to run unchanged under MPTCP \\[2ex]
\midrule
Mobility support & Fog nodes may exploit single-hop short-range \textit{IEEE802.11x} links for enabling VM migrations from the mobile devices & & 
Multi-RAT support & Multiple short/long-range WiFi/cellular networking technologies are simultaneously sustained, in order to support multi-NIC wireless devices & & 
Load balancing & By a balanced splitting of the overall traffic over multiple paths, energy saving may be attained \\[2ex]
\midrule
Support for live VM migration & Wireless devices may seamlessly offload to serving Fog nodes their running applications in the form of VMs & & 
  &   & & 
  &   \\[2ex]
\midrule
Energy efficiency & Wireless devices may save energy by exploiting nearby Fog nodes as surrogate servers & & 
  &   & & 
  &   \\
\bottomrule
\end{tabular}}
\caption{Native features of the pillar Fog, 5G and MPTCP paradigms and their synergic interplay.}
\label{tab:tabA}
\end{table*}

\textit{Fog Computing} (FC) is formally defined as a model for enabling the \textit{pervasive} local access to a centralized pool of configurable computing/networking resources that can be rapidly provisioned and released in an elastic (e.g., on-demand) basis. Proximate resource-limited mobile devices access these facilities over a \textit{single-hop} wireless access network, in order to reduce the communication delay and, then, support delay and delay-jitter sensitive VM migrations. As pointed in the first column of Table \ref{tab:tabA}, the main native features retained by the Fog paradigm are \cite{A1}: (i) edge location and context awareness; (ii) pervasive spatial deployment; and, (iii) support for mobile energy-saving live migration of VMs. 

Thank its sub-millisecond network latencies, the \textit{5G Communication} paradigm provides the ``right'' means to support the wireless access to the virtualized resources hosted by Fog nodes. In fact, the 5G paradigm retains, by design, the following main native features (see the second column of Table \ref{tab:tabA}) \cite{A4}: (i) ultra-low access latencies; (ii) on-demand provisioning of the wireless bandwidth; (iii) multiplexing, isolation and elastic scaling of the available physical communication resources; and, (iv) support for heterogeneous Radio Access Technologies (RATs) and multi-NIC mobile devices.

Thank to its support of multiple RATs, the 5G paradigm is the ideal partner of the \textit{MPTCP}, whose native goal is to provide multi-homing (e.g., radio bundling) to mobile devices equipped with multiple heterogeneous WNICs. This means that, by design, MPTCP supports simultaneous data paths over \textit{multiple} radio interfaces under the control of a \textit{single} Transport-layer connection. Doing so, MPTCP attains (see the third column of Table \ref{tab:tabA}) \cite{A2}: (i) bandwidth aggregation; (ii) robustness against mobility and/or fading-induced connection failures; (iii) backward compatibility with the legacy Single-Path TCP (SPTCP); and, (iv) load balancing of the migrated traffic.

\begin{figure*}[htb]
\centering
\includegraphics[width=0.85\columnwidth]{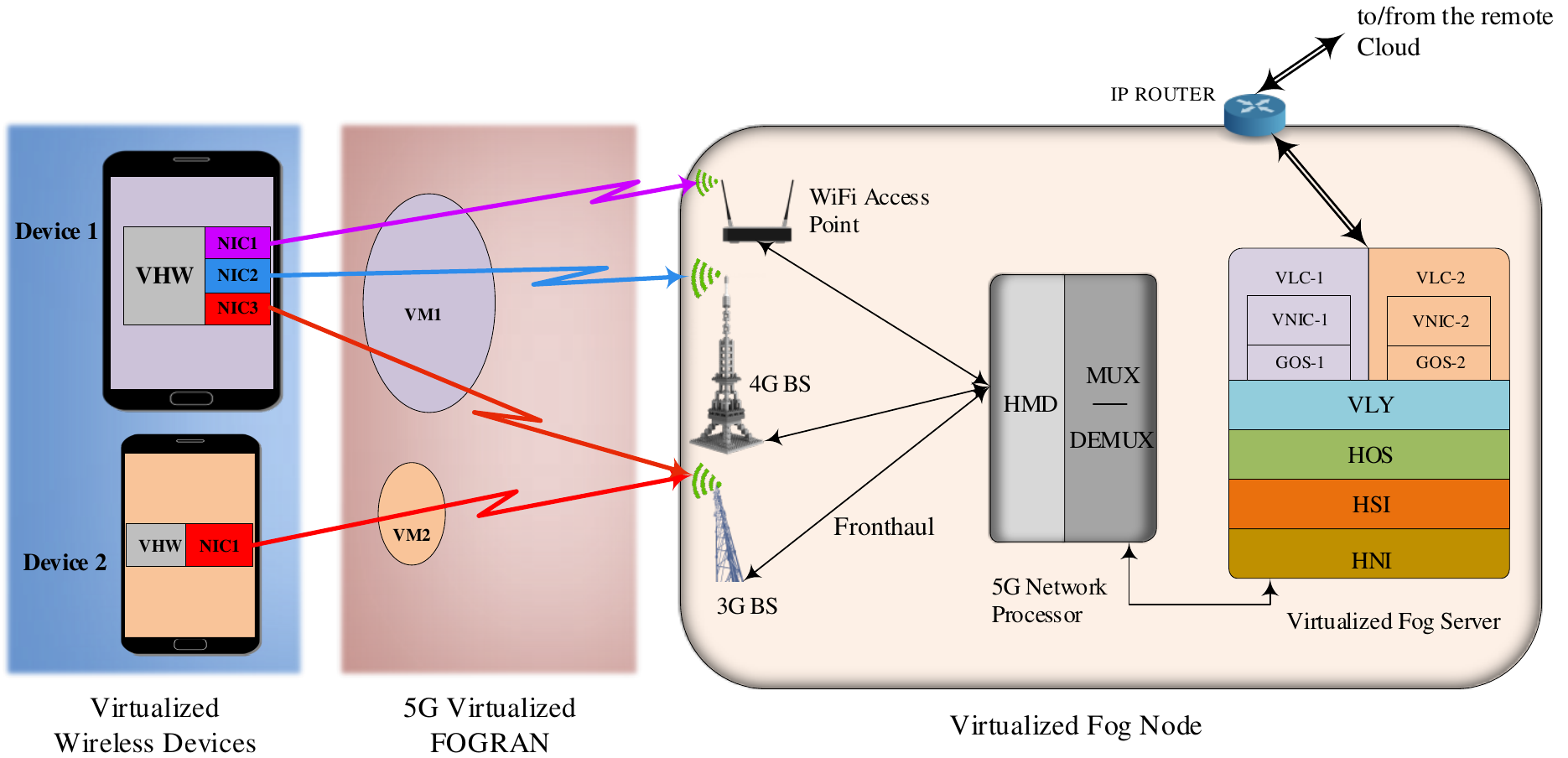}
\caption{The considered virtualized scenario over a 5G access network. VHW = Virtualized HardWare; NIC = Network Interface Card; VM = Virtual Machine; BS = Base Station; HMD = Heterogeneous Mo-Demodulators; MUX = MUltipleXer; DEMUX = DEMUltipleXer; HNI = Hardware Network Infrastructure; HSI = Hardware Server Infrastructure; HOS = Host Operating System; RAN = Radio Access Network; VLY = Virtualization LaYer; VNIC = Virtual Network Interface Card; VCL = Virtual Clone; GOS = Guest Operating System; $\boldsymbol{\longleftrightarrow}$ = Wired Connection.}
\label{fig:scenario}
\end{figure*}

The convergence of these three paradigms leads to the 5G FOGRAN technological platform of Fig. \ref{fig:scenario}, that constitutes the reference scenario of this paper.       

According to the emerging FOGRAN paradigm \cite{A6,A7,A8}, in this scenario, a mobile device that is equipped with the MPTCP protocol stack establishes multiple connections to the serving Fog node by turning-ON and using in parallel its native WNICs (see the device in the top part of Fig. \ref{fig:scenario}). Since the MPTCP is backward compatible with the SPTCP, a mobile device equipped with a single WNIC may continue to use legacy SPTCP connections (see the device in the bottom part of Fig. \ref{fig:scenario}). The serving Fog node employs hypervisor-based virtualization technology, in order to deliver computational resources in the form of a VM that runs atop a cluster of networked physical servers hosted by the Fog node \cite{A3}. The hosted VM and the hosting Fog node communicate through a Virtual NIC (VNIC), that typically emulates in software an Ethernet switch \cite{A5}. According to the FOGRAN paradigm, the receive antennas of Fig. \ref{fig:scenario} (also referred to as Remote Radio Heads (RRHs)) perform only the Radio Frequency (RF) processing of the received data streams (e.g., power amplification, A/D and D/A conversions and sampling), while: (i) mo/demodulation and co/decoding base-band operations; (ii) MAC multiplexing/de-multiplexing and forwarding of the data flows; (iii) Network-layer routing; and, (iv) elastic resource allocation of the available communication bandwidth and resources, are collectively performed by the 5G Network Processor of Fig. \ref{fig:scenario}. Specifically, according to the 5G paradigm (see Table \ref{tab:tabA}), the 5G Network Processor multiplexes among the migrating VMs the available network physical resources by: (i) performing Network Function Virtualization (NFV); and, then, (ii) instantiating Virtual Base Stations (VBSs) atop the underlying 5G physical network (see Section VIII of \cite{A8}). The connection between the distributed RRHs and the centralized 5G Network Processor is provided by the front-haul section of Fig. \ref{fig:scenario}, that is typically built up by ultra-broadband high-capacity optical-fiber channels \cite{A8}. 

In order to be robust against the connection failures (possibly) induced by wireless fading and/or device mobility, \textit{pre-copy} migration \cite{22} is the technique adopted by the wireless devices of Fig. \ref{fig:scenario} for performing VM offloading over the underlying 5G FOGRAN.

\subsection{Tackled problem, main contributions and organization of the paper}
\label{sec:ssec1.2}

By referring to the 5G FOGRAN scenario of Fig. \ref{fig:scenario}, the topic of this contribution is the settable-complexity energy-efficient dynamic management of the MPTCP bandwidth for the QoS pre-copy live migration of VMs. The target is the minimization of the network energy wasted by each wireless device for the migration of own VMs under six hard constraints, which are typically induced by the desired implementation complexity-vs.-performance tradeoff.

Specifically, the first constraint regards the settable implementation complexity of the proposed bandwidth manager. In order to (shortly) introduce it, we point out that the pre-copy migration technique requires, by design, that the memory footprint of the VM is migrated over a number $I_{MAX}$ of time rounds \cite{22}, whose transmission rates may be optimized in a more or less fine way, in order to save networking energy (see Fig. \ref{fig:PeC} in the sequel). A \textit{per-round} optimization of the migration rates leads to the \textit{maximum} saving of the network energy but also entails the \textit{maximum} implementation complexity of the resulting bandwidth manager. Hence, the first constraint concerns the fact that, in order to limit the resulting implementation complexity, \textit{only} a subset $Q$ out of the available $I_{MAX}$ migration rates is allowed to be dynamically updated.

The second and third constraints limit in a hard way the total migration time and downtime (e.g., the service interruption time) of the migrating VM. These constraints allow to support delay and delay-jitter sensitive applications run by the migrating VM. The forth constraint limits the maximum bandwidth available for the VM migration and it is enforced by the bandwidth allocation policy implemented by the 5G Network Processor of Fig. \ref{fig:scenario}. The fifth constraint upper bounds the maximum slowdown (e.g., the maximum stretching of the execution time) that is tolerated by the migrated application. Finally, the last constraint enforces the convergence of the iteration-based migration process by introducing a hard bound on the feasible speed-up factor, e.g., the minimum ratio between the volumes of data migrated over two consecutive rounds (see also Fig. \ref{fig:PeC} in the sequel).

As matter of these facts, we anticipate that the major contributions of this paper are the following ones:
\begin{enumerate}
	\item Being the MPTCP a novel paradigm that is not still fully standardized, various Congestion Control (CC) algorithms are currently proposed in the literature, in order to meet various application-depending tradeoffs among the contrasting targets of fairness, quick responsiveness and stable behavior \cite{35}. Hence, since the energy-vs.-transport rate profiles of the device-to-fog connections of Fig. \ref{fig:scenario} may depend also on the specifically adopted CC algorithm, we carry out a formal analysis of the power and energy consumptions of a number of MPTCP CC algorithms under the 5G FOGRAN scenario of Fig. \ref{fig:scenario}. This is done for both cases of unbalanced and balanced MPTCP connections. The key result of this analysis is a \textit{unified parametric} formula for the consumed dynamic power-vs.-transport rate profile that applies to \textit{all} considered MPTCP CC algorithms, as well as to the newReno CC algorithm of the standard SPTCP.
	\item On the basis of the obtained power profile of the wireless MPTCP connections, we develop a related delay-vs.-energy analysis, in order to formally characterize the operating conditions under which VM migration over the 5G FOGRAN of Fig. \ref{fig:scenario} allows the mobile device to really save energy. Interestingly, the carried analysis \textit{jointly} accounts for the CPU and WNIC energies, as well as for the migration time, computing time and migration bandwidth.
	\item By leveraging the obtained formulae for the energy-vs.-transport rate profile of the MPTCP wireless connections, we pass to formalize the problem of the minimum-energy live migration of VMs over the 5G FOGRAN of Fig. \ref{fig:scenario}. For this purpose, we formulate the problem of the minimum energy consumption of the wireless device under limited migration time and downtime in the form of settable-complexity (non-convex) optimization problem, e.g., the minimum-energy Settable Complexity Bandwidth Manager (SCBM) optimization problem. Its formulation is \textit{general enough} to embrace both MPTCP and SPTCP-based 5G FOGRAN scenarios, as well as in-band and out-band live migrations of VMs.
	\item The state of the utilized MPTCP connection of Fig. \ref{fig:scenario} may be subject to fast (and, typically, unpredictable) fluctuations, that may be induced by: (i) channel fading; (ii) device mobility; and: (iii) time-varying behavior of the migrating application. Hence, motivated by this consideration, we develop an \textit{adaptive} version of the resulting SCBM, which is capable to quickly react to the fluctuations of the state of the underlying MPTCP connection \textit{without} requiring any form of (typically, unreliable and error-prone) forecasting.   
	\item We present the results of extensive numerical tests, in order to check and compare the actual energy-vs.-migration delay performances of the proposed adaptive MPTCP-based SCBM. Overall, we anticipate that the carried out tests lead to two first main insights. First, the implementation complexity of the SCBM \textit{increases} in a linear way with the (aforementioned) number $Q$ of updated transmission rates. At the same time, both the energy consumptions and times of convergence to the steady-state \textit{decrease} for increasing values of $Q$ and approach their global minima for values of $Q$ quite low and typically limited up to $Q = 1 \div 3$. This supports the conclusion that the proposed adaptive SCBM is capable to attain a good performance-vs.-implementation complexity tradeoff. Second, in the carried out tests, the energy savings of the proposed SCBM over the state-of-the-art one in \cite{22} currently implemented by legacy Xen and KVM-based hypervisors \cite{19,42} are typically over 25\%, and reaches 70\% under strict limits on the allowed downtimes.
\end{enumerate}

A last contribution concerns the MPTCP-vs.-SPTCP energy performances under the considered hard constraints on the migration times and downtimes. In principle, due to the simultaneous powering of multiple WNICs, the network power required to sustain a MPTCP connection is (obviously) larger than the one needed by the corresponding SPTCP connection. However, due to the bandwidth aggregation effect, the opposite conclusion hold for the resulting migration time. Hence, since the migration energy equates the power-by-migration time product, an interesting still open question \cite{A10,A11} concerns the actual energy reductions possibly attained by MPTCP paradigm over the SPTCP one under the migration scenario of Fig. \ref{fig:scenario}. At this regard, we point out that:
\begin{enumerate}
\setcounter{enumi}{5}
	\item a last set of carry out numerical tests stress that the energy savings of the MPTCP connections with respect to the corresponding SPTCP ones may be significant in supporting the proposed SCBM. Specifically, the tested savings may reach 30\% and 70\% with respect to SPTCP setups that utilize only WiFi and 4G connections, when the tested migration scenario meets at least one of the following operating conditions: (i) the size of the migration VM is quite large (e.g., we say, of the order of some tens of Megabits); (ii) the tolerated downtimes are low (e.g., less than 200 ms); (iii) the dirty rate (that is, the rate of memory-write operations) of the migrated application is not negligible (e.g., the dirty rate-to-available migration bandwidth ratio is larger than 10\%).
\end{enumerate}

The rest of this paper is organized as follows. After a review of the main related work of Section \ref{sec:sec2}, in Section \ref{sec:sec3}, we shortly summarize the basic features of the pre-copy live migration, in order to point out its inherent failure recovery capability. By leveraging on the (aforementioned) unified characterization of the power-vs.-rate profile of the currently considered MPTCP CC algorithms of Section \ref{sec:sec4}, in Section \ref{sec:sec5}, we develop a delay-vs.-energy analysis, that aims to formally feature the operating conditions under which VM migration should be performed by the wireless device. Sections \ref{sec:sec6} and \ref{sec:sec7} are devoted to the formal presentation of the afforded constrained minimum-energy SCBM optimization problem, its feasibility conditions and the pursued solving approach, respectively. Afterward, in Section \ref{sec:sec8}, we develop the adaptive implementation of the proposed SCBM and discuss a number of related implementation aspects. Section \ref{sec:sec9} is devoted to the presentation of the carried out numerical tests and performance comparisons under synthetic/real-world applications and static/dynamic MPTCP wireless connections. The conclusive Section \ref{sec:sec10} reviews the main presented results and points out some hints for future research. Auxiliary analytical results are provided in the final Appendices.

Regarding the main adopted notation, the superscript: $\stackrel{\rightarrow}{\left(\cdot\right)}$ denotes vector parameters, the symbol: $\triangleq$ indicates a definition, $\log$ is the natural logarithmic, while $\delta(y)$ is the Kronecker's delta (i.g., $\delta(y) = 1$ and $\delta(y) = 0$, at $y = 0$ and $y \neq 0$, respectively). Table \ref{tab:acronyms} lists the main acronyms used in the paper.

\begin{table}[htb]
\centering
\begin{tabular}{ll}
\toprule
\textbf{Acronym}  &  \textbf{Description}  \\
\midrule
CC     &  Cloud Computing         \\
FC     &  Fog Computing           \\
MPTCP  &  MultiPath TCP           \\
PeCM   &  Pre-copy Migration      \\
PoCM   &  Post-copy Migration     \\
RAN    &  Radio Access Network    \\
RAT    &  Radio Access Technology \\
SaCM   &  Stop-and-Copy Migration \\
SCBM   &  Settable Complexity Bandwidth Manager \\
SPTCP  &  SinglePath TCP          \\
VM     &  Virtual Machine         \\
WNIC   &  Wireless Network Interface Card   \\
\bottomrule
\end{tabular}
\caption{List of the main acronyms used in the paper.}
\label{tab:acronyms}
\end{table}

\section{Related work}
\label{sec:sec2}

During the last years, the utilization of live VM migration as a network primitive functionality for attaining resource multiplexing, failure recovery and (possibly) energy reduction in centralized/distributed wired/wireless environments received an increasing attention both from industry and academy. An updated review on the main proposed solutions, open challenges and research directions is provided, for example, in \cite{A5}. However, at the best of the authors' knowledge, the wireless bandwidth management problem afforded by this paper under the 5G FOGRAN scenario of Fig. \ref{fig:scenario} deals with a still quasi unexplored topic, as also confirmed by an examination of \cite{A5} and references therein. In fact, roughly speaking, the published work mostly related to the afforded topic embraces four main research lines, namely: (i) the management of the network resources for the live VM migration; (ii) the utilization of the MPTCP for the live VM migration; (iii) the analysis and test of the MPTCP throughput in wireless/mobile scenarios; and, (iv) the development of Middleware platforms for the support of live VM migration.

Regarding the first research line, the current solution for the management of the migration bandwidth implemented by state-of-the-art hypervisors (like, for example, Xen, VMware and KVM) is the \textit{heuristic} one firstly proposed in \cite{22}. According to this heuristic solution, the migration bandwidth is linearly increased over consecutive migration rounds up to its maximum allowed value. Since the (single) goal of this heuristic is to reduce the final downtime, it neglects at all any energy-related performance index and does not enforces any constraints on the total migration time and/or downtime. Both these aspects are, indeed, accounted for by the authors of the (recent) contribution in \cite{32}. However, this last contribution focuses only on \textit{intra-data center} VM migrations over cabled Ethernet-type connections under legacy single-path newReno TCP. As a consequence, the solution in \cite{32}: (i) does \textit{not} allow to tradeoff the implementation complexity versus the resulting energy performance; and, (ii) it implements, by design, an adaptive mechanism that mainly aims to attain stable behavior in the steady-state \textit{instead of} quick reaction to the abrupt mobility-induced changes of the state of the underlying wireless connection. As consequence of these different design criteria pursued in \cite{32}, we anticipate that, in the considered wireless application scenario of Fig. \ref{fig:scenario}, the energy reductions attained by the here proposed adaptive SCBM over the solution in \cite{32} may be significant (e.g., up to 50\% under some harsh wireless scenarios).

Recent contributions along the (previously mentioned) second research line are reported in \cite{A12,A13,A14}. Specifically, the authors of \cite{A12} propose a nearby VM-based cloudlet for boosting the performance of real-time resource-stressing applications. In order to reduce the delay induced by the service initiation time, the authors of \cite{A12} develop an inter-cloudlet MPTCP-supported migration mechanism, that exploits the predicted mobility patterns of the served mobile devices in a pro-active way. Interestingly, almost zero downtimes are experienced at the destination cloudlets after the VM migrations are completed. The contribution in \cite{A13} proposes an adaptive queue-management scheduler, in order to exploit at the maximum the bandwidth aggregation capability offered by MPTCP when intra-data center live migration of VM must be carried out under delay-constraints. The topic of \cite{A14} is the exploitation of the aggregated bandwidth offered by MPTCP connections for the inter-fog wireless migration of Linux containers. For this purpose, a suitable version of the so-called Checkpoint/Restore In User-space (CRIU) migration technique is developed and its delay-performances are tested through a number of field trials. Overall, like our paper, the focus of all these contributions is on the effective exploitation of the MPTCP bandwidth aggregation capability, in order to reduce the VM migration times. However, unlike our paper, these contributions: (i) do \textit{not} afford the energy-related aspects; and, (ii) do \textit{not} consider the optimal management of the migration bandwidth under \textit{hard} QoS-induced delay-constraints.

Testing the MPTCP throughput under wireless/mobile scenarios is the main topic of the contributions in \cite{A10,A11,A15,A16,A17}. Specifically, the authors of \cite{A10} investigate the interplay between traffic partitioning and bandwidth aggregation under various MPTCP-supported wireless scenarios and, then, propose a per-component energy model, in order to quantify the energy consumed by the device's CPU during the migration process. The focus of \cite{A15} is on the 3G/WiFi handover under MPTCP and related energy measurements, while \cite{A11} and \cite{A16} report the results of a number of tests on the delay-vs.-energy MPTCP performance under different flow sizes and mobility scenarios. Lastly, \cite{A17} proposes a power model for MPTCP and tests its energy consumption under different file sizes to be transferred. Overall, like our contribution, the common topic of all these papers is on the measurements of the power/energy performance of MPTCP connections under wireless/mobile outdoor scenarios. However, unlike our work, the focus of these papers is \textit{not} on the live migration of VMs and/or on the effect of the various MPTCP CC algorithms on the energy performance of the underlying MPTCP connections.  

\begin{figure*}[htb]
\centering
\includegraphics[width=0.6\textwidth]{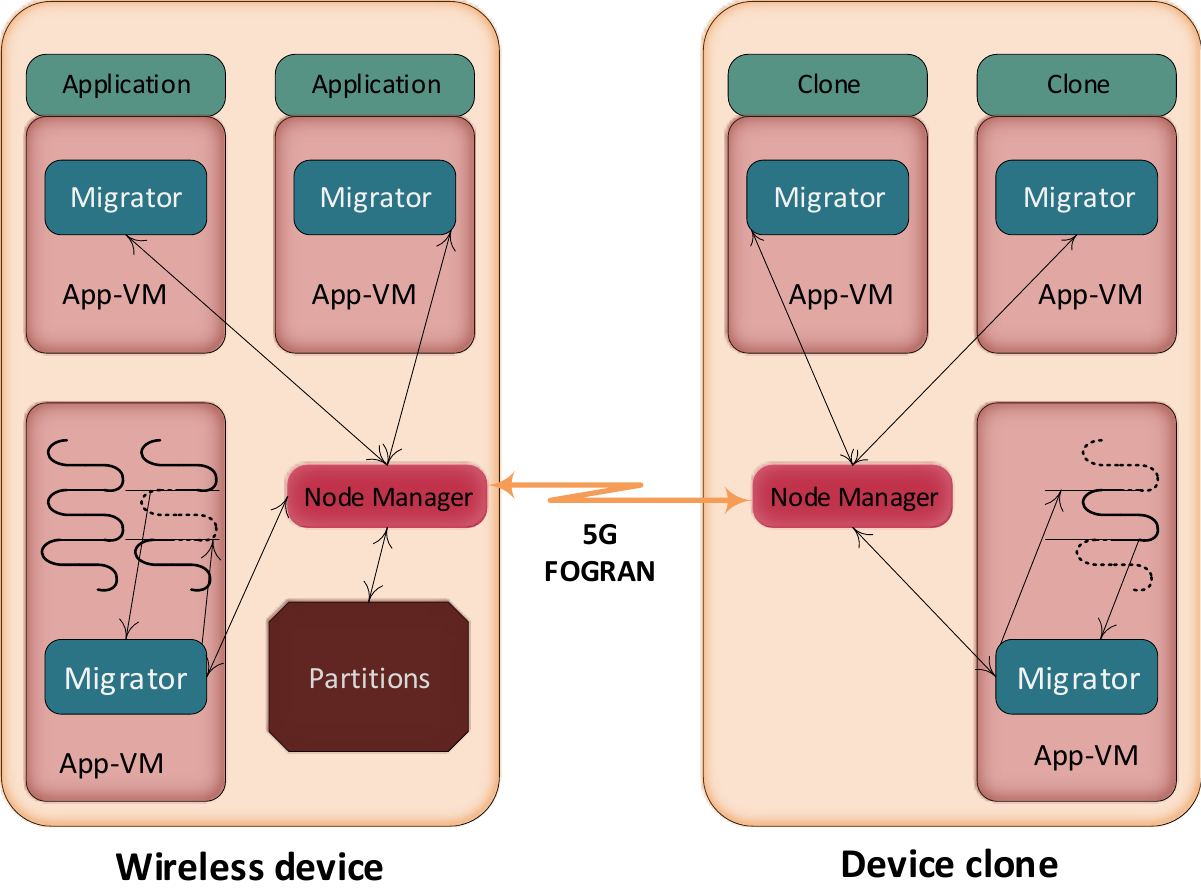}
\caption{The reference virtualized architecture considered for the middleware support of traffic offloading.}
\label{fig:Clone} 
\end{figure*}

Finally, \cite{A3} reports a quite comprehensive overview of a number of recent contributions, whose common topic is the development of software middleware platforms for the real-time support of traffic offloading. Among these contributions, we (shortly) review those reported in \cite{A18,A19,A20,A21,A22,A23,A24,A25}, that mainly match the considered virtualized scenario of Fig. \ref{fig:scenario}. A synoptic examination of these contributions allows us to enucleate the basic building blocks and related data paths that are shared by all them, as sketched by the reference virtualized architecture of Fig. \ref{fig:Clone}. 

Hence, by referring to Fig. \ref{fig:Clone}, the MAUI \cite{A18} and CloneCloud \cite{A19} frameworks develop and prototype in software two middleware platforms, that allow mobile devices to offload their virtualized tasks directly to public cloud data centers through cellular/WiFi connections. The common feature of the middleware virtualized platforms developed in \cite{A20,A21,A22,A23} is to be \textit{cloudlet}-oriented. This means that the middleware layers of all these platforms rely on single-hop WiFi-based links, in order to perform fine/coarse-grained traffic offloading to nearby small-size data centers, generally referred to as cloudlets. The Mobile Network Operator Cloud (MNOC) is, indeed, the common focus of the contributions in \cite{A24,A25}. These contributions consider a framework in which mobile devices exploit cellular 3G/4G connections, in order to offload their virtualized tasks to data centers that are directly managed by mobile network operators. Overall, like our work, all these contributions consider virtualized middleware-layer platforms for the resource augmentation of resource-limited wireless devices. However, unlike our work, the main focus of these contributions is on the design and implementation of middleware-layer software for the support of traffic migration, and they do not consider the problem of the delay-constrained and energy-efficient optimized management of the migration bandwidth.

\section{Basic live migration techniques -- A short overview}
\label{sec:sec3}

Live VM migration allows a running VM to be transferred between different physical machines without halting the migrated application. In principle, there are four main techniques for live VM migration, namely, Stop-and-Copy Migration (SaCM), Pre-Copy Migration (PeCM), and Post-Copy Migration (PoCM). They trade-off the volume of migrated data against the resulting downtime. These techniques rely on the implementation of at least one of the following three phases \cite{A5}:
\begin{itemize}
	\item[(i)] \textbf{Push phase}: the source device transfers to the destination device the memory image (e.g., the RAM content) of the migrating VM over consecutive rounds. To ensure consistency, the memory pages modified (e.g., dirtied) during this phase are re-sent over multiple rounds;
	\item[(ii)] \textbf{Stop-and-copy phase}: the VM running at the source device is halted and the lastly modified memory pages are transferred to the destination;
	\item[(iii)] \textbf{Pull phase}: the migrated VM begins to run on the destination device. From time to time, the access to the memory pages still stored by the source device is accomplished by issuing page-fault interrupts.
\end{itemize}
By design, the SaCM technique utilizes only the stop-and-copy phase. This guarantees that the volume of the migrated data equates the memory size of the migrated VM, but it generally induces long downtimes \cite{A5}. The PeCM technique implements both the push and stop-and-copy phases of the migration process. It guarantees finite migration times, tolerable downtimes and robustness against the (possible) failures of the communication link. However, it induces overhead in the total volume of the migrated data, which may be substantial under write-intensive applications \cite{A5}. The PoCM technique is composed by the stop-and-copy and pull phases of the migration process. Since only the I/O and CPU states of the source device are transferred to the destination device during the initial stop-and-copy phase, the experienced downtime is limited and no data overhead is induced. However, the resulting total migration time is, in principle, undefined and, due to the page-fault interrupts issued during the pull phase, the slowdown experienced by the migrated application may be substantial. We anticipate that the optimal bandwidth manager developed in this paper may be applied under all the mentioned migration techniques. However, in order to speed up its presentation, in the sequel, we focus on the PeCM as reference technique. The main reasons behind this choice are that: (i) PeCM is the migration technique currently implemented by a number of commercial hypervisors, such as, Xen, VMware and KVM; and, (ii) the bandwidth management framework of the PeCM technique is general enough to embrace those featured by the SaCM and PoCM techniques.

\begin{figure*}[htb]
\centering
\includegraphics[width=0.6\columnwidth]{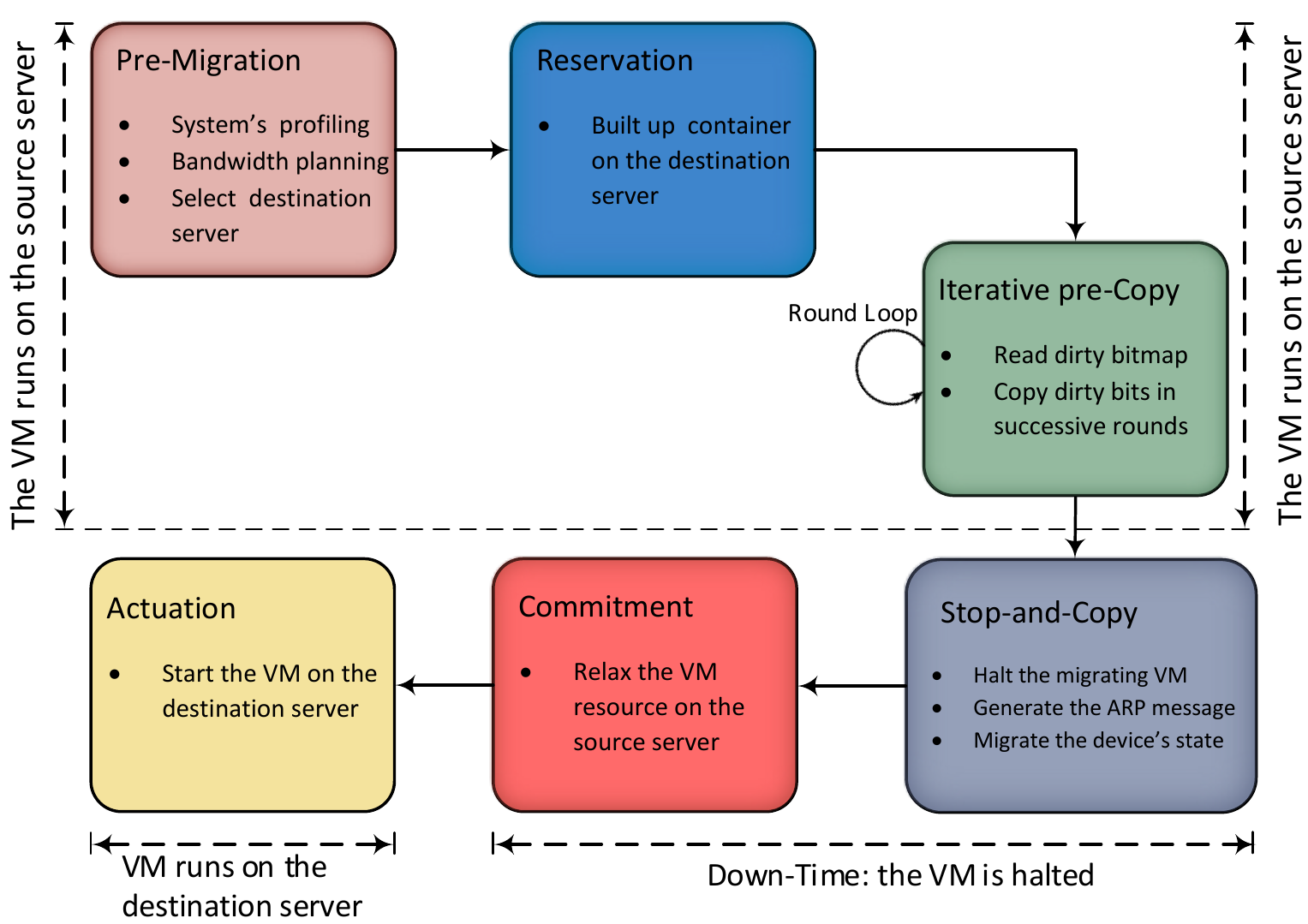}
\caption{The six stages of the Pre-Copy Migration (PeCM) technique.}
\label{fig:Live_migration}
\end{figure*}

\begin{figure}[htb]
\centering
	\includegraphics[width=0.7\textwidth]{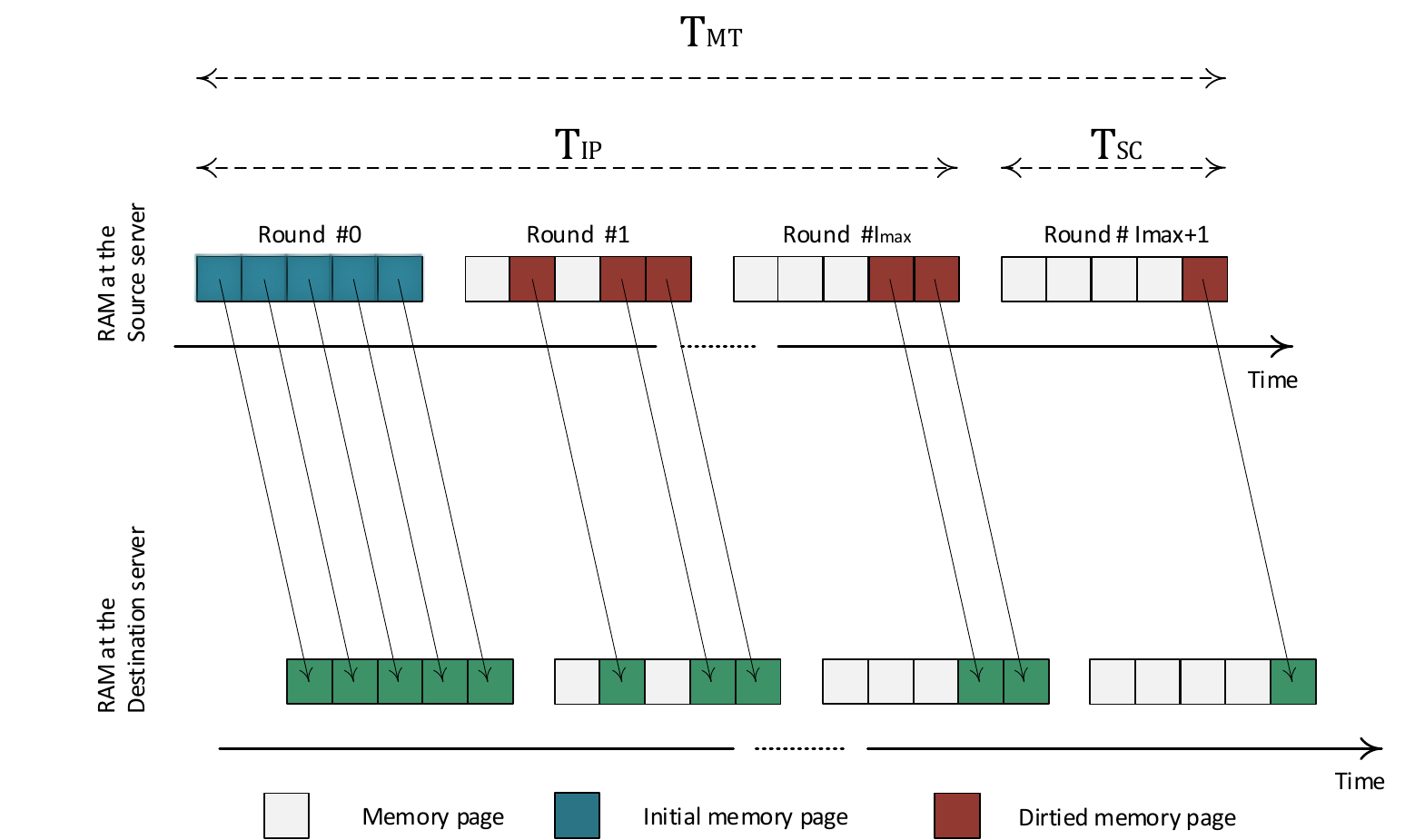}
\caption{An illustrative time-chart of the iterative pre-copy scheme of the PeCM technique.}
\label{fig:PeC} 
\end{figure}

From a formal point of view, the PeCM technique is the cascade of the six stages reported in Fig. \ref{fig:Live_migration}, namely:
\begin{enumerate}
	\item \textbf{Pre-migration}: the VM to be migrated is built up on the source device and the destination machine is selected on the destination server. This stage spans $T_{PM}$ seconds.
	\item \textbf{Reservation}: the computing/communication/storage/memory physical resources are reserved at the destination server by instantiating a large enough VM container. $T_{RE}$ (s) is the duration (in seconds) of this stage.
	\item \textbf{Iterative pre-copy}: this stage is composed by $\left( I_{MAX} + 1 \right)$ rounds and spans $T_{IP}$ seconds. During the initial round (e.g., at round \#0), the full memory content of the migrating VM is sent to the destination server. During the subsequent $I_{MAX}$ rounds (e.g., from round \#1 to round \#$I_{MAX}$), the memory pages modified (e.g., dirtied) during the previous round are re-transferred to the destination server (see Fig. \ref{fig:PeC}).
	\item \textbf{Stop-and-copy}: the migrating VM is halted and a final memory-copy round (e.g., round \#$(I_{MAX} + 1)$) is performed (see Fig. \ref{fig:PeC}). This last round spans $T_{SC}$ seconds.
	\item \textbf{Commitment}: the destination server notifies that it has received a right copy of the migrated VM. $T_{CM}$ (s) is the duration of this stage.
  \item \textbf{Re-activation}: the I/O resources and IP address are re-attached to the migrated VM on the destination server. $T_{AT}$ (s) is the needed time.
\end{enumerate}

\subsection{Formal definition of the migration delays}
\label{sec:ssec3.1}

From a formal point of view, the total migration time $T_{tot}$ (s) is the overall duration:
\begin{equation}
T_{tot} \triangleq T_{PM} + T_{RE} + T_{IP} + T_{SC} + T_{CM} + T_{AT},
\label{eq:eq3.1_P}
\end{equation}
of the (aforementioned) six stages of Fig. \ref{fig:Live_migration}, while the downtime:
\begin{equation}
T_{DT} \triangleq T_{SC} + T_{CM} + T_{AT},
\label{eq:eq3.2_P}
\end{equation}
is the time required for the execution of the corresponding last three stages. From an application point of view, $T_{tot}$ in Eq. \eqref{eq:eq3.1_P} is the time over which the source and destination servers must be synchronized, while $T_{DT}$ in Eq. \eqref{eq:eq3.2_P} is the period over which the migrating VM is halted and the user experiences a service outage. Let $R$ (Mb/s) be the transmission rate (measured at the Transport layer) during the third and fourth stages of the migration process, that is, the migration bandwidth. Since, by definition, only $T_{IP}$ and $T_{SC}$ depend on $R$, while all the remaining migration times in Eqs. \eqref{eq:eq3.1_P} and \eqref{eq:eq3.2_P} play the role of constant parameters, in the sequel, we focus on the evaluation of the (already defined) stop-and-copy time $T_{SC}$ and the resulting memory migration time $T_{MT}$, formally defined as:
\begin{equation}
T_{MT}\equiv T_{MT}(R) \triangleq T_{IP}(R) + T_{SC}(R).
\label{eq:eq3.3_P}
\end{equation}
Hence, $T_{MT}$ is the time needed for completing the memory transferring of the migrating VM, e.g., the duration of the performed $(I_{MAX} + 2)$ memory-copy rounds of Fig. \ref{fig:PeC}.  

Since the PeCM technique performs the iterative pre-copy of dirtied memory pages over consecutive rounds (see Fig. \ref{fig:PeC}), let $V_{i}$ (Mb) and $T_{i}$ (s), $i = 0, \ldots, I_{MAX} + 1$, be the volume of the migrated data and the time duration of the $i$-th migration round of Fig. \ref{fig:PeC}, respectively. By definition, $V_0$ and $T_0$ are the memory size $M_0$ (Mb) of the migrating VM and the time needed for migrating it during the $0$-th round, respectively (see the leftmost part of Fig. \ref{fig:PeC}). Hence, after indicating by $\overline{w}$ (Mb/s) the (average) memory dirty rate of the migrating application (e.g., the per-second average number of memory bits which are modified by the migrating application), from the reported definitions we have that:
\begin{equation}
V_i \triangleq \overline{w} \: T_{i-1}, \quad i = 1, \ldots, I_{MAX}+1,
\label{eq:eq3.4_P}
\end{equation} 
with $V_0 \equiv M_0$, and also:
\begin{equation}
T_i \triangleq \frac{V_i}{R_i} \equiv \frac{\overline{w}}{R_i} \: T_{i-1} = M_0 \: \overline{w}^{i} \left( \prod\limits_{m=0}^i R_m^{-1} \right), \quad i = 0, \ldots, I_{MAX}+1,
\label{eq:eq3.5_P}
\end{equation}
with $T_0 \equiv \frac{M_0}{R_0}$. As a consequence, we have that (see Eq. \eqref{eq:eq3.3_P}):
\begin{equation}
T_{MT}(R) \equiv \sum\limits_{i=0}^{I_{MAX}+1}T_i = M_0 \left[ \sum\limits_{i=0}^{I_{MAX}+1} \overline{w}^i \left( \prod\limits_{l=0}^i R_l^{-1} \right) \right],
\label{eq:eq3.6_P} 
\end{equation}
and then (see Eq. \eqref{eq:eq3.5_P}):
\begin{equation}
T_{SC}(R) \equiv T_{I_{MAX}+1} = M_0 \: \overline{w}^{I_{MAX}+1} \left( \prod\limits_{m=0}^{I_{MAX}+1} R_m^{-1} \right).
\label{eq:eq3.7_P}  
\end{equation}
In order to speed up the paper readability, the main used symbols and their meanings are listed in Table \ref{tab:taxonomy}.

\begin{table*}[htb]
\centering
\begin{tabular}{ll}
\toprule
\textbf{Symbol}            &  \textbf{Meaning}  \\
\midrule
$N$                        &  Number of WNICs equipping the wireless device  \\
$I_{MAX}$                  &  Maximum number of migration pre--copy rounds   \\
$i$                        &  Round index, with $i = 0,\ldots,(I_{MAX}+1)$   \\
$j$                        &  Integer-valued path-index of the MPTCP connection, with $j=1,\ldots,N$ \\
$n$                        &  Integer-valued iteration index                 \\
$\overline{w}$ ~(Mb/s)     &  Average memory dirty-rate of the migrated application \\
$R$ ~(Mb/s)                &  Migration bandwidth utilized by the MPTCP      \\
$\mathcal{P}$ ~(W)         &  Dynamic network power consumed by the MPTCP    \\
$M_0$ ~(Mb)                &  Memory size of the migrated VM                 \\
$\Delta_{MT}$ ~(s)         &  Maximum tolerated migration time               \\
$\Delta_{DT}$ ~(s)         &  Maximum tolerated downtime                     \\
$\widehat{R}$ ~(Mb/s)      &  Maximum migration bandwidth of the MPTCP       \\
$\mathcal{E}_{setup}$ ~(J) &  Total energy consumed by the mobile device for the parallel setup of its WNICs \\
$\mathcal{E}_{tot}$ ~(J)   &  Total setup-plus-dynamic network energy consumed by the mobile device for the VM migration \\
\bottomrule
\end{tabular}
\caption{Main symbols used in the paper and their meanings.}
\label{tab:taxonomy}
\end{table*}

\section{Power and energy analysis of MPTCP wireless connections}
\label{sec:sec4}

With reference to the 5G FOGRAN environment of Fig. \ref{fig:scenario}, main goal of this section is to develop a formal analysis of the dynamic power-vs.-transport rate profile of the MPTCP congestions that accounts (in a unified way) for the effects of the various CC algorithms currently considered in the literature \cite{35}. We anticipate that the final formulae of the carried out analysis will be directly employed in Section \ref{sec:sec6}, in order to formally define the objective function of the considered SCBM problem. In order to put the performed analysis under the right reference framework and speed-up its development, in the next two sub-sections, we shortly review some basic architectural features of the MPTCP protocol stack and the related CC algorithms.

\subsection{A short review of some key features of the MPTCP protocol stack}
\label{sec:ssec4.1}

\begin{figure}[htb]
\centering
\includegraphics[width=0.6\columnwidth]{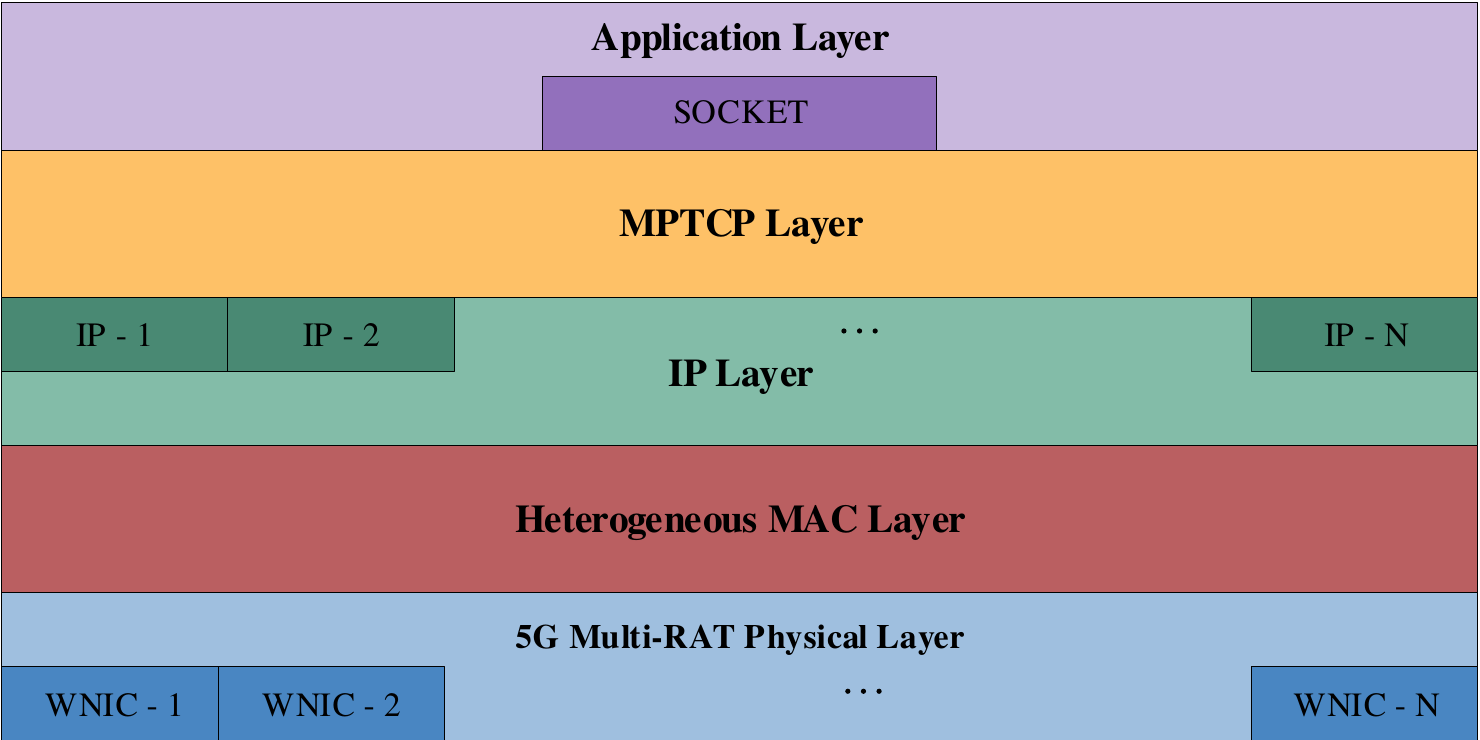}
\caption{A sketch of the MPTCP protocol stack.}
\label{fig:Protocol_Stack}
\end{figure}

Fig. \ref{fig:Protocol_Stack} sketches the main components of the MPTCP-compliant protocol stack implemented by a wireless device that is equipped with $N \geq 2$ heterogeneous WNICs \cite{A2}. Each WNIC posses an own IP address at the Network layer\footnote{Both the IPv4 and IPv6 Internet protocols may run under the MPTCP layer of Fig. \ref{fig:Protocol_Stack} \cite{A2}.}, with $IP(j)$, $j=1, \ldots, N$, denoting the IP address of the $j$-th WNIC. The MPTCP layer is a backward-compatible modification of the standard SPTCP that allows a single transport layer data flow (e.g., a single transport layer connection) to be split across the $N$ paths (also referred to as subflows) done available by the underlying Network layer. For this purpose, the Application and MPTCP layers communicate through a (single) socket that is labeled by \textit{single} port number \cite{A2}. After performing the connection setup through the exchange of SYN/ACK segments that carry out the MP\_CAPABLE option, the running application opens the socket, which, in turn, starts the Slow-Start phase of a first TCP subflow. If required by the application socket, more sub-flows can be dynamically added (resp., removed) later by issuing the MP\_JOIN/ADD\_ADDR commands of the MPTCP protocol. Each opened sub-flow is pinned to an own WNIC and labeled by the corresponding IP address (see Fig. \ref{fig:Protocol_Stack}).

The MPTCP protocol uses two levels of sequence number for guaranteeing the in-order delivering of the transported segments, namely, a single connection-level sequence number and a set of subflow-level sequence numbers. The single connection-level sequence number is the data sequence number utilized by the running application. When the MPTCP sender starts to transmit in parallel over multiple subflows, each connection-level sequence number is suitably mapped onto a corresponding subflow sequence number. Doing in so, each subflow can send own data and process the corresponding ACK messages as a regular standing-alone SPTCP connection. The MPTCP receiver uses the connection-level sequence number for orderly reassembling the received subflows. An arrived data segment is marked as ``in-sequence'' if both its subflow and connection sequence numbers are in-order. 

Out-of-sequence segments are temporarily stored by a connection-level buffer that equips the receive side of the MPTCP connection. This buffering introduces, in turn, harmful queue-induced delivering delays, that, in some limit cases, may fully offset the corresponding reduction in the transport delays gained by using MPTCP. Hence, out-of-sequence phenomena should be avoided as much as possible.

\subsection{A synoptic comparison of the MPTCP Congestion Control algorithms}
\label{sec:ssec4.2}

After exiting the setup phase, the state of the MPTCP connection enters the Slow Start (SS) phase. During this phase, each active subflow works independently from the others and applies the same SS algorithm as a regular SPTCP flow \cite{A2}. 

At the end of the SS phase, the MPTCP connection enters the Congestion Avoidance (CA) state, e.g., the steady-state. During this state, the behavior of the MPTCP connection is managed by the the corresponding CC algorithm, whose specific features make the MPTCP very different from the regular SPTCP. To begin with, the MPTCP sender implements and updates in parallel multiple mutually interacting Congestion WiNDows (CWNDs) for controlling the local traffic over each path, whilst the MPTCP receiver uses a single receive window \cite{A2}. The (possibly, coupled) updating of the CWNDs at the sender side is dictated by the actually adopted CC algorithm. A family of different MPTCP CC algorithms have been proposed over the last years, each one featured by a specific tradeoff among the contrasting targets of fair behavior, quick responsiveness in the transient-state and stable behavior in the steady-state (see \cite{35} for a formal presentation of this specific topic).
 
However, according to the analysis presented in \cite{35}, the basic control mechanism implemented by the currently proposed CC algorithms may be shortly described as follows. Let $W_d(j)$, $j=1, \ldots, N$, be the size (measured in Mb) of the CWND of the $j$-th subflow and let:
\begin{equation}
W_d \triangleq \sum_{j=1}^N W_d(j), \quad (Mb)
\label{eq:eqx.1}
\end{equation}
be the resulting size of the total CWND. On each subflow $j$, the MPTCP source increases (resp., decreases) the corresponding $j$-th CWND by a number of Megabits equal to: $I_c(j)$ (resp., $D_c(j)$) at the return of each ACK message (resp., at the detection of each segment loss). These increments and decrements are dictated by the following general expressions \cite{35}:
\begin{equation}
W_d(j) = W_d(j) + I_c(j),
\label{eq:eqx.2}
\end{equation}
for each ACK on the $j$-th subflow, and 
\begin{equation}
W_d(j) = W_d(j) - D_c(j),
\label{eq:eqx.3}
\end{equation}
for each segment loss on the $j$-th subflow.

After denoting by $RTT(j)$ the round-trip-time (measured in seconds) of the $j$-th subflow, let: $R(j) \triangleq W_d(j) / RTT(j)$ (Mb/s) be the corresponding net transmission rate (e.g., the throughput), so that:
\begin{equation}
R \triangleq \sum_{j=1}^N R(j), \quad (Mb/s)
\label{eq:eqx.4}
\end{equation}
is the resulting total transmission rate (e.g., the total throughput) of the overall MPTCP connection. Hence, as detailed in Table \ref{tab:tabB} \cite{35}, the specific feature of each CC algorithm depends on the way in which the increment/decrement terms in \eqref{eq:eqx.2} and \eqref{eq:eqx.3} are actually computed. For comparison purpose, the last row of Table \ref{tab:tabB} reports the basic CC features of the commodity NewReno SPTCP.

\begin{table*}[htb]
\centering
\begin{tabular}{llll}
\toprule
\textbf{Acronym of the CC algorithm} &  \textbf{Expression of} $\mathbf{I_c(j)}$  &  \textbf{Expression of} $\mathbf{D_c(j)}$  &  \textbf{Auxiliary notations}  \\
\midrule
\textbf{EWTCP} \cite{A26}             & $a/W_d(j)$  &  $W_d(j)/2$  &  $a = 1$  \\[2ex]
\textbf{Semicoupled MPTCP} \cite{A27} & $a/W_d$, $\forall j = 1, \ldots, N$ & $W_d(j)/2$ & $a = N$  \\[3ex]
\textbf{Max MPTCP} \cite{50}          & $\min\left\{ \frac{1}{W_d(j)}; \: \frac{\max_k \left\{ W_d(k) / RTT^2(k) \right\}}{\left( \sum_{k=1}^N W_d(k) / RTT(k) \right)^2}  \right\}$  &  $W_d(j)/2$  &  $a = N^2$  \\[3ex]
\textbf{Balia MPTCP} \cite{35}        & $\left( \frac{R(j)}{RTT(j) \left( \sum_{k=1}^N R(k) \right)^2} \right) \times \left( \frac{1 + \gamma(j)}{2} \right) \times \left( \frac{4 + \gamma(j)}{5} \right)$  &  $\left( W_d(j)/2 \right) \times \min\left\{\gamma(j); \: 1.5 \right\}$  & $a = N^2$; ~$\gamma(j) = \frac{\max_k\left( R(k) \right)}{R(j)}$  \\[3ex]
\textbf{NewReno SPTCP}                & $1/W_d$  &  $W_d(j)/2$  &  $N = 1$  \\[0.5ex]
\bottomrule
\end{tabular}
\caption{Expressions of the increments/decrements of the per-flow congestion windows of some MPTCP CC algorithms \cite{35}. SPTCP mantains, by design, a single subflow.}
\label{tab:tabB}
\end{table*}

Detailed analysis and comparisons of the fairness/responsiveness/stability properties of the MPTCP CC algorithms of Table \ref{tab:tabB} are reported in \cite{35} and, then, they will be not replicated here.  We shortly limit to remark that: (i) all the considered CC algorithms react against per-subflow segment loss events by halving the current size of the corresponding per-subflow congestion window; (ii) the EWTCP algorithm applies the NewReno SPTCP increment policy on each subflow independently. For this reason, as also pointed out in \cite{35} and \cite{50}, it is the most reactive in coping with the abrupt changes typically affecting the state of wireless/mobile MPTCP connections; (iii) the Semicoupled, Max and Balia algorithms introduce various forms of coupling in updating the sizes of the congestion windows, in order to be more fair (like as the Balia CC algorithm) and/or more stable in the steady-state (like as the Semicoupled CC algorithm). Hence, these last CC algorithms appear to be more suitable for managing wired (possibly, multi-hop) MPTCP connections (see \cite{A2} and Section V of \cite{35}).

\subsection{Unified analysis of the power-vs.-rate performance of the MPTCP under 5G scenarios}
\label{sec:ssec4.3}

The total network energy $\mathcal{E}_{tot}$ (J) consumed by the multiple WNICs equipping the wireless device of Fig. \ref{fig:scenario} during each VM migration process is the summation of a static (e.g., setup) part: $\mathcal{E}_{setup}$ (J), and a dynamic part: $\mathcal{E}_{dyn}$ (J). By design, $\mathcal{E}_{setup}$ does not depend on the set of the per-subflow transmission rates and equates the summation:
\begin{equation}
\mathcal{E}_{setup} \triangleq \sum_{j=1}^N \mathcal{E}_{setup}(j),
\label{eq:eqx.5}
\end{equation}
of the setup energies needed by the involved WNICs to establish and maintain the MPTCP connection \cite{A11}. The dynamic portion $\mathcal{E}_{dyn}$ of the consumed energy depends on the per-subflow transmission rates in \eqref{eq:eqx.4} through the corresponding total dynamic power $\mathcal{P}$ (W), that, in turn, is given by the following summation:
\begin{equation}
\mathcal{P} \triangleq \sum_{j=1}^N \mathcal{P}(j),
\label{eq:eqx.6}
\end{equation}
where $\mathcal{P}(j)$, $j = 1, \ldots, N$, is the dynamic power consumed by the $j$-th WNIC for sustaining its transport rate $R(j)$ (see Eq. \eqref{eq:eqx.4}). Now, the key point to be remarked is that the (aforementioned) MPTCP CC algorithms may introduce coupling effects, so that each $\mathcal{P}(j)$ may be a function of the \textit{overall} spectrum: $\left\{ R(k), k = 1, \ldots, N \right\}$ of transport rates, with the analytical form of the function that may depend, in turn, on \textit{both} the subflow index $j$ and the actually considered CC algorithm. Interestingly, a suitable exploitation of the expressions of Table \ref{tab:tabB} allows us to arrive at the following \textit{unified} formula for the profile of the total dynamic power $\mathcal{P}$ in \eqref{eq:eqx.6} as a function of transport rates $\left\{ R(k), k = 1, \ldots, N \right\}$ (see the Appendix \ref{sec:appA} for the derivation): 
\begin{equation}
\mathcal{P} = \mathcal{A} \left( \sum_{j=1}^N \tau(j) \left( R_j \right)^c \right).
\label{eq:eqx.7}
\end{equation}
In Eq. \eqref{eq:eqx.7}, the expressions of $\mathcal{A}$, $c$ and $\left\{ \tau(j), j = 1, \ldots, N \right\}$ depend on the considered CC algorithms. They are detailed in Table \ref{tab:tabC}, where: (i) $MSS$ (Mb) is the maximum size of a MPTCP segment; (ii) $\alpha > 1$ is a dimensionless power exponent; (iii) the constant $a$ is given by the forth column of the above Table \ref{tab:tabB}; and, (iv) $\Omega(j)$, $j = 1, \ldots, N$, represents the noise power-to-coding gain ratio of the transmission path sustained by the $j$-th WNIC. According to Eq. \eqref{eq:eqA.3} of the Appendix \ref{sec:appA}, it is measured in ($W^{1/\alpha}$) and it is formally defined as follows:
\begin{equation}
\Omega(j) \triangleq Pr_{loss}(j) \left( \mathcal{P}(j)^{1/\alpha} \right),
\label{eq:eqx.8}
\end{equation}
where $Pr_{loss}(j)$ is the (dimensionless) segment loss probability experienced by the $j$-th subflow.

\begin{table*}[htb]
\centering
\begin{tabular}{lllc}
\toprule
\textbf{Acronym of the CC algorithm} &  \textbf{Expression of} $\mathcal{A}$  &  \textbf{Expression of} $\boldsymbol{\tau(j)}$  &  \textbf{Expression of} $\mathbf{c}$  \\
\midrule\\[-2ex]
\textbf{EWTCP} \cite{A26}             & $1$  &  $\left( \left( RTT(j) \times \Omega(j) \right) / \left( MSS \sqrt{2a} \right) \right)^\alpha$  &  $2 \alpha$  \\[4ex]
\textbf{Semicoupled MPTCP} \cite{A27} & $\left( \frac{\sum_{k=1}^N RTT(k) R(k)}{2a MSS^2} \right)^\alpha$  & $\left( RTT(j) \times \Omega(j) \right)^\alpha$ & $\alpha$  \\[4ex]
\textbf{Max MPTCP} \cite{50}          & $\left( \frac{ \left( \sum_{k=1}^N RTT(k) \right)^2}{2a MSS^2 \times \max_k \left( \frac{R(k)}{RTT(k)} \right)} \right)^\alpha$  & $\left( RTT(j) \times \Omega(j) \right)^\alpha$  &  $\alpha$  \\[4ex]
\textbf{Balia MPTCP} \cite{35}        & $\left( \frac{  \left( \sum_{k=1}^N RTT(k) \right)^2}{0.4a MSS^2 \times \max_k \left( R(k) \right)} \right)^\alpha$  & $\left( RTT(j)^2 \times \Omega(j) \right)^\alpha$  & $\alpha$  \\[4ex]
\textbf{NewReno SPTCP}                & $1$  &  $\left( \left( RTT/MSS \right) \times \left( \Omega/2 \right) \right)^\alpha$  &  $\alpha$  \\[0.5ex]
\bottomrule
\end{tabular}
\caption{Expressions of the parameters involved by the unified power-vs.-rate formula in \eqref{eq:eqx.7} for the MPTCP.}
\label{tab:tabC}
\end{table*}

Regarding the derived dynamic power-vs.-transport rate formula in \eqref{eq:eqx.7}, two main remarks are in order. First, it is composed by a linear superposition of $N$ power terms, each one involving the transport rate of a MPTCP subflow. This power-like form is compliant, indeed, with the results reported by a number of measurement/test-based studies \cite{A10,A11,A16,A17,44,45,43}. Second, the key features of our formula in \eqref{eq:eqx.7} are that: (i) it holds in general, e.g., regardless from the specifically considered CC algorithm; and, (ii) its parametric form allows us to formally account in a direct way for the effect induced on the consumed power by the actually considered CC algorithm (see Table \ref{tab:tabC}).

\subsection{The case of load-balanced MPTCP connections}
\label{sec:ssec4.4}

A direct inspection of Eqs. \eqref{eq:eqx.7} and \eqref{eq:eqx.8} leads to the conclusion that, when the $N$ subflows are load balanced, e.g.:
\begin{equation}
R(1) = R(2) = \ldots = R(N) \equiv \frac{R}{N},
\label{eq:eqx.9}
\end{equation}
then, the power-rate formula in \eqref{eq:eqx.7} reduces to the following \textit{monomial} one:
\begin{equation}
\mathcal{P} = K_0 \left( R \right)^\alpha.
\label{eq:eqx.10}
\end{equation}
In Eq. \eqref{eq:eqx.10}, $K_0$ is measured in $\left( \frac{W}{\left( Mb/s \right)^\alpha} \right)$ and its formal expressions are reported in Table \ref{tab:tabD} for the (above considered) CC algorithms.

\begin{table}[htb]
\centering
\begin{tabular}{ll}
\toprule
\textbf{Acronym of the CC algorithm}  &  \textbf{Expression of} $\mathbf{K_0}$ (W/(Mb/s)$^\alpha$)   \\
\midrule\\[-2ex]
\textbf{EWTCP} \cite{A26}             & $\left( \frac{1}{\sqrt{2a} \times N \times MSS} \right)^\alpha \times \left( \sum_{k=1}^N \left( RTT(k) \times \sqrt{\Omega(k)} \right)^\alpha \right)$   \\[4ex]
\textbf{Semicoupled MPTCP} \cite{A27} & $\left( \frac{\sqrt{\sum_{k=1}^N RTT(k)}}{N \times MSS \times \sqrt{2a}} \right)^\alpha \times \left( \sum_{k=1}^N \left( RTT(k) \times \Omega(k) \right)^{\alpha/2} \right)$    \\[4ex]
\textbf{Max MPTCP} \cite{50}          & $\left( \frac{\min_k \left( RTT(k) \right)}{2a \times MSS^2} \right)^{\alpha/2} \times \left( \sum_{k=1}^N \left( RTT(k) \times \Omega(k) \right)^{\alpha/2} \right)$   \\[4ex]
\textbf{Balia MPTCP} \cite{35}        & $\left( \frac{1}{0.4a \times MSS^2} \right)^{\alpha/2} \times \left( \sum_{k=1}^N \left( RTT(k) \times \sqrt{\Omega(k)} \right)^\alpha \right)$    \\[4ex]
\textbf{NewReno SPTCP}                & $\left( \left( RTT/MSS \right) \times \left( \Omega/2 \right) \right)^\alpha$    \\[0.5ex]
\bottomrule
\end{tabular}
\caption{Expressions of $K_0$ in Eq. \eqref{eq:eqx.10} for the the case of load-balanced MPTCP.}
\label{tab:tabD}
\end{table}

The expression in \eqref{eq:eqx.10} is central for the future developments of our paper, and merits, indeed, three main remarks. 

First, from a formal point of view, an examination of the general expressions reported in the 3-rd column of Table \ref{tab:tabC} points out that the equal-balanced condition in \eqref{eq:eqx.9} is met when the values assumed by the $j$-indexed products: $RTT(j) \times \sqrt{\Omega(j)}$ (resp., the products: $RTT(j) \times \Omega(j)$) do \textit{not} depend on the subflow index $j$ under the EWTCP and Balia MPTCP (resp., under the Semicoupled and Max MPTCP). The measurement-based studies reported, for example, in \cite{A11,A15,A17,50,44,45} support the conclusion that, in practice, these formal conditions are quite well met under 3G/4G cellular and WiFi connections, mainly because the average segment loss probability (resp., the average round-trip-time) of WiFi connections is typically one order of magnitude larger (resp., lower) than the corresponding one of 3G/4G cellular connections.

Second, an inspection of the expressions of Table \ref{tab:tabD} leads to the conclusion that, at $\alpha > 1$, the $K_0$ power-coefficient in \eqref{eq:eqx.10} scales down for increasing number $N$ of the available load-balanced subflows as:
\begin{equation}
K_0 \propto \left( \frac{1}{N^{\alpha - 1}} \right).
\label{eq:eqx.11}
\end{equation}
In the sequel, we refer to the scaling down behavior in \eqref{eq:eqx.11} as the ``multipath gain'' of the equal-balanced MPTCP. Intuitively, it arises from the fact that, at $\alpha > 1$, the dependence of the dynamic power $\mathcal{P}$ in \eqref{eq:eqx.7} on each subflow transport rate is of convex-type, so that equal-balanced transport rates reduce the resulting total power consumption. 

Third, unbalanced transport rates increase the chance of out-of-sequence segment arrivals at the receive end of the MPTCP connection, that, in turn, give arise to harmful queue-induced delivering delays (see the last remark of the above Section \ref{sec:ssec4.1}). 

Overall, motivated by the above remarks, in the sequel, we directly focus on the equal-balanced case, and, then, we exploit the expression in \eqref{eq:eqx.10} for the formal characterization of the resulting total migration energy. At this regard, we anticipate that the topic regarding the on-line (e.g., real-time) profiling of the $K_0$ and $\alpha$ parameters present in \eqref{eq:eqx.10} will be afforded in Section \ref{sec:ssec8.1}, whilst some additional remarks on the case of unbalanced MPTCP connections will be reported in Section \ref{sec:sec10}.

\section{To migrate or not to migrate -- A delay-vs.-energy analysis}
\label{sec:sec5}

The two-fold task of the migration module at the mobile device of Fig. \ref{fig:Clone} is to plan: (i) ``when'' (e.g., under which operating conditions) performing the migration; and, (ii) ``how'' manage the planned migration. Although the focus of this paper is on the second question, in this section we shortly address the first one. At this regard, we observe that the final goal of the mobile-to-fog migration process would be to reduce as much as possible both the execution time of the migrating application and the corresponding energy consumed by the mobile device.

In the sequel, we develop a time-energy analysis of the mobile-to-fog migration costs under the considered 5G FOGRAN scenario of Fig. \ref{fig:scenario}, in order to formally characterize the traded-off operating conditions under which the VM migration is worth.

In order to characterize and compare the local-vs.-remote execution times of the migrating VM, let $\gamma \triangleq \left( w_{VM}/M_0 \right)$ be the ratio between the workload $w_{VM}$ (bit) to be processed by the migrating VM and its (previously) defined size $M_0$ (bit). Hence,  after indicating by $s_{com}^{Mob}$ (bit/s) the processing (e.g., computing) speed at the mobile device, the resulting time $T_{exe}^{Mob}$ (s) for the local execution of the VM at the mobile device equates:
\begin{equation}
T_{exe}^{Mob} \triangleq \frac{w_{VM}}{s_{com}^{Mob}} \equiv \frac{\gamma M_0}{s_{com}^{Mob}}.
\label{eq:eq3.13}
\end{equation}
However, after indicating by $s_{com}^{Fog}$ (bit/s) the processing speed of the device clone at the Fog node of Fig. \ref{fig:scenario}, when the migration is performed, the resulting execution time $T_{exe}^{Fog}$ (s) at the Fog node is the summation:
\begin{equation}
T_{exe}^{Tog} = T_{tot} + \frac{\gamma M_0}{s_{com}^{Fog}},
\label{eq:eq3.14}
\end{equation}
of the total migration time in Eq. \eqref{eq:eq3.1_P} and the execution time: $w_{VM}/s_{com}^{Fog} \equiv \gamma M_0/s_{com}^{Fog}$, of the migrated VM at the Fog node. As a consequence, the VM migration would reduce the execution time when $T_{exe}^{Mob} \geq T_{exe}^{Fog}$, that is, when the corresponding migration time $T_{MT}$ in Eq. \eqref{eq:eq3.3_P} meets the following upper bound: 
\begin{equation}
T_{MT} \leq \max\left\{ 0; \gamma M_0 \left( \frac{1}{s_{com}^{Mob}} - \frac{1}{s_{com}^{Fog}} \right) - \left( T_{PM} + T_{RE} + T_{CM} + T_{AT} \right) \right\},
\label{eq:eq3.15}
\end{equation}
where the $\max\left\{\cdot\right\}$ operator in \eqref{eq:eq3.15} accounts for the fact that, by definition, $T_{MT}$ is non-negative.

Passing to consider the energy consumptions induced by the local and remote workload processing, let $\mathcal{P}_{com}^{Mob}$ (Watt) be the power consumed by the mobile device when it runs at the (aforementioned) processing speed $s_{com}^{Mob}$. Hence, the energy $\mathcal{E}_{com}^{Mob}$ (J) wasted by the mobile device for the local execution of the workload $w_{VM}$ equates:
\begin{equation}
\mathcal{E}_{com}^{Mob} = \mathcal{P}_{com}^{Mob} \left( \frac{w_{VM}}{s_{com}^{Mob}} \right) \equiv \mathcal{P}_{com}^{Mob} \left( \frac{\gamma M_0}{s_{com}^{Mob}} \right).
\label{eq:eq3.16}
\end{equation}
The corresponding energy $\mathcal{E}_{com}^{Fog}$ (J) consumed by the mobile device when the VM is migrated and remotely processed at the Fog node is the summation of three contributions. The first one is the already considered energy $\mathcal{E}_{tot}$ consumed by the device-to-fog migration. The second one accounts for the idle energy: $\mathcal{E}_{idle}^{Mob}$ (J) consumed by the mobile device in the idle state during the remote execution of the migrated workload at the Fog node. It equates: $\mathcal{E}_{idle}^{Mob} = \mathcal{P}_{idle}^{Mob} \left( w_{VM}/s_{com}^{Fog} \right) \equiv \mathcal{P}_{idle}^{Mob} \left( \gamma M_0/s_{com}^{Fog} \right)$, where $\mathcal{P}_{idle}^{Mob}$ (W) is the power consumed by the mobile device in the idle state. The third component accounts for the energy: $\mathcal{E}_R^{Mob}$ (J) consumed by the mobile device, when it receives the processed workload sent back by the Fog node. It equates: $\mathcal{E}_R^{Mob} = \mathcal{P}_R^{Mob} \left( \mathcal{T} M_0 / R_{D} \right)$, where: (i) $\mathcal{P}_R^{Mob}$ (W) is the network power consumed by all WNICs of the mobile device under the receive operating mode; (ii) the positive and dimensionless coefficient $\mathcal{T}$ is the relative size of the processed data, so that the product: $\mathcal{T} M_0$ (bit) is the size of the processed data sent back by the Fog node; and, (iii) $R_D$ (bit/s) is the aggregated download bandwidth used by the Fog node for the fog-to-device data transfer over the 5G FOGRAN of Fig. \ref{fig:scenario}. Overall, we have that:
\begin{equation}
\mathcal{E}_{com}^{Fog} = \mathcal{E}_{tot} + \mathcal{E}_{idle}^{Mob} + \mathcal{E}_R^{Mob} \equiv \mathcal{E}_{tot} + \mathcal{P}_{idle}^{Mob} \left( \frac{\gamma M_0}{s_{com}^{Fog}} \right) + \mathcal{P}_R^{Mob} \left( \frac{\mathcal{T} M_0}{R_D} \right).
\label{eq:eq3.17}
\end{equation}
Therefore, the VM migration is energy saving when: $\mathcal{E}_{com}^{Mob} \geq \mathcal{E}_{com}^{Fog}$, that is, when the migration energy $\mathcal{E}_{tot}$ meets the following inequality (see Eqs. \eqref{eq:eq3.16} and \eqref{eq:eq3.17}):
\begin{equation}
\mathcal{E}_{tot} \leq \max\left\{ 0; \: M_0 \left[ \left( \frac{\mathcal{P}_{com}^{Mob}}{s_{com}^{Mob}} - \frac{\mathcal{P}_{idle}^{Mob}}{s_{com}^{Fog}} \right) \gamma - \frac{\mathcal{T} \mathcal{P}_R^{Mob}}{R_D} \right] \right\},
\label{eq:eq3.18}
\end{equation}
where the $\max\left\{\cdot\right\}$ operator in \eqref{eq:eq3.18} accounts for the non-negativity of $\mathcal{E}_{tot}$. As a consequence, VM migration is time (resp., energy) efficient when the memory migration time $T_{MT}$ (resp., the migration energy $\mathcal{E}_{tot}$) meets the upper bound in Eq. \eqref{eq:eq3.15} (resp., in Eq. \eqref{eq:eq3.18}).

An examination of the bounds in \eqref{eq:eq3.15} and \eqref{eq:eq3.18} allow us to address three questions about ``which'' VM to migrate, ``when'' performing the migration and ``how'' manage the (planned) migration. At this regard,  three main remarks are in order.

First, we observe that the (previously defined) parameters $M_0$, $\gamma$ and $\mathcal{T}$ depend only on the migrating VM and the corresponding migrated application. Hence, about the question concerning ``which'' VM to migrate, the lesson supported by the bounds in \eqref{eq:eq3.15} and \eqref{eq:eq3.18} is that the migrated VM should offer large workload $w_{VM}$ (e.g., large values of the product: $M_0 \gamma$), but quite limited values of the size of the processed data sent back from the Fog node (e.g, low values of the $\mathcal{T}$ coefficient in Eq. \eqref{eq:eq3.18}).

Second, about the question concerning ``when'' performing the migration, the reported bounds point out that the 5G FOGRAN infrastructures which are more effective for supporting VM migration should be equipped with (very) speed computing servers (i.e., $s_{com}^{Fog} >> s_{com}^{Mob}$) and broad Fog-to-mobile download bandwidths (i.e., large $R_D$, see Eq. \eqref{eq:eq3.18}). Furthermore, the mobile devices that receive more benefit from the migration process are those equipped with power-consuming slow CPUs (e.g., low processing speeds $s_{com}^{Mob}$ and high computing powers $\mathcal{P}_{com}^{Mob}$; see Eqs. \eqref{eq:eq3.15} and \eqref{eq:eq3.18}) and power-efficient WNICs (e.g., low idle and receive powers $\mathcal{P}_{idle}^{Mob}$ and $\mathcal{P}_R^{Mob}$, respectively; see Eq. \eqref{eq:eq3.18}). Typical values of $\mathcal{P}_{com}^{Mob}$, $\mathcal{P}_{idle}^{Mob}$, $\mathcal{P}_R^{Mob}$, $s_{com}^{Mob}$ and $R_D$ for a spectrum of mobile devices and wireless access technologies are reported, for example, in \cite{44,45,43}.

Finally, about the question concerning ``how'' managing the planned VM migration, we observe that, in principle, under a given operating scenario (e.g., at fixed values of the bounds in Eqs. \eqref{eq:eq3.15} and \eqref{eq:eq3.18}), task of the migration module of Fig. \ref{fig:Clone} is to manage the device-to-Fog migration bandwidth of Fig. \ref{fig:scenario}, in order to reduce as much as possible both the memory migration time and migration energy in Eqs. \eqref{eq:eq3.15} and \eqref{eq:eq3.18}. However, an examination of the (previously reported) Eq. \eqref{eq:eq3.6_P} points out that lower values of $T_{MT}$ require higher migration bandwidths that, in turn, give arise to higher migration energies. This forbids to pursue the \textit{simultaneous} minimization of $T_{MT}$ and $\mathcal{E}_{tot}$. Hence, after observing that stretching as much as possible the battery life of the wireless devices remains a main target of the 5G paradigm, in the sequel, we approach the question concerning ``how'' managing the planned VM migration by formulating and solving a settable-complexity constrained optimization problem that \textit{dynamically} minimizes at run-time the migration energy $\mathcal{E}_{tot}$, while simultaneously upper bounding (in a hard way) the resulting memory migration time in \eqref{eq:eq3.6_P}. In principle, the resulting SCBM provides the (formally optimal) response to the question about ``how'' manage the planned migration in an energy and time efficient way.

\section{The afforded settable-complexity minimum-energy bandwidth optimization problem}
\label{sec:sec6}

\begin{figure}[htb]
\centering
	\includegraphics[width=0.7\columnwidth]{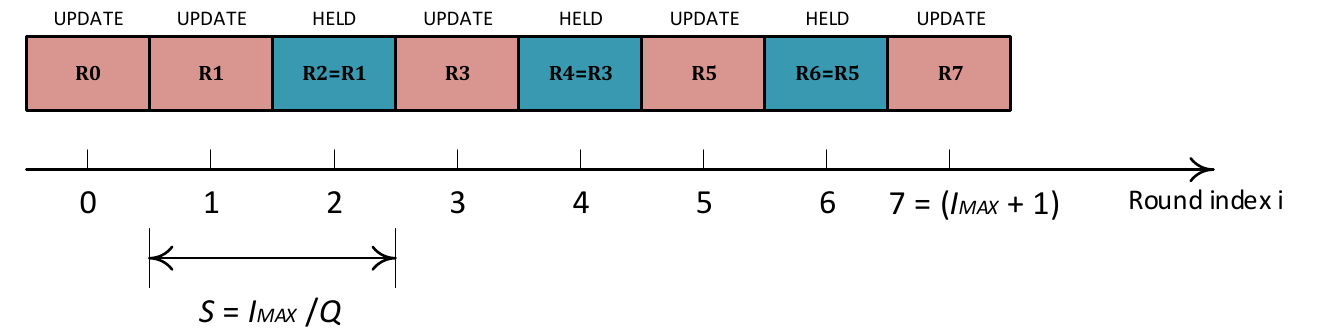}
\caption{A sketch of the main idea behind the proposed SCBM.}
\label{fig:Migration_Ov} 
\end{figure}

We pass, now, to formalize the afforded SCBM constrained optimization problem. The idea behind the proposed SCBM is quite simple. By referring to Fig. \ref{fig:PeC}, in addition to $R_0$ and $R_{I_{MAX}+1}$, we update $Q$ out of the $I_{MAX}$ rates of the pre-copy rounds that are evenly spaced apart by $S \triangleq \frac{I_{MAX}}{Q}$ rounds over the round index set: $\left\{ 1, 2, \ldots, I_{MAX} \right\}$. For this purpose, we perform the partition of the round index set: $\left\{ 1, 2, \ldots, I_{MAX} \right\}$ into $Q$ not overlapping contiguous clusters of size $S$, as shown in Fig. \ref{fig:Migration_Ov}.

The first rate $R_{jS + 1}$, $j = 0,\ldots,Q-1$, of each cluster is updated, while the remaining $(S - 1)$ rates are set to: $R_{jS + 1}$, that is $R_i \equiv R_{jS + 1}$, for $i = jS+2, jS+3, \ldots, (j + 1)S$. Fig. \ref{fig:Migration_Ov} illustrates the framework of the updated/held migration rates for the case of $I_{MAX} = 6$ and $Q = 3$. In this example, $R_0, R_1, R_3, R_5$ and $R_7$ are the: $Q + 2 = 5$ migration rates to be updated, while $R_2, R_4$ and $R_6$ are the: $(I_{MAX} - Q) = 3$ migration rates which are held, e.g., $R_2 \equiv R_1$, $R_4 \equiv R_3$ and $R_6 \equiv R_5$.

In order to formally introduce the SCBM, let $I_{MAX}$ be the maximum number of the performed pre-copy rounds, so that the overall set of the $(I_{MAX} + 2)$ migration rates is (see Fig. \ref{fig:Migration_Ov}): $\left\{ R_i, 0 \leq i \leq I_{MAX} + 1 \right\}$. Let $Q$ be the integer-valued number of the pre-copy migration rates we select to update and let $S \equiv \frac{I_{MAX}}{Q}$ be the resulting integer-valued size of the rate clusters (see Fig. \ref{fig:Migration_Ov}). Formally speaking, $Q$ and $S$ are to be selected according to the following two rules:
\begin{itemize}
	\item[i)] if $I_{MAX} \geq 1$, $Q$ must be integer-valued and falling into the interval: $1 \leq Q \leq I_{MAX}$. Furthermore, $Q$ must be selected so that the resulting ratio: $S \equiv \frac{I_{MAX}}{Q}$ is also integer-valued;
	\item[ii)] if $I_{MAX} = 0$, the set $\left\{ R_1, R_2, \ldots, R_{I_{MAX}} \right\}$ of the pre-copy rates is the empty one, so that we must pose: $Q = 1$ and $S = 0$.
\end{itemize} 
Under these settings, the proposed SCBM is formally defined as follows: 
\begin{itemize}
	\item[i)] the SCBM updates the following set of $(Q + 2)$ migration rates: $\Xi \triangleq \left\{ R_{0}; R_{jS+1}, \: j = 0, 1, \ldots, (Q - 1);\: R_{I_{MAX}+1} \right\}$;
	\item[ii)] the SCBM sets the remaining $(I_{MAX} - Q)$ migration rates as follows: $R_i \equiv R_{jS+1}$, $\forall \, i = (jS + 2),\: \ldots, \:(j + 1)S$, and: $\forall \, j = 0, 1, 2, \ldots, (Q - 1)$.
\end{itemize}	
Doing so, the expression in Eq. \eqref{eq:eq3.2_P} for the downtime becomes:
\begin{equation}
T_{DT} = M_{0} \left\{ \frac{1}{R_0} \delta\left( 1 + I_{MAX} \right) + \left\{ \left( \overline{w} \right)^{\left(1 + I_{MAX} \right)} \left( \frac{1}{R_0  R_{I_{MAX}+1}} \right) \left[ \prod\limits_{k=0}^{Q-1} \left( \frac{1}{R_{kS+1}} \right)^{S} \right] \right\} \times \left( 1 - \delta\left( 1+I_{MAX} \right) \right) \right\}.
\label{eq:eq3.29_P}
\end{equation}
Furthermore, the resulting total migration time $T_{TM}$ of Eq. \eqref{eq:eq3.6_P} assumes the following closed-form expression under the SCBM:
\begin{equation}
\begin{split}
T_{TM} &= M_{0} \left\{ \frac{1}{R_0} + \left( \overline{w} \right)^{1 + I_{MAX}} \left( R_0 R_{I_{MAX}+1} \right)^{-1} \left[ \prod\limits_{m=0}^{Q-1} \left( R_{mS+1} \right)^{-S} \right] \right. \\
 & \hphantom{=} + \left. \left( 1 - \delta\left( I_{MAX} \right) \right) \frac{1}{R_0} \left\{ \sum\limits_{k=0}^{Q-1} \sum\limits_{l=kS+1}^{(k+1)S} \left( \overline{w} \right)^{l} \left\{ \delta(k) \left( R_1 \right)^{-l} + \left( 1 - \delta(k) \right) \left[ \prod\limits_{p=0}^{k-1} \left( R_{pS+1} \right)^{-S} \right] \left( R_{kS+1} \right)^{-l+kS} \right\} \right\} \right\}.
\end{split}
\label{eq:eq3.31_P}
\end{equation}
As a consequence, the expression in Eq. \eqref{eq:eqx.10} for the total energy $\mathcal{E}_{tot}$ wasted by the SCBM reads as in:
\begin{equation}
	\begin{split}
		\mathcal{E}_{tot} &= \mathcal{E}_{setup} + \mathcal{P} T_{MT}  \equiv \mathcal{E}_{setup} + K_0 M_0 \left( R_0 \right)^{\alpha-1} +  K_0 M_0 \left( R_0 \right)^{-1} \left\{ \left( \overline{w} \right)^{1 + I_{MAX}} \left( R_{I_{MAX}+1} \right)^{\alpha-1} \times \left[ \prod\limits_{k=0}^{Q-1} \left( R_{kS+1} \right)^{-S} \right] \right.  \\
		& \hphantom{=} + \left. \left( 1 - \delta \left( I_{MAX} \right) \right) \left\{ \sum\limits_{m=0}^{Q-1} \left\{ \sum\limits_{l=mS+1}^{(m+1)S} \left( \overline{w} \right)^{l}  \left( \delta(m) \left( R_{1} \right)^{\alpha-l} + \left( 1 - \delta(m) \right) \left[ \prod\limits_{p=0}^{m-1} \left( R_{pS+1} \right)^{-S} \right] \left( R_{mS+1} \right)^{\alpha+mS-l} \right) \right\} \right\} \right\}.
	\end{split}
	\label{eq:eq3.34_P}
\end{equation}
Interestingly, the reported expressions in \eqref{eq:eq3.29_P}, \eqref{eq:eq3.31_P} and \eqref{eq:eq3.34_P} for the SCBM depend only on $Q$ and the $(Q + 2)$ migration rates to be updated (see the above defined set $\Xi$).

\subsection{The afforded QoS minimum-energy bandwidth optimization problem}
\label{sec:ssec6.1}

In order to properly account for the metrics commonly considered for measuring the QoS of live VM migrations, we introduce four sets of hard constraints in the formulation of the minimum-energy SCBM optimization problem.

The first two constraints enforce QoS-dictated upper bounds: $\Delta_{MT}$ (s) and $\Delta_{DT}$ (s) on the tolerated migration time $T_{MT}$ and downtime $T_{DT}$, respectively. Hence, they read as:
\begin{equation}
\varphi_1 \triangleq \frac{T_{MT}}{\Delta_{MT}} - 1 \leq 0,
\label{eq:eq3.39_P}
\end{equation}
and
\begin{equation}
\varphi_2 \triangleq \frac{T_{DT}}{\Delta_{DT}} - 1 \leq 0,
\label{eq:eq3.40_P}
\end{equation}
where $T_{MT}$ and $T_{DT}$ depend on the per-round aggregated migration rates of the underlying (equal-balanced) MPTCP connection (see Eqs. \eqref{eq:eq3.29_P} and \eqref{eq:eq3.31_P}).  

The third constraint arises from the consideration that, in principle, the pre-copy stage in Fig. \ref{fig:Live_migration} of the PeCM technique could run indefinitely, if suitable stop conditions are not imposed. Commonly considered stop conditions account for \cite{22}: (i) the maximum number $I_{MAX}$ of the allowed migration rounds of Fig. \ref{fig:PeC}; and, (ii) the maximum tolerated value: $\beta \geq 1$ of the ratio $\left( V_i/V_{i+1} \right)$ of the migrated data over two consecutive rounds. Hence, since the above constraints in Eqs. \eqref{eq:eq3.39_P} and \eqref{eq:eq3.40_P} already account for $I_{MAX}$ through the corresponding expressions of $T_{MT}$ and $T_{DT}$ in Eqs. \eqref{eq:eq3.29_P} and \eqref{eq:eq3.31_P}, we leverage Eq. \eqref{eq:eq3.4_P}, in order to formulate the constraints on the ratio $\left( V_i/V_{i+1} \right)$ in the following form:
\begin{equation}
\varphi_3 \left( R_i \right) \triangleq \frac{\beta \overline{w}}{R_i} - 1 \leq 0, \quad \text{for } i=0, \:\text{and}\: i = jS+1, \: \forall \, j=0,\ldots, (Q-1),
\label{eq:eq3.41_P}
\end{equation}
where $R_i$ denotes the aggregated bandwidth done available by the considered MPTCP connection during the $i$-th migration round of Fig. \ref{fig:Live_migration}.

The last constraint accounts for the maximum uplink bandwidth that the 5G Network Processor of Fig. \ref{fig:scenario} allocates to the requiring device for the VM migration. In principle, depending on the bandwidth allocation policy actually implemented by the 5G Network Processor, the bandwidth assigned to the device could be exclusively allotted for the migration process (e.g., out-band migration) or it could be shared with the migrating application through statistical multiplexing (in-band migration; see \cite{32}). In the first case, the bandwidth costraint reads as in: $R \leq R_{MAX}$, whilst, in the second case, we have: $R \leq \rho_{mgr} R_{AGR}$, where: (i) $R_{MAX}$ (Mb/s) is the uplink bandwidth that the 5G Network Processor of Fig. \ref{fig:scenario} reserves for the exclusive support of the migration process; (ii) $R_{AGR}$ (Mb/s) is the total aggregate bandwidth allocated by the 5G Network Processor for both the migration of the VM and the support of the migrating application; and, (iii) $\rho_{mgr} \in \left[ 0, \, 1 \right]$ is the (dimensionless) fraction of the aggregate bandwidth $R_{AGR}$ that is reserved by the wireless device for the migration process. Hence, after introducing the auxiliary position:
\begin{equation}
\widehat{R} \triangleq \min \left\{ R_{MAX}; \: \rho_{mgr} R_{AGR} \right\},
\label{eq:eqx.IV}
\end{equation}
the following set of bandwidth constraints:
\begin{equation}
R_i \leq \widehat{R}, \quad \text{for} \quad i=0, \:\text{and}\: i=jS+1, \: \forall \, j= 0, \ldots, (Q-1),
\label{eq:eq3.42_P}
\end{equation}
apply both cases of in-band (e.g., $R_{MAX} = \infty$) and out-band (e.g., $\rho_{mgr} R_{AGR} = \infty$) migration over the 5G FOGRAN of Fig. \ref{fig:scenario}. 

Overall, as a matter of these considerations, the afforded SCBM constrained optimization problem is formally stated as follows:
\begin{equation}
\begin{split}
\min_{ \stackrel{\{ R_0, R_{jS+1}, R_{I_{MAX}+1} \}}{j = 0, 1, \dots, (Q-1)} } \mathcal{E}_{tot}, \\
s.t.: \quad \text{Eqs. \eqref{eq:eq3.39_P}, \eqref{eq:eq3.40_P}, \eqref{eq:eq3.41_P}, and \eqref{eq:eq3.42_P}},
\end{split}
\label{eq:eq3.38_P}
\end{equation}
with $\mathcal{E}_{tot}$ given by Eq. \eqref{eq:eq3.34_P} under the considered MPTCP-supported SCBM.

In the last part of this section, we discuss some possible generalizations and refinements of the stated SCBM optimization problem.

\paragraph{Accounting for wireless connection failures} 
According to the time chart of Fig. \ref{fig:Live_migration} and the related text, PeCM guarantees, by design, that the migrating VM continues to run on the wireless device till the end of the Commitment phase, so that the possible occurrence of failures of the device-to-fog MPTCP connection \textit{do not} cause service interruption \cite{22}. This inherent robustness of the PeCM technique against connection failure events makes it the preferred candidate for performing VM migration in failure-prone wireless environments \cite{A5}. However, nothing is for free, so that connection failure events still waste energy resources of the wireless device. In order to formally quantify the (average) energy loss caused by failures of the used MPTCP connection, we need to characterize the connection failure probability: $Pr_{cf} \left(\Delta_{MT} \right)$, formally defined as the probability that a connection failure event occurs during the migration interval, e.g., over the time: $\left[ 0, \Delta_{MT} \right]$. In general, in the considered mobile Fog scenario of Fig. \ref{fig:scenario}, this probability may depend of a (large) number of (possibly, unpredictable) factors, such as, mobility speed and mobility trajectory of the device, radio coverage radius, utilized inter-cell handover mechanism, statistics of the wireless fading and so on, just to cite a few. However, the formal general analysis carried out in \cite{A33} supports the conclusion that the failure probability of a mobile connection is typically well described by the following Pareto-like expression: 
\begin{equation}
Pr_{cf} \left( \Delta_{MT} \right) = \begin{cases}
1,  &  \text{for} \quad \Delta_{MT} > T_{con},  \\
\left( 1 + \sigma \right) \left( \frac{\Delta_{MT}}{T_{con}} \right) - \sigma \left( \frac{\Delta_{MT}}{T_{con}} \right)^3,  &  \text{for} \quad \Delta_{MT} \leq T_{con}.
\end{cases}
\label{eq:eqx.12}
\end{equation}
In the above relationship, we have that: (i) $\sigma$ is a dimensionless non-negative shaping factor, that accounts for the handover and/or mobility-induced heavy-tailed behavior of the connection failure probability; and, (ii) $T_{con}$ (s) is the maximum expected duration of an on-going MPTCP connection. Although it may be hard to develop general computing formulae for these parameters, we note that, in our framework, they may be profiled on-line by the 5G Network Processor of Fig. \ref{fig:scenario}, which typically records the statistics of the sustained connections \cite{A8}. Therefore, after profiling the connection failure probability and under the worst-case assumption that all the already migrated data are lost when a connection failure event happens, the resulting failure-induced energy loss suffered by the wireless device reads as the following product: 
\begin{equation}
\overline{N}_{mgr} Pr_{cf} \left( \Delta_{MMT} \right) \overline{\mathcal{E}}_{tot},
\label{eq:eqx.13}
\end{equation}
where $\overline{N}_{mgr} \geq 0$ is the (possibly, profiled) average number of VM migrations attempted by the wireless device of Fig. \ref{fig:scenario} during the time interval $\Delta_{MT}$, and $\overline{\mathcal{E}}_{tot}$ is the average per-migration energy consumed by the wireless device.

\paragraph{Accounting for time-varying dirty rates of the migrating application} 
Let us consider the case in which the migrating application performs so many memory writing operations that the resulting dirty-rate changes during the migration time. Hence, after indicating by: $\overline{w}_i$ (Mb/s), and $w_{max} \triangleq \max_{1 \leq i \leq I_{MAX}+1} \left( \overline{w}_i \right)$ the dirty rate over the $i$-th migration round and the resulting maximum dirty rate respectively, let us introduce the following dummy positions: $\Gamma_i \triangleq \prod_{k=1}^i \overline{w}_i$, for $i \geq 1$, and $\Gamma_0 \triangleq 0$. Hence, after replacing: $\left( \overline{w}/R \right)^i$, $\left( \overline{w}/R \right)^{I_{MAX}+1}$ and $\overline{w}~$ by: $\Gamma_i/\left( R \right)^i$, $\Gamma_i/\left( R \right)^{I_{MAX}+1}$ and $w_{max}$ respectively, the resulting  formulation of the considered bandwidth optimization problem directly applies to the case of time-varying dirty rates.

\paragraph{Accounting for the stretching of the execution times and memory compression of the migrating VM} 
During the execution of in-band VM migrations, the (previously introduced) aggregated bandwidth $R_{AGR}$ available at the wireless device of Fig. \ref{fig:scenario} is partially utilized for migration purpose, so that the execution speed of the corresponding migrating application may decrease due to bandwidth contention phenomena \cite{A34}. In order to quantify the resulting stretching of the execution time of the migrating application, let $\overline{T}_{ex-mgr}$ and $\overline{T}_{ex}$ be the (average) execution times of the considered application in the presence/absence of VM migration, respectively. Hence, the queue analysis reported in \cite{A34} leads to the conclusion that the execution time stretching ratio: $\overline{T}_{ex-mgr} / \overline{T}_{ex}$ scales as
\begin{equation}
\frac{\overline{T}_{ex-mgr}}{\overline{T}_{ex}} = \frac{1}{1 - \rho_{mgr}},
\label{eq:eqx.14}
\end{equation}
where $\rho_{mgr}$ is the (previously defined) fraction of the overall aggregated bandwidth that is reserved by the device for the VM migration. The above stretching ratio may be lowered by reducing the size of the migrated VM through compression coding (e.g., delta coding, run-length coding and similar). Hence, in order to simultaneously account for the contrasting effects on $\rho_{mgr}$ of both compression coding and ARQ/FEC-based error-protection coding, let $\widetilde{M}_0$ (Mb) be the size of the uncompressed and uncoded VM, and let: $cp \in \left[ 0, \, 1 \right]$, and $cr \in \left[ 0, \, 1 \right]$ be the compression ratio and the overall coding rate of the ARQ/FEC-based error protection mechanisms implemented at both the MAC and Physical layers of the wireless device. Hence, the actual size $M_0$ of the VM to be migrated is directly computed as follows:
\begin{equation}
M_0 = \frac{cp}{cr} \widetilde{M}_0.
\label{eq:eqx.15}
\end{equation}
%

\section{Feasibility conditions and optimized tuning of the maximum number of migration rounds}
\label{sec:sec7}

Regarding the feasibility conditions of the stated bandwidth optimization problem, we observe that the involved constraint functions $\varphi_1$ and $\varphi_2$ in Eqs. \eqref{eq:eq3.39_P} and \eqref{eq:eq3.40_P} strictly decrease for increasing value of the aggregate transport rate $R$. therefore, as a quite direct consequence, the following formal result holds.
\begin{proposition}\label{prop:prop1}
The SCBM optimization problem in Eq. \eqref{eq:eq3.38_P} is feasible if and only if the following three conditions are simultaneously met:
\begin{equation}
\frac{M_0}{\Delta_{TM}} \left[ \frac{1}{\widehat{R}} \left( I_{MAX} + 2 \right) \delta\left(\frac{\overline{w}}{\widehat{R}} - 1 \right) + \left( \frac{1 - \left( \overline{w}/\widehat{R} \right)^{I_{MAX} + 2} }{\widehat{R}- \overline{w}} \right) \left( 1 - \delta\left( \frac{\overline{w}}{\widehat{R}} - 1 \right) \right) \right]  \leq 1,
\label{eq:eq3.45_P}
\end{equation}
\begin{equation}
\frac{M_0}{\Delta_{DT}}  \frac{1}{\widehat{R}} \left( \frac{\overline{w}}{\widehat{R}} \right)^{I_{MAX}+1} \leq 1,
\label{eq:eq3.46_P}
\end{equation}
and,
\begin{equation}
\beta \: \left( \frac{\overline{w}}{\widehat{R}} \right) \leq 1.
\label{eq:eq3.47_P}
\end{equation}
\end{proposition}

Interestingly, the above feasibility conditions lead to some first insights about the effects of the parameters of $I_{MAX}$, $\widehat{R}$ and $\overline{w}$ on the (expected) behavior of the resulting SCBM. At this regard, three main remarks are in order. First, at fixed $I_{MAX}$, the expressions at the l.h.s. of Eqs. \eqref{eq:eq3.45_P}, \eqref{eq:eq3.46_P} and \eqref{eq:eq3.47_P} strictly increase (resp., decrease) for increasing values of the dirty rate $\overline{w}$ (resp., the maximum transport rate $\widehat{R}$). Second, for increasing values of $I_{MAX}$, the function at the l.h.s. of Eq. \eqref{eq:eq3.46_P}: (i) remains unchanged; (ii) decreases; and, (iii) increase, at $\left( \overline{w} / \widehat{R} \right) = 1$, $\left( \overline{w} / \widehat{R} \right) < 1$, and $\left( \overline{w} / \widehat{R} \right) > 1$, respectively. Third, for increasing values of $I_{MAX}$, the l.h.s. of Eq. \eqref{eq:eq3.45_P} increases, regardless from the value assumed by the ratio $\overline{w} / \widehat{R}$. The consequences are that: (i) increasing values of $\widehat{R}$ are welcome, because they always reduce both the resulting total migration time and downtime; but, (ii) increasing values of $I_{MAX}$ lower the resulting downtimes at $\left( \overline{w} / \widehat{R} \right) < 1$, whilst increase them at $\left( \overline{w} / \widehat{R} \right) > 1$.

These considerations open the doors to the question regarding the optimized setting of $I_{MAX}$. Unfortunately, this is still an \textit{open} question even in the case of state-of-the-art hypervisors, whose typically adopted application-oblivious default setting \cite{22,42}: $I_{MAX} = 29$ does \textit{not} guarantee, indeed, the convergence of the iterative pre-copy migration process and does \textit{not} assure the minimization of the overall consumed energy (see, for example, \cite{A5,A37} and references therein). However, we anticipate that the carried out tests support the conclusion that, under the following setting of $I_{MAX}$:
\begin{equation}
\widetilde{I}_{MAX} \equiv \left\lceil \frac{\log\left( \frac{M_0}{\Delta_{DT} \widehat{R}} \right)}{\log \left( \frac{\widehat{R}}{\overline{w}} \right)} - 1 \right\rceil, \quad \text{for} \quad (\overline{w}/\widehat{R}) < 1,
\label{eq:eq3.48_P}\vspace{0.5ex}
\end{equation}
the resulting total migration energy wasted by the proposed SCBM typically attains its minimum. Intuitively, this is due to the fact that the expression in \eqref{eq:eq3.48_P} is obtained by calculating the value of $I_{MAX}$ that meets the feasibility constraint in \eqref{eq:eq3.46_P} with the strict equality. Hence, under the setting in \eqref{eq:eq3.48_P}, the maximum tolerated downtime $\Delta_{DT}$ is fully exploited by the SCBM. This leads to a reduction of the average transport rate $\overline{R} \triangleq E\left\{ R \right\}$ utilized during the overall migration process, that lowers, in turn, the resulting total migration energy. On the basis of this consideration, in the sequel, we refer to $\widetilde{I}_{MAX}$ in \eqref{eq:eq3.48_P} as the optimized setting of $I_{MAX}$ under the proposed SCBM.

\section{Proposed Settable-Complexity Bandwidth Manager and related implementation aspects}
\label{sec:sec8}

From a formal point of view, the energy function $\mathcal{E}_{tot}$ in \eqref{eq:eq3.34_P} is a superposition of power-fractional terms of the type $\left( 1/R_i \right)^{m_i}$, which involve the rate variables. Hence, $\mathcal{E}_{tot}$ is \textit{not} a convex function in the transmission rates to be optimized, so that the resulting optimization problem in \eqref{eq:eq3.38_P} is \textit{not} a convex optimization problem. However, since all the $(Q + 2)$ rate variables to be optimized are, by design, non-negative, we may introduce the following $\log$-transformations: 
\begin{equation}
\widetilde{R}_i \triangleq \log (R_{i}),  \quad i = 0; \: i = jS+1, \: \forall \, j = 0, \ldots, (Q-1); \: i = (I_{MAX} + 1), 
\label{eq:eq3.49_P}
\end{equation}
so that:
\begin{equation}
R_i = e^{\widetilde{R}_i}, \quad i = 0; \: i = jS+1, \: \forall \, j = 0, \dots, (Q-1); \: i = (I_{MAX} + 1).
\label{eq:eq3.50_P}
\end{equation}
Furthermore, after collecting the $(Q + 2)$ log-rates in \eqref{eq:eq3.49_P} in the following $(Q + 2)$-dimensional (column) vector:
\begin{equation}
\overrightarrow{\widetilde{R}} \triangleq  \left[ \widetilde{R_0}, \:\widetilde{R_1}, \:\widetilde{R}_{S+1}\:, \ldots, \: \widetilde{R}_{(Q-1)S+1}, \: \widetilde{R}_{I_{MAX}+1} \right]^T,
\label{eq:eq3.51_P}
\end{equation}
we indicate as:
\begin{equation}
\begin{split}
\mathcal{E}_{tot}\left( \overrightarrow{\widetilde{R}} \right) &= \mathcal{E}_{setup} + K_0 M_0 \:e^{(\alpha-1)\widetilde{R}_0} +  K_0 M_0 \:e^{-\widetilde{R}_0} \left\{ \left( \overline{w} \right)^{I_{MAX}+1} \times  e^{\left[ (\alpha-1)\widetilde{R}_{I_{MAX}+1}  -  S\widetilde{R}_{1} - \left( 1 - \delta\left(Q - 1\right) \right) S \left( \sum\limits_{k=1}^{Q-1}\widetilde{R}_{kS+1} \right) \right]} \right. \\
  & \hphantom{=} + \left. \left(1 - \delta\left( I_{MAX} \right) \right)  \left\{ \sum\limits_{m=0}^{Q-1} \left\{ \sum\limits_{l=mS+1}^{(m+1)S}\left( \overline{w} \right)^{l} \left\{ \delta(m) e^{(\alpha-l)\widetilde{R}_1} + \left( 1 - \delta(m) \right) \:e^{\left[ (\alpha +mS-l)\widetilde{R}_{mS+1}-S\widetilde{R}_{1} - \left(1 - \delta(m-1) \right) \left( \sum\limits_{p=1}^{m-1}\widetilde{R}_{pS+1} \right) \right]} \right\} \right\} \right\} \right\}, 
\end{split}
\label{eq:eq3.52_P}
\end{equation}
\begin{equation}
T_{DT} \left(\overrightarrow{\widetilde{R}} \right) = M_0 \:e^{- \widetilde{R}_0} \left( \overline{w} \right)^{1 + I_{MAX}}  \times e^{\left[ -\widetilde{R}_{I_{MAX}+1} - SR_{1} - \left(1 - \delta(Q - 1) \right) S \left( \sum\limits_{k=1}^{Q-1}\widetilde{R}_{kS+1} \right) \right]}, 
\label{eq:eq3.53_P}
\end{equation}
\begin{equation}
\begin{split}
T_{TM} \left( \overrightarrow{\widetilde{R}} \right) &= M_0 \:e^{-\widetilde{R}_0} \left\{ 1 + \left( \overline{w} \right)^{1 + I_{MAX}} \: e^{\left[ -\widetilde{R}_{I_{MAX}+1} - S \widetilde{R}_{1} - \left( 1 - \delta(Q-1) \right) S \right]} \times e^{ \left( \sum\limits_{k=1}^{Q-1} \widetilde{R}_{kS+1} \right) }  \right. \\
  & \hphantom{0} \left. +  \left( 1 - \delta\left(I_{MAX}\right) \right) \left\{ \sum\limits_{k=0}^{Q-1}  \left\{ \sum\limits_{l=kS+1}^{(k+1)S} \left( \overline{w} \right)^{l}  \left\{ \delta(k) \: e^{-l\widetilde{R}_{1}} + \left( 1 - \delta(k) \right) \times \: e^{ \left[ (kS-l)\widetilde{R}_{kS+1} - S \widetilde{R}_1 -\left( 1 - \delta(k-1) \right) S \left( \sum\limits_{p=1}^{k-1} \widetilde{R}_{pS+1} \right) \right] } \right\} \right\} \right\} \right\},
\end{split}
\label{eq:eq3.54_P}
\end{equation}
the resulting expressions for the total energy, downtime and migration time that are obtained by introducing the positions in \eqref{eq:eq3.50_P} into Eqs. \eqref{eq:eq3.29_P}, \eqref{eq:eq3.31_P} and \eqref{eq:eq3.34_P}, respectively. Hence, after posing: $R_i = e^{\widetilde{R}_i}$ into Eq. \eqref{eq:eq3.38_P} and related constraints, it may be directly viewed by inspection that the resulting log-transformed problem is \textit{now jointly convex} in the involved log-transformed transmission rates: $\left\{ \widetilde{R}_i \right\}$\footnote{Formally speaking, this is a consequence of the fact that each term: $e^{\mp b_i \widetilde{R}_i}$ is a convex function in $\widetilde{R}_i$ for any value of the coefficient $b_i$. This is the formal reason that motivates the introduction of the log-transformations in Eq. \eqref{eq:eq3.49_P}.}. Furthermore, under the assumption that the feasibility conditions in Eqs. \eqref{eq:eq3.45_P} -- \eqref{eq:eq3.47_P} are satisfied with the strict inequalities, the so-called Slater's qualification condition also holds \cite{20}, so that the solution of the considered optimization problem may be computed by applying the Karush-Khun-Tucker (KKT) optimality conditions (see, for example, Chapter 4 of \cite{20} for a detailed presentation of this formal topic). Finally, due its convex form, the (log-transformed) box constraints in \eqref{eq:eq3.42_P} on the maximum aggregate bandwidth of the underlying MTCP connection may be managed as implicit ones by performing orthogonal projections \cite{20}. 

According to these observations, let:
\begin{equation}
\overrightarrow{\lambda} \triangleq \left[ \lambda_{1}, \: \lambda_{2}, \: \lambda_{30}, \: \lambda_{31}, \: \lambda_{3(S + 1)}, \: \lambda_{3(2S + 1)}, \: \dots, \: \lambda_{3((Q - 1)S + 1)} \right]  \: \in \mathbb{R}^{Q + 3},
\label{eq:eq3.61_P}
\end{equation}
be the non-negative $(Q + 3)$-dimensional vector of the Lagrange multipliers associated to the $(Q + 3)$ log-transformed inequality constraints in \eqref{eq:eq3.39_P} -- \eqref{eq:eq3.41_P}. Hence, the Lagrangian function: $\mathcal{L} \left( \overrightarrow{\widetilde{R}}, \: \overrightarrow{\lambda} \right)$ of the resulting log-transformed convex optimization problem reads as in:
\begin{equation}
\begin{split}
\mathcal{L} \left( \overrightarrow{\widetilde{R}},\: \overrightarrow{\lambda} \right) &\triangleq \mathcal{E}_{tot} \left( \overrightarrow{\widetilde{R}} \right)  +  \lambda_1  \left\{ \left[ \frac{1}{\Delta_{TM}} \:T_{TM} \left(\overrightarrow{\widetilde{R}}\right) \right] - 1 \right\} + \lambda_2 \left\{ \left[ \frac{1}{\Delta_{DT}} T_{DT} \left( \overrightarrow{\widetilde{R}} \right) \right] - 1 \right\} \\
		& \hphantom{=} +  \left\{ \lambda_{30} \left[ \beta \:\overline{w} \: e^{-\widetilde{R}_0}  -  1 \right] + \left\{ \sum\limits_{m=0}^{Q - 1} \lambda_{3(mS+1)} \left[ \beta \: \overline{w} \: e^{-\widetilde{R}_{Sm+1}}  -  1 \right] \right\} \right\},
\end{split}
\label{eq:eq3.62_P}
\end{equation}
and, then, the minimization to be carried out is the following one:
\begin{equation}
\max_{\overrightarrow{\lambda} \geq \overrightarrow{0} } \:\:\ \left\{\min_{ \left\{ \overrightarrow{R}_{i} \leq \log \widehat{R},\: i= \:0,\: (S+1),\: \dots,\: (Q-1)S+1,\: (I_{MAX}+1) \right\} } \left\{ \mathcal{L} \left( \overrightarrow{\widetilde{R}}, \: \overrightarrow{\lambda} \right) \right\} \right\}.
\label{eq:eq3.63_P}
\end{equation}
The latter may be, in turn, computed by performing the orthogonal projection on the box sets in \eqref{eq:eq3.63_P} of the vector which solves the nonlinear algebraic system that is obtained by equating to zero the $(2Q + 5)$-dimensional vector gradient of the Lagrangian function in \eqref{eq:eq3.63_P} with respect to both the primal $\overrightarrow{\widetilde{R}}$ and dual $\overrightarrow{\lambda}$ variables in \eqref{eq:eq3.51_P} and \eqref{eq:eq3.61_P}, that is:
\begin{equation}
\overrightarrow{\nabla} \mathcal{L} \left( \overrightarrow{\widetilde{R}}, \: \overrightarrow{\lambda} \right) = \overrightarrow{\mathbf{0}}_{(2Q + 5) \times 1}.
\label{eq:eq3.64_P}
\end{equation}
In this sequel, we indicate by $\left\{ \overrightarrow{\widetilde{R}}^{\ast}, \: \overrightarrow{\lambda}^{\ast} \right\}$ the $(2Q + 5)$-dimensional projected solution of this algebraic equations system. Being nonlinear, it does not admit a closed-form solution. However, its solution may be iteratively computed through a suitable set of projected gradient-based primal-dual iterations. At this regard, we note that, as pointed out in \cite{35}, the primal-dual algorithm is an iterative procedure for solving convex optimization problems, which applies quasi-Newton methods for updating the primal-dual variables simultaneously, in order to move toward the saddle-point \eqref{eq:eq3.63_P} of the underlying Lagrangian function at each iteration.

In the proposed framework, the $(2Q + 5)$ scalar updatings to be carried out at the $n$-th iteration read as follows\footnote{By definition, $\left[x\right]^\alpha$ means: $\min\left\{ x; \: \alpha \right\}$, while $\left[x\right]_+$ denotes: $\max\left\{ x; \: 0 \right\}$.}:
\begin{equation}
\overrightarrow{\widetilde{R}}_i^{(n+1)} = \left[ \overrightarrow{\widetilde{R}_i}^{(n)} - \omega_i^{(n)} \: \nabla_{\widetilde{R}_i} \mathcal{L} \left( \overrightarrow{\widetilde{R}}^{(n)}, \overrightarrow{\lambda}^{(n)} \right) \right]^{\log\left(\widetilde{R}\right)}, \quad n \geq 0, \quad \widetilde{R}_i^{(0)} = \log\left(\overline{w}\right),
\label{eq:eq3.67_P}
\end{equation}
\begin{equation}
\lambda_{i}^{(n+1)} = \left[ \lambda_i^{(n)} + \xi_i^{(n)} \: \nabla_{\lambda_i} \mathcal{L} \left( \overrightarrow{\widetilde{R}}^{(n)}, \overrightarrow{\lambda}^{(n)} \right) \right]_{+}, \quad n \geq 0, \:\: \lambda_i^{(0)} = 0, \:\: i = 1,2,
\label{eq:eq3.68_P}
\end{equation}
and,
\begin{equation}
\lambda_{3i}^{(n+1)} = \left[ \lambda_{3i}^{(n)} + \Psi_{3i}^{(n)} \: \nabla_{\lambda_{3i}} \mathcal{L} \left( \overrightarrow{\widetilde{R}}^{(n)}, \overrightarrow{\lambda}^{(n)} \right) \right]_{+}, \quad n \geq 0, \:\: \lambda_{3i}^{(0)} = 0,
\label{eq:eq3.69_P}
\end{equation}
for $i = 0$, and $i = jS + 1$, $\forall \, j = 0, 1, \ldots, (Q - 1)$, where $\left\{ \omega_i^{(n)}, \: n \geq 0 \right\}$, $\left\{ \xi_i^{(n)}, \: n \geq 0 \right\}$ and $\left\{ \Psi_{3i}^{(n)}, \: n \geq 0 \right\}$ are $(2Q + 5)$ suitable non-negative sequences of $n$-varying step-sizes. The projections in \eqref{eq:eq3.67_P}, \eqref{eq:eq3.68_P} and \eqref{eq:eq3.69_P} account for the box-type constraints in \eqref{eq:eq3.63_P}. The Appendix \ref{sec:appB} details the analytical expressions of the partial derivatives $\nabla_{\widetilde{R}} \mathcal{L}(\cdot)$ and $\nabla_{\lambda_i} \mathcal{L}(\cdot)$ present in Eqs. \eqref{eq:eq3.67_P} -- \eqref{eq:eq3.69_P}.

\subsection{Adaptive implementation and settable implementation complexity of the proposed SCBM}
\label{sec:ssec8.1}

It is expected that, due to fast fading phenomena and/or device mobility, the state of the MPTCP connection of Fig. \ref{fig:scenario} may experience abrupt time fluctuations during the migration process. In order to effectively cope with these (typically unpredictable) fluctuations, the gain sequences in Eqs. \eqref{eq:eq3.67_P}, \eqref{eq:eq3.68_P} and \eqref{eq:eq3.69_P} must be adaptively updated, in order to allow the SCBM to quickly response in a reactive way, e.g., without the support of (typically unreliable and error-prone) forecasting of the fading levels and/or device mobility platform. 

At this regard, we note that the Lyapunov-based formal analysis of the stability of MPTCP connections recently reported in \cite{35} guarantees that the asymptotical global stability of the MPTCP CC algorithms of Section \ref{sec:sec4} is attained when each gain sequence: $\left\{ g_y^{(n)},\: n \geq 1 \right\}$ required to update the corresponding sequence: $\left\{ y^{(n)}, \: n \geq 1 \right\}$ of the associated primal/dual variable is computed step-by-step on the basis of the following (memoryless and uncoupled) relationship (see Theorem 3.3 of \cite{35} for a formal proof of this result):
\begin{equation}
g_y^{(n)} = \frac{1}{2} \left( y^{(n)} \right)^2.
\label{eq:eqx.16}
\end{equation}
The scenario considered in \cite{35} is of wired-type, so that the attainment of quick responsiveness is typically not a major concern. However, in the considered wireless/mobile environment of Fig. \ref{fig:scenario}, the following two additional goals should be, indeed, also attained. First, $g_y^{(n)}$ must remain sufficiently large for vanishing $y^{(n)}$, in order to allow the SCBM to fast react to abrupt environmental changes. Second, $g_y^{(n)}$ must not grow too much for increasing values of $y^{(n)}$, in order to limit the occurrence of unstable oscillatory phenomena. Overall, motivated by these considerations, we propose to update the gain sequences in Eqs. \eqref{eq:eq3.67_P}, \eqref{eq:eq3.68_P} and \eqref{eq:eq3.69_P} according to the following ``clipped'' $n$-indexed relationships:
\begin{equation}
\omega_i^{(n)} = \max \left\{ \frac{a_{MAX}}{10}; \: \min \left\{ a_{MAX}; \:\frac{1}{2} \left( \widetilde{R}_i^{(n)} \right)^2 \right\} \right\}, \quad n \geq 1, \quad \omega_{i}^{(0)} = a_{MAX},
\label{eq:eq3.71_P}
\end{equation}
\begin{equation}
\xi_i^{(n)} = \max \left\{ \frac{a_{MAX}}{10}; \: \min \left\{ a_{MAX}; \:\frac{1}{2} \left( \lambda_i^{(n)} \right)^2 \right\} \right\}, \quad n \geq 1, \quad \xi_{i}^{(0)} = a_{MAX}, \:  i = 1, 2,
\label{eq:eq3.72_P}
\end{equation}
and:
\begin{equation}
\Psi_{3i}^{(n)} = \max \left\{ \frac{a_{MAX}}{10}; \: \min \left\{ a_{MAX}; \:\frac{1}{2} \left( \lambda_{3i}^{(n)} \right)^2 \right\} \right\}, \quad n \geq 1, \quad \Psi_{3i}^{(0)} = a_{MAX}.
\label{eq:eq3.73_P}
\end{equation}
In the above relationships, $a_{MAX}$ is a positive parameter, whose goal is to clip from below and above the instantaneous values assumed by the gain sequences, in order to assure: (i) fast responsiveness in the transient-state; (ii) fast convergence to the steady-state, and, (iii) stability in the steady-state. At this regard, we anticipate that the carried out numerical tests of Section \ref{sec:sec9} show that the performances of the resulting adaptive SCBM remain nearly unchanged over a quite broad range of values of $a_{MAX}$. This supports, in turn, the conclusion that the behavior of the adaptive SCBM is not too much sensitive with respect to the actual setting of $a_{MAX}$.

\paragraph{Implementation complexity aspects} 

From an implementation point of view, the primal-dual iterations in Eqs. \eqref{eq:eq3.67_P}, \eqref{eq:eq3.68_P} and \eqref{eq:eq3.69_P} and the associated ones in Eqs. \eqref{eq:eq3.71_P}, \eqref{eq:eq3.72_P} and \eqref{eq:eq3.73_P} must be carried out by the wireless device of Fig. \ref{fig:scenario} at the end of the Pre-migration phase of Fig. \ref{fig:Live_migration}. Hence, the duration $T_{IT}$ (s) of each n-indexed iteration should be small enough to allow the convergence to the (global) optimum within a small portion of the Pre-migration phase. At this regard, we have numerically ascertained that it suffices to set $T_{IT}$ to twenty times the inverse of the maximum clock frequency of the device CPU. 

Passing to consider the implementation complexity of the adaptive SCBM, an inspection of Eqs. \eqref{eq:eq3.67_P}, \eqref{eq:eq3.68_P} and \eqref{eq:eq3.69_P} points out that it is \textit{independent} from the allowed maximum number of round $I_{MAX}$, and it is affected by two factors, namely, the number of gradient iterations to be updated at each step $n$ and the number of the $n$-indexed steps needed for their convergence. Since the number of iterations equates (see Eqs. \eqref{eq:eq3.67_P} -- \eqref{eq:eq3.69_P} and \eqref{eq:eq3.71_P} -- \eqref{eq:eq3.73_P}):
\begin{equation}
2 \times \left( Q + 5 \right),
\label{eq:eqx.17}
\end{equation}
we conclude that the dependence of the implementation complexity of the adaptive SCBM on the (settable) number $Q$ of migration rates is of \textit{linear} type. At this regard, we anticipate that the carried out numerical tests show that values of $Q$ as small as 2 -- 3 suffice for attaining the minimum migration energy. The implementation complexity also scales up, by design, in a \textit{linear} way with the number of the $n$-indexed iterations requested for the convergence of Eqs. \eqref{eq:eq3.67_P} -- \eqref{eq:eq3.69_P}. Although this number may depend, in turn, on the selected value of $Q$, we have numerically ascertained that, in general, the clipping action of the $a_{MAX}$ parameter in \eqref{eq:eq3.71_P} -- \eqref{eq:eq3.73_P} allows to attain small convergence times of the order of about 5 -- 6 $n$-indexed iterations (see Section \ref{sec:sec9}).

\paragraph{Real-time profiling of the application dirty-rate and MPTCP connection}

The memory size $M_0$ of the migrating VM may be measured during the Pre-migration phase of Fig. \ref{fig:Live_migration} by issuing the \textit{xenstore} command of the legacy Xen hypervisor \cite{42}. In order to profile on-line the average dirty rate $\overline{w}$ of the migrating application, we point out that state-of-the-art hypervisors (such as, for example, the Xen hypervisor) periodically report the map of the changed (e.g., dirtied) memory cells of the migrating VM, in the form of dirty bitmaps \cite{42}. Hence, the actual dirty rate $\overline{w}$ of the migrating VM may be profiled on-line by averaging over the migration time the number of memory cells marked as dirtied in the reported bitmaps.

Passing to consider the online power-profiling of the (equal-balanced) MPTCP connection of Section \ref{sec:ssec4.4}, we begin to observe that the subflow round-trip-times and the corresponding maximum available bandwidths may be directly profiled at the Transport layer of Fig. \ref{fig:Protocol_Stack} by issuing, for example, the \textit{iperf} command of Linux \cite{42}. In order to profile online the setup energy in \eqref{eq:eqx.5} and the $K_0$ and $\alpha$ parameters in \eqref{eq:eqx.10}, the OS of the wireless device may issue the \textit{ifconfig} command of Xen on a per-WNIC basis, in order to acquire the actual power states of the utilized WNICs \cite{42}. Specifically, under the equal-balanced condition of Eq. \eqref{eq:eqx.9}, by issuing the \textit{ifconfig} command at vanishing and unit-value transport rate $R$, the device directly measures $\mathcal{E}_{setup}$ in \eqref{eq:eqx.5} and $K_0$ in \eqref{eq:eqx.10}, respectively. Afterward, by issuing once a time the \textit{ifconfig} command at $R = \widehat{R}$, the device measures the total static-plus-dynamic power $\mathcal{P}_{tot}\left( \widehat{R} \right)$ consumed by the underlying WNICs. Finally, since the corresponding dynamic power $\mathcal{P}\left( \widehat{R} \right)$ equates, by definition, the difference: $\mathcal{P}_{tot}\left( \widehat{R} \right) - \mathcal{E}_{setup} \widehat{R}$, by solving Eq. \eqref{eq:eqx.10} with respect to $\alpha$, we obtain the following formula for the online profile of the $\alpha$ exponent: 
\begin{equation}
\alpha = \frac{\log \left( \left( \mathcal{P}_{tot}\left( \widehat{R} \right) - \mathcal{E}_{setup} \widehat{R} \right)/K_0 \right)}{\log \widehat{R}}.
\label{eq:eqx.18}
\end{equation}
%

\paragraph{Virtualization-vs.-Containerazation technologies}

\begin{figure}[htb]
\centering
	\subfloat[VM-based server virtualization]
	{\label{fig:VM_virtualization}
		\includegraphics[width=.25\columnwidth]{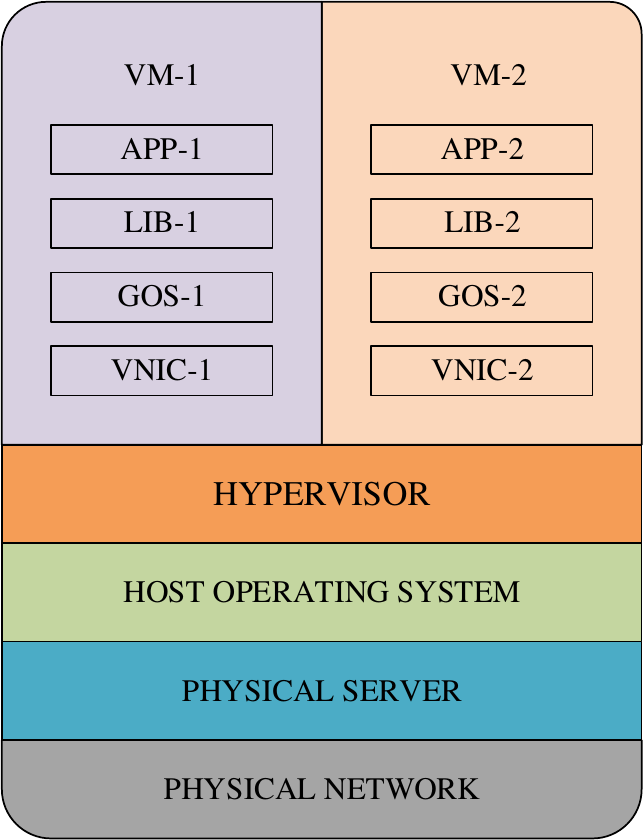}} 
	\hspace{3em}
	\subfloat[Container-based server virtualization]
	{\label{fig:CNT_virtualization}
		\includegraphics[width=.25\columnwidth]{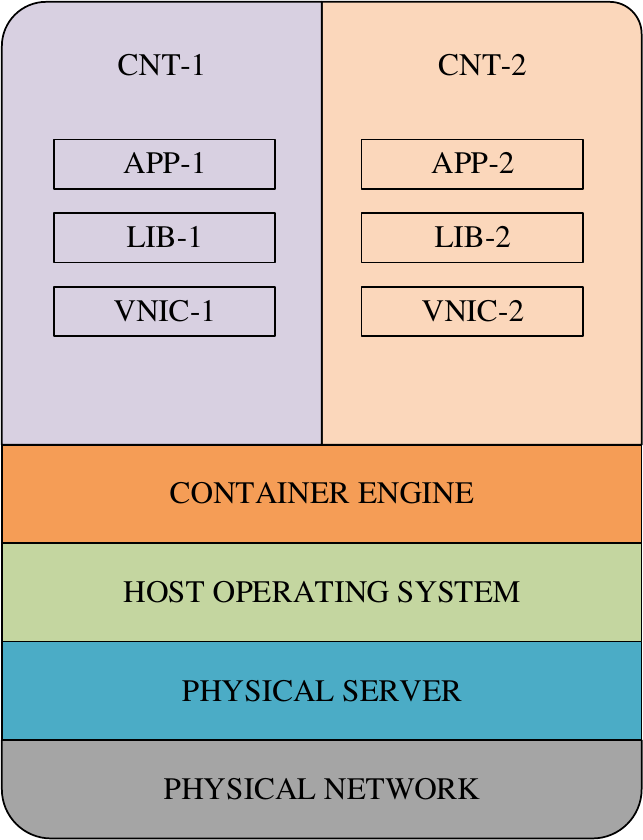}}
\caption{VM-vs.-Container based server virtualization. VM = Virtual Machine; LIB = LIBrary; GOS = Guest Operating System; VNIC = Virtual Network Interface Card; CNT = CoNTainer.}
\label{fig:VM_CNT_virtualization}
\end{figure}

The basic feature of the virtualized Fog node of Fig. \ref{fig:scenario} is that each served wireless device is mapped into a virtual clone that acts as a virtual processor and executes the migrated application on behalf of the wireless device. In addition to the here considered VM-based virtualization technology, the emerging CoNTainer (CNT)-based technology is quickly gaining momentum \cite{A35}. Figure \ref{fig:VM_CNT_virtualization} reports ``win-to-win'' the basic architectures of these two technologies.

Shortly, their main architectural differences are that \cite{A35}: (i) the VM technology relies on a middleware software layer (e.g., the so called Hypervisor), that statically  performs hardware virtualization, while the CNT technology uses an Execution Engine, in order to dynamically carry out resource scaling and multiplexing; and, (ii) each VM is equipped with an own (typically, heavy-weight) Guest Operating System (GOS), while a container comprises only application-related (typically, light-weight) libraries, and shares with the other containers the Host Operating System (HOS) provided by the physical server. The resulting pros and cons of these two virtualization technologies are summarized in Table \ref{tab:tabE}.

\begin{table*}[htb]  
\caption{Comparisons of the main native features of the VM and Container virtualization technologies.}
\label{tab:tabE}
\centering
\begin{tabular}{p{3.2cm}c
>{\columncolor[HTML]{E6E6E6}}p{5.2cm}
cp{5.2cm}}
 		\toprule
 		\textbf{Attributes}    &  & \textbf{Virtual Machine}   &  &  \textbf{Container}     \\
 		\midrule
 		\textbf{Guest OS}      
		&  &  Each VM is equipped with an own Guest OS that runs atop the hardware abstraction provided by an underlying hypervisor
		&  &  All the containers share a same OS that is directly stored by the physical memory of the hosting server   \\[2ex] 
		\midrule
    \textbf{Migration support} 
		&  &  Middleware support is provided for live migration of VMs 
		&  &  Live container migration is not provided. Only Stop-and-Copy type migration is currently supported        \\[2ex]
		\midrule
		\textbf{Virtualization-induced slowdown} 
		&  &  Due to the presence of both the Guest and Host operating systems, VMs suffer from virtualization-induced slowdowns 
		&  &  Containers offer quasi native computing speeds as compared to the underlying Host OS                      \\[2ex]
		\midrule
		\textbf{Resource isolation}  
		&  &  Quasi perfect isolation between VMs running on a  same server is attained by forbidding the sharing of files and libraries  
		&  &  Inter-container file sharing is allowed that reduces resource isolation \vspace{3.5ex}                    \\[2ex]
		\midrule
		\textbf{Bootstrap delay}
		&  &  The bootstrap delay of a VM is of the order of some minutes
		&  &  The bootstrap delay of a Container is of the order of some seconds                                        \\[2ex]
		\midrule
		\textbf{Storage requirement}
		&  &  VMs demand for not negligible storage space, in order to instal the whole Guest OS and associated libraries
		&  &  Containers require less storage space than VMs, because the base OS is shared                             \\[1ex]
 		\bottomrule     
 	\end{tabular}
\end{table*}

In a nutshell, the main pros of the CNT-based virtualization are that: (i) containers are light-weight and can be bootstrapped significantly quicker than VMs; and, (ii) the physical resources required by a container can be scaled up/down in real-time by the corresponding Execution Engine. However, since all containers hosted by a same physical server share the same HOS, the main cons of the CNT-based virtualization are that: (i) the level of resource isolation offered by the CNT-based virtualization is lower than the corresponding one of the VM-based technology; and, (ii) at the present time, the CNT technology does not support \textit{live} migration (see the second row of Table \ref{tab:tabE}). This the reason why in this contribution we directly refer to the VM technology.

At this regard, we remark that the so-called Checkpoint/Restore in Userspace (CRIU) technique is today utilized as tool for the migration of containers \cite{A36}. However, the CRIU technique is of Stop-and-Copy type, so that it is not suitable for the migration of delay-sensitive real-time applications \cite{A35}. Motivated by this consideration, container developers are currently working on a new project\footnote{\url{https://criu.org/P.Hau}}, in order to develop a middleawre platform for the support of live migration of containers. Since the proposed SCBM does not exploit, by design, primitives that are specific of the VM virtualization technology, we expect that it will be able to run unchanged even atop future middleware layers that support live migration of containers.

\section{Performance tests}
\label{sec:sec9}

The simulated scenario is the 5G FOGRAN of Fig. \ref{fig:scenario}. However, due to its multi-facet nature, it is (very) challenging to build up a testbed and, then, carry out repeatable field trials. Hence, motivated by this consideration, we chose numerical simulations as a means to evaluate the performance of the proposed adaptive SCBM, in order to assure both the repeatability of the performed tests and carry out fair performance comparisons. The MATLAB toolkit is used as the simulation platform, mainly because it natively supports a number of primitives for the numerical implementation of optimization solvers. The simulated scenario refers to the aforementioned in-band PeCM VM migration running atop the end-to-end virtualized architecture of Fig. \ref{fig:Clone}.

In detail, the simulated platform emulates the wireless interconnection between an iPhone7-Plus smartphone and a virtualized Dell Power Edge server that acts as a Fog node. The smartphone is assumed to be equipped with the iOS7 operating system, that already supports MPTCP for enabling its SIRI-based voice recognition applications\footnote{\url{https://support.apple.com/lv-lv/HT201373}}. 
The smartphone is assumed to be also equipped with: (i) 1.4 GHz CPU and 128 MB RAN; (ii) IEEE802.11b, 3G-UTRA and 4G-LTE WNICs; and, (iii) a Xen 3.3-like hypervisor \cite{42}, that provides the basic virtualization functions and manages them according to the reference architecture of Fig. \ref{fig:Clone}. The Dell Power Edge server at the simulated Fog node is assumed to be equipped with 3.06 GHz Intel Xeon CPU and 4GB  RAM, and its hardware resources are virtualized by a Xen-compliant hypervisor. WiFi connectivity is assumed to be guaranteed by a WRT54GL-type access point that is co-located with the server, while national-wide cellular operators provide  mobile 3G/4G cellular coverage. The proposed adaptive SCBM is assumed to be implemented in software at the driver domain (e.g., Dom-0; see \cite{42}) of the (aforementioned) legacy Xen 3.3 hypervisor. Furthermore, in order to avoid performance interferences, we also assume that the driver domain is forced to use a single physical core of the wireless device, whilst the migrating VM runs atop a second physical core. The device-to-fog simulated wireless connection is impaired by frequency-flat Rice fading with Rice factor of 6.5, and the coverage radius of the WiFi link is 300 (m). Finally, in order to account for the signaling overhead introduced by the protocol stack of Fig. \ref{fig:Protocol_Stack}, the maximum rates available at the MPTCP layer are assumed to be limited up to 90\% of the corresponding raw bit rates measured at the Physical layer.

Table \ref{tab:tabF} reports the values of the main simulated parameters, that are compliant with those typically considered by the open literature (see, for example, in \cite{44,43}). It is understood that the setup energies of Eq. \eqref{eq:eqx.5} are evaluated by multiplying the corresponding setup powers in the last row of Table \ref{tab:tabF} by the considered total migration times $\Delta_{TM}$.

\begin{table}[htb]  
\caption{Main simulated parameters \cite{44,43}.}
\label{tab:tabF}
\centering
 	\begin{tabular}{lcccccc}
 		\toprule
 		\textbf{Parameter}     &  & \textbf{3G}              &  &  \textbf{4G}           &  &   \textbf{IEEE802.11b WiFi}   \\
 		\midrule
 		$\alpha$               &  &  2                       &  &  2                     &  &  2                            \\[0.5ex] 
 		$MSS$ ~(Mb)            &  &  $8 \times 10^{-3}$      &  &  $8 \times 10^{-3}$    &  &  $8 \times 10^{-3}$           \\[0.5ex]
 		$R_{MAX}$ ~(Mb/s)      &  &  $0.9 \times 2$          &  &  $0.9 \times 50$       &  &  $0.9 \times 11$              \\[0.5ex]
 		$\overline{RTT}$ ~(ms) &  &  $250$                   &  &  $35$                  &  &  $25$                         \\[0.5ex]
 		$\Omega$ ~(W$^{0.5}$)  &  &  $10^{-4}$               &  &  $5.1 \times 10^{-3}$  &  &  $10^{-2}$                    \\[0.5ex]
 		$\mathcal{P}_{setup}$ ~(mW) &  & 87.83               &  &  137.83                &  &  159.46                       \\
 		\bottomrule     
 	\end{tabular}
\end{table}

\subsection{Benchmark bandwidth managers}
\label{sec:ssec9.1}

The benchmark bandwidth managers considered in the following subsections for performance comparisons are those described in \cite{22} and \cite{32}, that are also the only ones currently available into open literature.

Specifically, on the basis of some (recent) surveys, as well as at the best of the authors' knowledge, the bandwidth manager originally proposed in \cite{22} is the ``de facto'' standard one that currently equips a number of commercial hypervisors, like Xen, KVM and WMware (see \cite{A5} and references therein). Roughly speaking, it implements a heuristic best effort policy, in which the total migration bandwidth\footnote{In the sequel, we denote by the upper-scripts: $\left(\cdot\right)^{XEN}$ and $\left(\cdot\right)^{LIV\_MIG}$ the performance metrics and working parameters of the considered benchmark bandwidth managers in \cite{22,32}, respectively. The corresponding performance metrics and working parameters of the proposed adaptive SCBM are marked as: $\left(\cdot\right)^{SCBM}$.}: $R^{XEN}$ is incremented from the minimum: $R^{XEN}_{MIN} \equiv \overline{w}$ to the maximum: $R^{XEN}_{MAX} \equiv \widehat{R}$ by passing from the $0$-th initial migration round to the $\left( I^{XEN}_{MAX} + 1 \right)$-th final round (see Fig. \ref{fig:PeC}). For this purpose, at the beginning of each round, the migration bandwidth is incremented by the quantity \cite{22}: $\Delta R^{XEN} \equiv \left( R^{XEN}_{MAX} - \overline{w} \right) / \left( I^{XEN}_{MAX} + 1 \right)$. The rationale behind this heuristic approach is to shorten as much as possible the final downtime by allowing to migrate at the maximum transmission rate: $R^{XEN}_{MAX} \equiv \widehat{R}$ during the final Stop-and-Copy phase of the migration process (see Figs. \ref{fig:Live_migration} and \ref{fig:PeC}). Doing in so, this heuristic does not consider at all the resulting network energy consumption induced by the migration process, and the numerical results reported in the sequel confirm, indeed, its energy sub-optimality.

In order to improve the energy performance of the bandwidth manager in \cite{22}, the recent contribution in \cite{32} develops a formal approach that is based on the constrained optimization of the total energy: $\mathcal{E}_{tot}$ wasted by the migration process. However, unlike the here proposed SCBM, the bandwidth manager in \cite{32} is designed for operating under wired intra-data center environments, and, then (see \cite{32}): (i) it allows the optimization of only a \textit{single} migration rate, e.g., it uses a single optimized transmission rate over all the performed migration rounds; (ii) it is not optimized for working under MPTCP, e.g., only the case of SPTCP is considered in \cite{32}; and, (iii) its adaptation mechanism is optimized for attaining a stable behavior in the steady-state, so that its responsiveness is, indeed, quite limited. As a matter of these facts, the numerical results reported in the sequel point out that the energy performance of the bandwidth manager in \cite{32} is no so good under the here considered FOGRAN mobile scenarios.

\subsection{Description of the obtained numerical results and performance comparisons}
\label{sec:ssec9.2}

Goal of this subsection is threefold. First, we present a number of numerical results that aim at giving insights into the implementation complexity-vs.-energy performance tradeoff and the adaptive capability of the proposed SCBM. Second, we present some comparative results about the energy performance of the SCBM with respect to the (previously described) benchmark bandwidth managers in \cite{22} and \cite{32} under both synthetic and real-world workloads. Third, we compare the energy performances of the proposed SCBM under EWTCP MPTCP and newReno SPTCP. At this regard, we remark that a number of formal and measurement-based studies confirm that the enhanced aggregation bandwidth and responsiveness capabilities exhibited by the EWTCP CC algorithm make it (very) suitable for wireless/mobile applications (see, for example, \cite{A11,A15,A16,A17} and references therein). Therefore, unless otherwise stated, in the sequel, we focus on the numerical results for the case of equal-balanced EWTCP connections that operate at (see Eq. \eqref{eq:eqx.10}) $K_0 = 2.5 \times 10^{-2}$  (W $\times$ (Mb/s)$^2$) by simultaneously utilizing the 3G/4G/WiFi transmission technologies featured in Table \ref{tab:tabF}. However, we explicitly remark that, thanks to the unified power analysis of Section \ref{sec:sec4}, the corresponding steady-state energy performance of the other MPTCP CC algorithms may be obtained by scaling the corresponding expressions of $K_0$ of Table \ref{tab:tabD}.

\paragraph{Considered synthetic and real-world test applications}

Both synthetic and real-world applications have been selected, in order to simulate the workloads actually generated by the migrating VMs. Specifically, we have used the \textit{memtester} application for the generation of synthetic workloads\footnote{Memtester-Memory Diagnostic tool, available at: \url{http://pyropus.ca/software/memtester}}. This is a write-intensive application that it is currently used to detect faults in RAM by allowing the programmer to set the desired value of the average memory dirty rate $\overline{w}$. As a set of real-world benchmark applications, the \textit{bzip2}, \textit{mcf} and \textit{memcached} programs have been selected from the SPEC CINT2000 benchmark tool \footnote{SPEC CPU2000. Standard Performance Evaluation Corporation, available at: \url{http://www.spec.org./cpu2000/CINT200}}. The first one is a CPU and memory read-intensive application, that is featured by a quite low memory dirty rate. The second one is characterized by a larger dirty rate, because it presents a balanced mix of read and write memory operations. Finally, the third one is a write-intensive memory program, that emulates the rate-intensive caching of multiple key/value pairs in the RAM of the migrating applications.

\paragraph{Implementation complexity-vs.-energy performance trade-off under the proposed SCBM}

\begin{figure*}[htb]
\centering
 	\subfloat[]
 	{\label{fig:fig8a}
 		\includegraphics[width=.45\columnwidth]{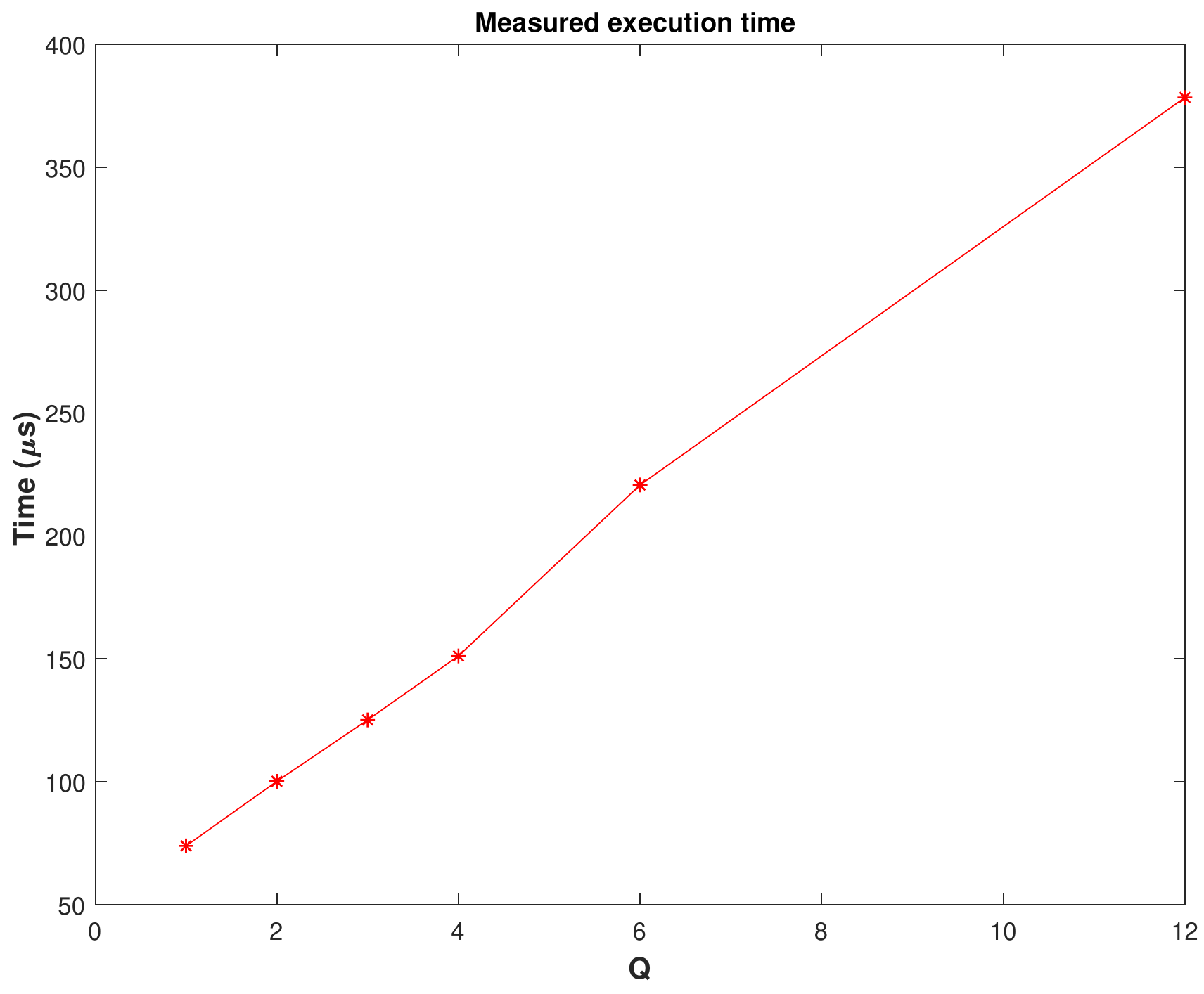}} 
		\hspace{3em}
	 \subfloat[]
 	{\label{fig:fig8b}%
 		\includegraphics[width=.45\columnwidth]{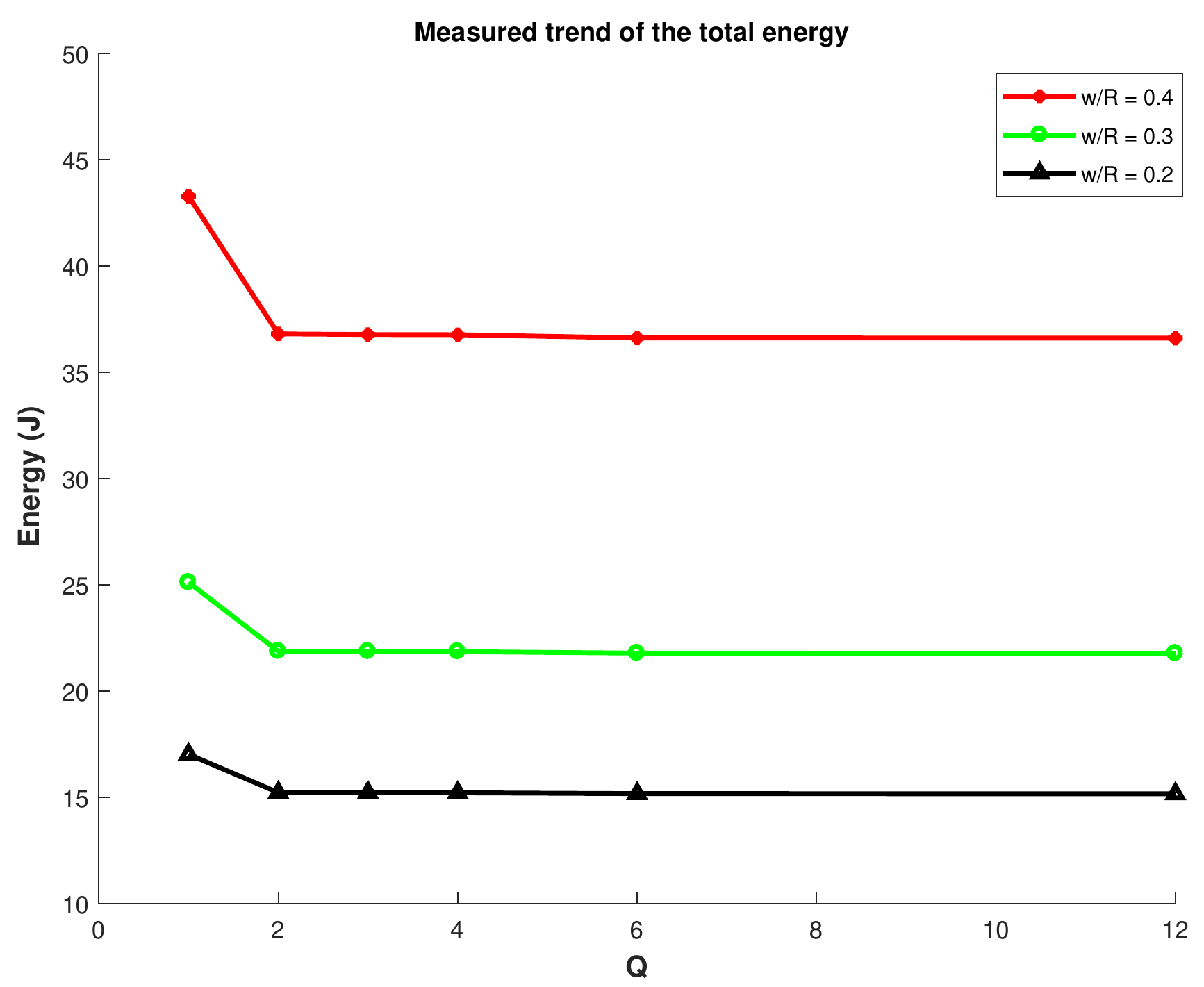}}	
\caption{(a) Measured execution times of the proposed SCBM for increasing values of $Q$. (b) Behaviors of the total energy $\mathcal{E}_{tot}^{SCBM}$ consumed by the proposed SCBM for increasing values of $Q$ at $\left( \overline{w}/\widehat{R} \right) = 0.2, \:0.3, \:0.4$ under the considered EWTCP-based scenario.}
\label{fig:figure8}
\end{figure*}

The numerical plots of Fig. \ref{fig:figure8} are obtained under the synthetic workloads generated by the \textit{memtester} application. Their goal is to give some first insights about the interplay between settable implementation complexity and energy performance of the proposed bandwidth manager. Specifically, Fig. \ref{fig:fig8a} reports the numerically measured execution times of the proposed SBCM for increasing values of the settable number $Q$ of the transport rates to be optimized.

Interestingly, the measured trend of the execution times is nearly linear and, then, it is compliant with the implementation complexity formula of Eq. \eqref{eq:eqx.17}. Regarding the actual values of $Q$ needed to drive the total energy $\mathcal{E}_{tot}^{SCBM}$ consumed by the proposed SCBM to its global minimum, the behaviors of the plots of Fig. \ref{fig:fig8b} support two main conclusions. First, since the volume of the data to be migrated increases for increasing values of the ratio $\left( \overline{w}/\widehat{R} \right)$ (e.g., for increasing values of the dirty rate of the migrating application), the plots of Fig. \ref{fig:fig8b} scale up for growing values of $\left( \overline{w}/\widehat{R} \right)$. Second, the semi-flat behaviors of the plots of Fig. \ref{fig:fig8b} support the conclusion that small values of $Q$ ranging from 1 to 3 are suffice, in order to drive the total energies consumed by the SCBM to their global minima. Since analogous trends have been, indeed, obtained under all the simulated scenarios, we may conclude that, in practice, the implementation complexity of the $Q$-optimized SCBM is limited and of the order of (see Eq. \eqref{eq:eqx.17} with $Q = 1, \,3$): $\mathcal{O}(12)$ -- $\mathcal{O}(16)$.

\paragraph{Responsiveness of the proposed SCBM}

\begin{figure}[htb]
\centering
\includegraphics[width=.425\columnwidth]{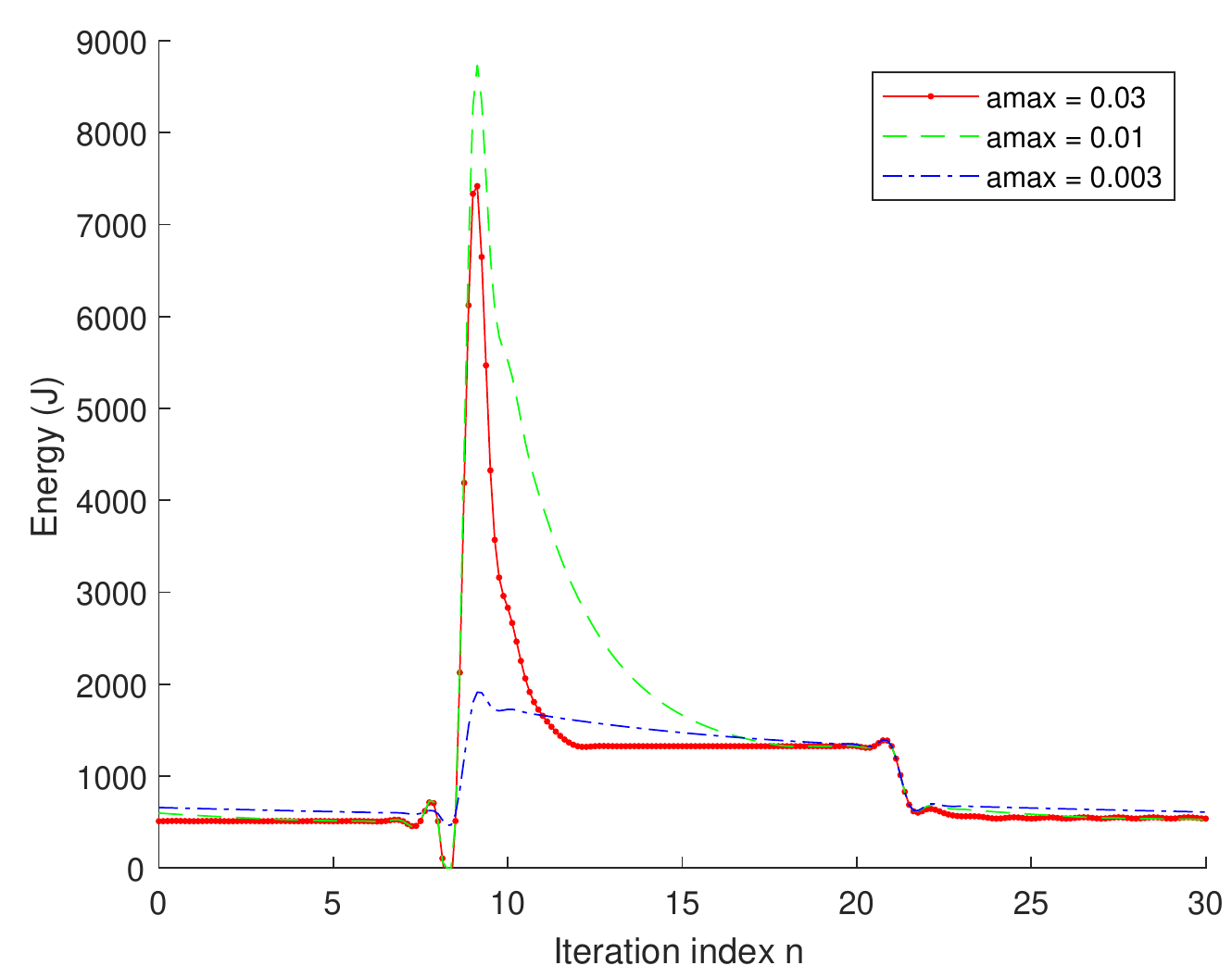}
\caption{Time evolutions (in the $n$ index) of the energy consumption of the adaptive SCBM under the considered EWTCP-based scenario. Case of time-varying dirty rate $\overline{w}$ at: $\widehat{R} = 18$ (Mb/s), $M_0 = 64$ (Mb), $\beta = 2$, $\Delta_{TM} = 50.5$ (s), and $\Delta_{DT} = 5.6$ (ms).}
\label{fig:fig9}
\end{figure}

\begin{figure}[htb]
\centering
\includegraphics[width=.425\columnwidth]{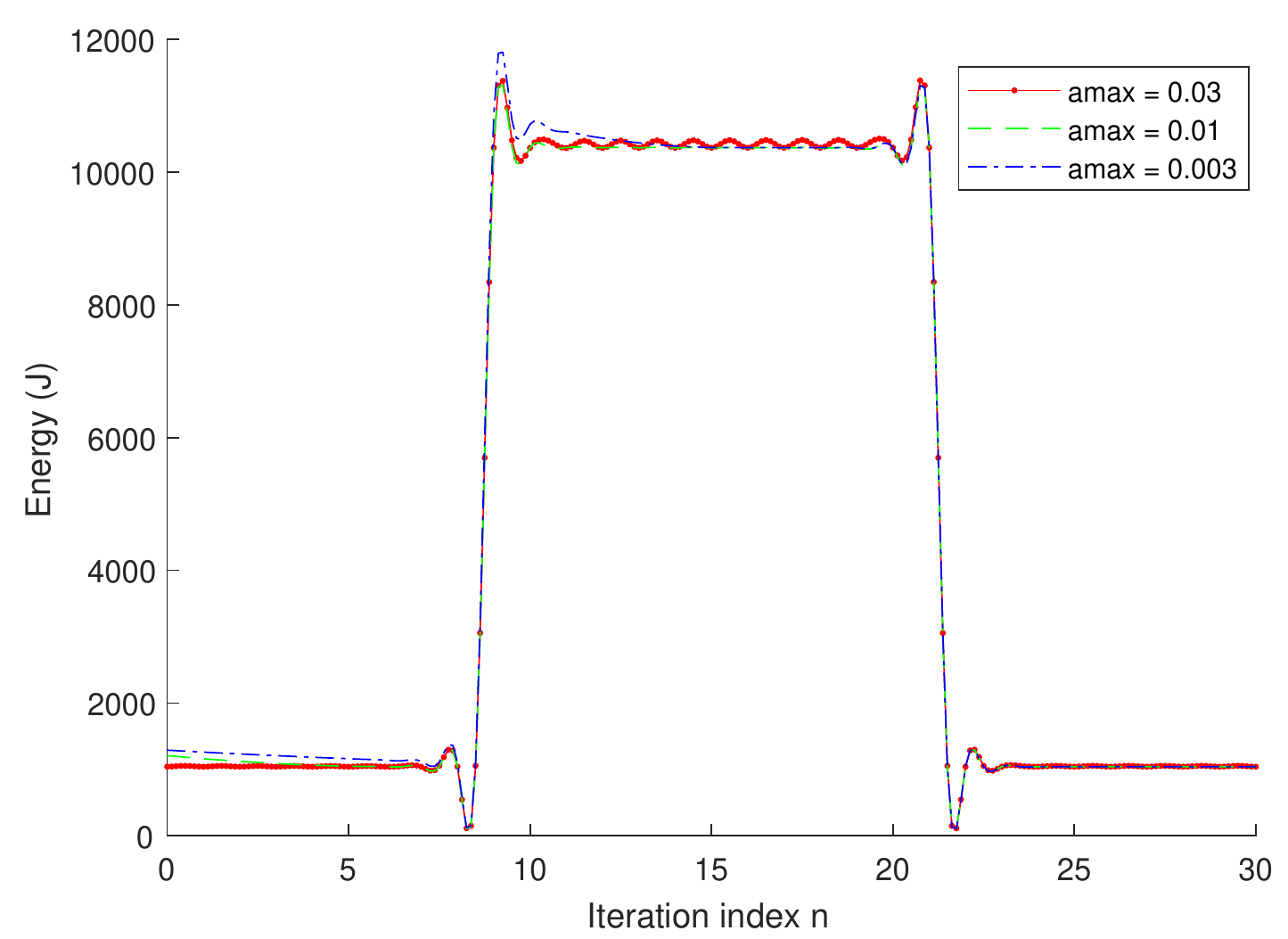}
\caption{Time evolutions (in the $n$ index) of the energy consumption of the adaptive SCBM under the considered EWTCP-based scenario. Case of time-varying $K_0$ of the EWTCP connection at: $\widehat{R} = 18$ (Mb/s), $M_0 = 64$ (Mb), $\beta = 2$, $\Delta_{TM} = 50.5$ (s), $\Delta_{DT} = 5.6$ (ms).}
\label{fig:fig10}
\end{figure}

Goal of the plots of Figures \ref{fig:fig9} and \ref{fig:fig10} is to numerically check the actual responsiveness of the adaptive SCBM when, due to memory-contention phenomena possibly and/or fluctuations of the round-trip-times of the utilized transport connections, the dirty rate of the migrated VM and/or the state of the EWTCP connection undergo unpredicted quick changes. At this regard, the plots of Figure \ref{fig:fig9} (resp., Figure \ref{fig:fig10}) report the numerically evaluated trajectories of the $n$-indexed sequence: $\left\{ \mathcal{E}_{tot}^{SCBM(n)}, \: n \geq 1 \right\}$ of the total energy consumed by the SBCM when the initial value of the dirty rate: $\overline{w} = 1$ (Mb/s) of the migrated VM (resp., the initial value: $K_0 = 2.5 \times 10^{-2}$ (W $\times$ (Mb/s)$^2$) of the power-profile of the underlying EWTCP connection) jumps to: $2 \times \overline{w}$ (resp., to: $10 \times K_0$) at $n = 8$, and, then, comes back to its initial values at $n = 20$. The workload of the \textit{memtester} application has been generated, in order to simulate the migrating VMs.

An examination of these plots leads to two main insights. First, the relationships in Eqs. \eqref{eq:eq3.71_P} -- \eqref{eq:eq3.73_P} designed for updating the gain sequences of the gradient-based iterations of Eqs. \eqref{eq:eq3.67_P} -- \eqref{eq:eq3.69_P} allow the SCBM to be very responsive, with convergence times that are typically limited up to 4 -- 6 $n$-indexed iterations. Second, the sensitivity of these convergence times on the actual value assumed by the ``clipping'' parameter $a_{MAX}$ in Eq. \eqref{eq:eq3.71_P} -- \eqref{eq:eq3.73_P} is not substantial, at least for values of $a_{MAX}$ falling into the (quite broad) interval: $3 \times 10^{-3}$ -- $3 \times 10^{-2}$.

\paragraph{Performance comparisons under synthetic workloads}

Goal of this subsection is to compare the energy performance of the proposed SCBM against the corresponding ones of the (previously described) benchmark bandwidth managers of Section \ref{sec:ssec9.1}, when the migrating VM is running the synthetic workload generated by the \textit{memtester} application. The obtained numerical results are reported in Table \ref{tab:tab4.2_P}, together with the corresponding \textit{numerically optimized} values of $I_{MAX}^{XEN}$, $~\widetilde{I}_{MAX}$ and $Q$.

\begin{table}[htbp]  
\caption{Energy performances of the proposed SCBM and the considered benchmark bandwidth managers under the considered EWTCP-based scenario at: $M_0 = 128$ (Mb) and $\overline{w}/\widehat{R} = 0.33$. The (numerically evaluated) reported values of $I_{MAX}^{XEN}$, $\widetilde{I}_{MAX}$ and $Q$ are the optimal ones, e.g., they minimize the energy consumptions of the corresponding bandwidth managers.}
\label{tab:tab4.2_P}
\centering
	\begin{tabular}{lccc}
		\toprule
		$\Delta_{DT}$ (s)                   & 0.103 &  $5.42 \times 10^{-4}$ &  $4.03 \times 10^{-5}$  \\[2ex]
		$\Delta_{TM}$ (s)                   & 46.9  &  65.2 &  83.6         \\[1.5ex]
		$\beta$                             & 2     &  2    &  2            \\[0.5ex]
		\midrule
		$I_{MAX}^{XEN}$                     & 6     &  14   &  29           \\[1.5ex]
		$\widetilde{I}_{MAX}$               & 4     &  9    &  11           \\[1.5ex]
		$Q$                                 & 1     &  1    &  1            \\[1.5ex]
		$\mathcal{E}_{tot}^{XEN}$ (J)       & 1880  &  2150 &  2470         \\[1.5ex]
		$\mathcal{E}_{tot}^{LIV\_MIG}$ (J)  & 1550  &  1557 &  1560         \\[1.5ex]
		$\mathcal{E}_{tot}^{SCBM}$ (J)      & 1366  &  1373 &  1373         \\[0.5ex]
    \midrule
		Energy saving over XEN (\%)         & 27.3  & 36.1  &  44.4         \\[1.5ex]
		Energy saving over LIV\_MIG (\%)    & 11.8  & 11.8  &  11.9         \\[0.5ex]
		\bottomrule     
	\end{tabular}
\end{table}		

An examination of this Table \ref{tab:tab4.2_P} leads to five main conclusions. 

First, at assigned values of the downtime, migration time and speed up factor (see the first three rows of Table \ref{tab:tab4.2_P}), we have numerically ascertained that the number of the maximum migration rounds that minimizes the energy consumptions of both the proposed SCBM and the benchmark one in \cite{32} is given by the relationship of Eq. \eqref{eq:eq3.48_P}. Second, the corresponding (numerically evaluated) optimized values of the maximum number of migration rounds: $I_{MAX}^{XEN}$ that allow the Xen manager to attain its minimum energy consumptions are quite \textit{larger} than the corresponding ones: $\widetilde{I}_{MAX}$ of Eq. \eqref{eq:eq3.48_P} (compare the 4-th and 5-th rows of Table \ref{tab:tab4.2_P}). Third, as it could be expected, decreasing values of the tolerated downtimes lead to larger values of the energies consumed by all tested bandwidth managers (see the 7-th, 8-th and 9-th rows of Table \ref{tab:tab4.2_P}). Forth, the per-cent energy savings: $\left( 1 - \left( \mathcal{E}_{tot}^{SCBM}/\mathcal{E}_{tot}^{XEN} \right) \right)$ (\%) of the proposed SCBM over the benchmark Xen one in \cite{22} are over 25\% and approach 45\% under strict downtimes (see the 10-th row of Table \ref{tab:tab4.2_P}). At this regard, we have numerically ascertained that about 25\% -- 30\%  of these attained energy savings are induced by the application-aware setting of $\widetilde{I}_{MAX}$ of Eq. \eqref{eq:eq3.48_P}, and this supports the actual effectiveness of this setting. Fifth, the corresponding per-cent energy savings: $\left( 1 - \left( \mathcal{E}_{tot}^{SCBM}/\mathcal{E}_{tot}^{LIV\_MIG} \right) \right)$ (\%) of the proposed SCBM over the benchmark bandwidth manager in \cite{32} maintain around 12\% (see the last row of Table \ref{tab:tab4.2_P}). Since these last energy savings are obtained at $Q = 1$ (see the 6-th row of Table \ref{tab:tab4.2_P}), we conclude that small increments in the implementation complexity of the proposed SCBM (e.g., small values of the settable $Q$ parameter) over the benchmark manager in \cite{32} suffice, in order to attain noticeable energy reductions.

\paragraph{Performance comparisons under real-world workloads}

These conclusions are corroborated by  the additional numerical results of this section. Specifically, its goal is to numerically test and compare the energy performance of the proposed SCBM under \textit{hard} operating conditions featured by \textit{large} sizes of the migrating VMs and \textit{low} values of the tolerated downtimes. In order to (attempt to) to provide a comprehensive picture covering heterogeneous application scenarios, all three the (aforementioned) real-world workloads: \textit{bzip2}, \textit{mcf} and \textit{memcached} have been considered in the carried out tests. The upper part of Table \ref{tab:tabG}  reports the main parameters of the simulated real-world workloads, whilst the middle and bottom parts of this table point out both the energy consumptions of the considered bandwidth managers and the resulting per-cent energy savings of the proposed SCBM over the benchmarks ones. In all the carried out tests, the sizes of the migrated VMs scaled up to $M_0 = 128$ (Mb), whilst the tolerated downtimes scaled down to $\Delta_{DT} = 1.0$ (ms).

\begin{table}[htbp] 
\caption{Simulated parameters, obtained energy consumptions and resulting per-cent energy savings under the considered EWTCP-based scenario for the real-world workloads \textit{bzip2}, \textit{mcf} and \textit{memcached} at: $\beta = 2$, $\Delta_{TM} = 81.5$ (s), and $\Delta_{DT} = 1.0$ (ms). The reported results of the SCBM refer to $Q = 3$.} 
\label{tab:tabG}
 	\centering
 	\begin{tabular}{lcccccc}
	\toprule
	\textbf{Parameter}  & &  \textbf{\emph{bzip2}}  & &  \textbf{\emph{mcf}}  & &  \textbf{\emph{memcached}}  \\
	\midrule
	$M_{0}$   ~(Mb)                            & & 128    					 & & 128               & & 128        \\[1ex]
	$\overline{w}$  ~(Mb/s)                    & & 0.72              & & 1.08              & & 1.26       \\[1ex]
	$\overline{w}/\widehat{R}$                 & & 0.4               & & 0.6               & & 0.7        \\[0.5ex]
  \midrule
	$\mathcal{E}_{tot}^{XEN}$  ~(J)            & & 107    					 & & 196               & & 228        \\[1ex]
	$\mathcal{E}_{tot}^{LIV\_MIG}$  ~(J)       & & 100    					 & & 128               & & 135        \\[1ex]
	$\mathcal{E}_{tot}^{SCBM}$  ~(J)           & & 39.4    					 & & 60.34             & & 66.4       \\[0.5ex]
  \midrule
	Energy saving respect XEN (\%)             & & 60.50             & & 69.21             & & 70.84      \\[1ex]
	Energy saving respect LIV\_MIG (\%)        & & 50.17             & & 52.86             & & 63.75      \\[0.5ex]
	\bottomrule  
 	\end{tabular}
\end{table}

Overall, a synoptic examination of the numerical data of Table \ref{tab:tabG} leads to three main conclusions. First, it is confirmed that the energy consumptions of all the considered bandwidth managers increase for increasing values of the ratio $\left( \overline{w}/\widehat{R} \right)$. Second, under the here considered ``harsh'' operating conditions, the per-cent energy savings of the proposed SCBM over the benchmark XEN one (resp., the benchmark LIV\_MIG one) maintain over 60\%  (resp., over 50\%) and approach 70\% (resp., 64\%) for values of the ratio $\left( \overline{w}/\widehat{R} \right)$ of the order of 0.7. Third, in order to attain these (large) energy reductions, the $Q$ parameter of the SCBM scaled up to: $Q = 3$.

\paragraph{SCBM atop EWTCP and newReno SPTCP: performance comparisons}

MPTCP (and, then, EWTCP) utilizes multiple WNICs in parallel. Therefore, in principle, it could happen that the energy reductions provided by the bandwidth aggregation is offset by the increment of the static energy in Eq. \eqref{eq:eqx.5} needed to simultaneously turning-ON the utilized WNICs, so that the resulting total energy consumed by MPTCP could be larger than the corresponding ones wasted by commodity SPTCP \cite{A11,A17}. However, it is also reasonable to expect that, in our framework, this energy offset may (eventually) happen when the constraints on the allowed downtimes are (very) relaxed, and/or the dirty rates and sizes of the migrating VMs are (very) low. 

Motivated by these considerations, goal of the last set of numerical results reported in this subsection is to give some insight about the actual energy consumptions of the proposed SCBM when it runs atop EWTCP and newReno SPTCP, respectively. For this purpose, a number of application scenarios featured by: (i) different sizes of the migrated VMs; (ii) migration downtimes; and, (iii) different values of the ratio $\left( \overline{w}/\widehat{R} \right)$ have been considered. In order to bypass the possible bias in the attained results induced by the reduced transmission bandwidth of the 3G WNIC (see the 3-rd row of Table \ref{tab:tabF}), the equal-balanced EWTCP connections simulated in this subsection utilize only 4G and WiFI WNICs. As a matter of this fact, the performance comparisons are carried out with respect to the newReno SPTCP that utilizes \textit{only} the 4G and \textit{only} the WiFi WNIC. Furthermore, in order to cover a broad spectrum of application environments, the synthetic workload of the \textit{memtester} tool has been generated for values of the ratio $\left( \overline{w}/\widehat{R} \right)$ ranging from 0.1 to 0.4. Table \ref{tab:tab5.1_P} recaps in a synoptic way the main parameters of the simulated WiFi-SPTCP, 4G-SPTCP and WiFi/4G-EWTCP connections.

\begin{table}[htb]  
\caption{Main parameters of the simulated WiFi-SPTCP, 4G-SPTCP and equal-balanced WiFi/4G-EWTCP connections. The reported values are compliant with the power analysis of Section \ref{sec:sec4}.}
\label{tab:tab5.1_P}
\centering
 	\begin{tabular}{lcccccc}
 		\toprule
 		\textbf{Parameter}     &  & \textbf{WiFi-SPTCP}      &  &  \textbf{4G-SPTCP}     &  &   \textbf{WiFi/4G-EWTCP}  \\
 		\midrule
 		$\alpha$               &  &  2                       &  &  2                     &  &  2                        \\[0.5ex] 
		$K_0$ ~(W/(Mb/s)$^2$)  &  &  $5 \times 10^{-2}$      &  &  $5 \times 10^{-2}$    &  &  $2.5 \times 10^{-2}$     \\[0.5ex]
 		$R_{MAX}$ ~(Mb/s)      &  &  $0.9 \times 11$         &  &  $0.9 \times 50$       &  &  $2 \times 9.9$           \\[0.5ex]
 		$\mathcal{P}_{setup}$ ~(mW) &  & 159.46              &  &  137.83                &  &  297.29                   \\
 		\bottomrule     
 	\end{tabular}
\end{table}

The obtained numerical results are reported in Table \ref{tab:tabH}. At this regard, we begin to point out that the size of the migrated VM (resp., the tolerated downtime) increases (resp., decreases) by crossing Table \ref{tab:tabH} from the top to the bottom (see the caption of Table \ref{tab:tabH}), so that harder operating conditions are considered by moving from the top to the bottom of this table. Therefore, on the basis of this remark, a comparative examination of the numerical results of Table \ref{tab:tabH} supports three main insights. First, as it could be expected, in all considered cases, the energy consumptions of the SCBM atop the WiFi-SPTCP, 4G-SPTCP and WiFi/4G-EWTCP connections increase for increasing values of the ratio: $\left( \overline{w}/R_{MAX} \right)$. Second, a comparative examination of the attained energy consumptions: $\mathcal{E}_{tot}^{4G}$ and $\mathcal{E}_{tot}^{WiFi}$ of the 4G-SPTCP and WiFi-SPTCP connections confirms that the 4G transmission technology is less energy efficient than the WiFi one\footnote{Obviously, this drawback is counter-balanced by the long-range high-broadband mobile cellular coverages supported by the 4G transmission technology.}. As a matter this fact, in the carried out tests, the per-cent energy savings: $Energy~saving_{4G}$ of the WiFi/4G-EWTCP over the 4G-SPTCP are quite large and span the interval (see the 5-th column of Table \ref{tab:tabH}): 52\% -- 73\%, whilst the corresponding energy savings: $Energy~saving_{WiFi}$ over the WiFi-SPTCP are somewhat more limited and range over the interval (see the last column of Table \ref{tab:tabH}): 17\% -- 33\%. Third, in all three operating cases featured by the top, middle and bottom sections of Table \ref{tab:tabH}, the attained per-cent energy savings are maximum at $\left( \overline{w}/R_{MAX} \right) = 0.25$,  whilst they tend to decrease at $\left( \overline{w}/R_{MAX} \right) = 0.1$ and at $\left( \overline{w}/R_{MAX} \right) = 0.4$. Intuitively, this non monotonic ``bell-shaped'' behavior is due to the fact that the bandwidth aggregation capability of the EWTCP is less effective when the volume of the migrated data decreases, whilst it somewhat saturates when the data to be migrated grow too much.

\begin{table}[htbp]
\caption{Comparative energy performances of the proposed SCBM under the simulated WiFi-SPTCP, 4G-SPTCP and equal-balanced WiFi/4G-EWTCP connections at: $\widehat{R} = R_{MAX}$ and $\Delta_{TM} = 37$ (s). $Energy~saving_{4G}$ (\%) and $Energy~saving_{WiFi}$ (\%) are the attained per-cent energy savings of the WiFi/4G-EWTCP solution over the 4G-SPTCP and WiFi-SPTCP ones, respectively. \textbf{Top}: $M_0 = 64$ (Mb) and $\Delta_{DT} = 200$ (ms); \textbf{Middle}: $M_0 = 128$ (Mb) and $\Delta_{DT} = 150$ (ms); \textbf{Bottom}: $M_0 = 256$ (Mb) and $\Delta_{DT} = 75$ (ms).}
\label{tab:tabH}
\centering
	\begin{tabular}{cccccc}
	\toprule
	$\left( \frac{\overline{w}}{R_{MAX}} \right)$  &  $\mathcal{E}_{tot}^{4G}$ (J)  &  $\mathcal{E}_{tot}^{WiFi}$ (J)  &  $\mathcal{E}_{tot}^{EWTCP}$ (J)  &  $Energy~saving_{4G}$ (\%)  &  $Energy~saving_{WiFi}$ (\%)   \\
	\midrule
	0.10   & ~\:73.95  & ~\:42.93  & ~\:35.42  &  52.10  &   17.49  \\[0.25ex]	
	0.25   &   108.17  & ~\:51.30  & ~\:40.77  &  62.30  &   20.52  \\[0.25ex]
	0.40   &   113.75  & ~\:55.68  & ~\:44.87  &  60.50  &   19.40  \\[0.25ex]
	\midrule
	0.10   &   178.90  & ~\:76.13  & ~\:60.05  &  66.40  &   21.12  \\[0.25ex]	
	0.25   &   214.85  & ~\:84.86  & ~\:64.31  &  70.06  &   24.21  \\[0.25ex]	
	0.40   &   240.72  & ~\:96.75  & ~\:75.46  &  68.65  &   22.00  \\[0.25ex]	
	\midrule
	0.10   &   323.84  &   145.52  &   101.16  &  68.76  &   30.48  \\[0.25ex]		
	0.25   &   413.14  &   163.88  &   109.45  &  73.50  &   33.21  \\[0.25ex]	
	0.40   &   552.90  &   215.51  &   160.57  &  70.96  &   25.49  \\[0.25ex]	
	\bottomrule     
	\end{tabular}
\end{table}

Overall, the final conclusion stemming from the carried out comparative tests is that the WiFi/4G-EWTCP solution is more energy efficient in supporting the proposed SCBM than the corresponding WiFi and 4G-SPTCP ones when the sizes of the migrated VMs are larger enough (we say, over 60 (Mb)) and the tolerated downtimes are strict (we say, below 250 -- 200 (ms)).

\section{Conclusion and hints for future research}
\label{sec:sec10}

In this paper, we developed an adaptive settable-complexity bandwidth manager (SCBM) for the constrained minimization of the network energy consumed by the live migration of VMs over wireless (possibly, mobile) 5G FOGRAN connections that utilize the MPTCP as the transport protocol. Hard constraints on both the total migration time and downtime are guaranteed by the proposed SCBM, as well as quick responsiveness to (possibly, unpredictable) fading and/or mobility-induced abrupt changes of the state of the underlying MPTCP connection. After discussing a number of related implementation aspects, we have numerically checked the implementation complexity-vs.-energy performance tradeoff attained by the proposed SCBM, as well as its energy performances  under both MPTCP and SPTCP transport protocols. Finally, we have numerically compared the average energy consumptions of the SCBM against the corresponding ones of the benchmark bandwidth managers in \cite{22,32} under both synthetic and real-world workloads. In a nutshell, the carried out comparisons point out that the average energy savings of the proposed SCBM over the one in \cite{22} (resp., \cite{32}) may approach 70\%  (resp., 50\%) under strict constraints on the tolerated downtimes.

The presented results may be extended over four main research directions. 

First, in this contribution, we focus on the aggregation bandwidth-capability of the MPTCP, in order to cope with hard constraints on the migration delays. However, MPTCP may be also effectively used to cope with the failure of (some of) the underling paths by shifting the transported flows towards the still active paths. How the proposed SCBM could effectively exploit the native failure recovery capability of the MPTCP may be, indeed, a first topic for future research.

Second, since the current version of the proposed SCBM selects the dirtied memory pages to be migrated in a random way, it may happen that a same memory page is dirtied multiple times and, then, it is  migrated several times during the VM migration process. In principle, multiple migrations of the dirtiest memory pages could be avoided by exploiting the information on the dirtied memory pages that is periodically provided by the hypervisor's bitmap \cite{42}. How this information could be effectively used by the proposed SCBM is a second topic of potential interest. 

Third, although the equal-balanced operating condition for the MPTCP is well motivated in the considered all-wireless FOGRAN scenario of Fig. \ref{fig:scenario}, it may become more questionable when heterogeneous mixed wireless/wired scenarios are considered. This may be, for example, the case of some emerging wireless/wired network infrastructures for the implementation of the front-haul section of Fig. \ref{fig:scenario} \cite{A8}. However, the generalization of the here developed SCBM to the case of un-balanced MPTCP connections is expected to introduce two additional non-minor challenges. First, the number of migration rates to be updated scales up by a factor $N$ equal to the number of used WNICs, so that it is expected that the implementation complexity of the resulting SCBM will also increase by the same factor. Second, under the general power formula in \eqref{eq:eqx.7} for the unbalanced MPTCP, the resulting (log-trasformed) objective function in Eq. \eqref{eq:eq3.52_P} is \textit{no longer convex}. This means, in turn, that gradient-based solving approaches do no longer guarantee the convergence to the global minimum of the energy objective function $\mathcal{E}_{tot}$. In principle, metaheuristic-based optimization methods could be utilized, in order to escape local minima. This approach has been, indeed, recently pursued in \cite{A38} for solving the (somewhat related) non-convex problem of the energy-efficient VM placement and server consolidation in large-scale data centers. How to apply this meta-heuristic based approaches to the case of the energy-efficient management of the migration bandwidths of un-balanced wireless MPTCP connections is an additional topic for future research. 

Finally, the performance results presented in this paper rely on numerical simulations. At this regard, we note that a final goal of the on-going project: GAUChO\footnote{\url{https://www.gaucho.unifi.it/}} funded by the Italian MIUR is to prototype a testbed that implements the main building blocks of the 5G FOGRAN infrastructure of Fig. \ref{fig:scenario}.

\section*{Acknowledgments}

This work has been supported by the project: ``GAUChO -- A Green Adaptive Fog Computing and networking Architectures" funded by the MIUR Progetti di Ricerca di Rilevante Interesse Nazionale (PRIN) Bando 2015 -- grant 2015YPXH4W\_004, and by the projects V-FOG and V-FOG2: ``Vehicular Fog energy-efficient QoS mining and dissemination of multimedia Big Data streams" funded by Sapienza University of Rome, Bando 2016 and 2017.

%

%

%

%

\appendix

\section{Derivation of the dynamic power-vs.-transport rate formula for wireless MPTCP connections}
\label{sec:appA}

The MPTCP connection works in the staedy-state under the Congestion Avoidance phase, so that the average sizes: $W_d(j)$, $j = 1, \ldots, N$, of the per-subflow CWNDs must not vary. Hence, the following sets of relationships must simultaneously hold:
\begin{equation}
\left( 1 - Pr_{loss}(j) \right) \times I_c(j) = Pr_{loss}(j) \times D_c(j), \quad j = 1, \ldots, N,
\label{eq:eqA.1}
\end{equation}
where $Pr_{loss}(j)$ is the loss segment probability of the $j$-th subflow. Eq. \eqref{eq:eqA.1} guarantees that the ACK-induced average increment: $\left( 1 - Pr_{loss}(j) \right) \times I_c(j)$ of the $j$-th window size $W_d(j)$ equates the corresponding loss segment-induced average decrement: $Pr_{loss}(j) \times D_c(j)$. Since $Pr_{loss}(j)$ remains, by design, largely less than the unit during the Congestion Avoidance phase\footnote{Typical values of $Pr_{loss}(j)$ during the Congestion Avoidance phase are below 1\% \cite{35}.}, we have that: $\left( 1 -Pr_{loss}(j) \right) \approx 1$, so that Eq. \eqref{eq:eqA.1} may be replaced by by the following one:
\begin{equation}
I_c(j) = Pr_{loss}(j) \times D_c(j), \quad j = 1, \ldots, N.
\label{eq:eqA.2}
\end{equation}
Now, according to Table \ref{tab:tabA}, the virtualization of the network resources performed by the 5G Network Processor of Fig. \ref{fig:scenario} leads, by design, to the isolation of the bandwidths allocated to the migrating VMs. Hence, it is reasonable to assume that the segment loss events affecting the virtualized 5G FOGRAN of Fig. \ref{fig:scenario} are mainly due to fading and/or device mobility \cite{A4,A16,A17}. This means, in turn, that, according to well established results reported, for example, in \cite{A28,A29,A32}, $Pr_{loss}(j)$ in \eqref{eq:eqA.2} scales down as a power of the per-flow dynamic power $\mathcal{P}(j)$, so that we can write: 
\begin{equation}
Pr_{loss}(j) = \frac{\Omega(j)}{\mathcal{P}(j)^m}.
\label{eq:eqA.3}
\end{equation} 
In Eq. \eqref{eq:eqA.3}, $\Omega(j)$ depends on the transmission technology employed by the $j$-th WNIC and the (dimensionless) positive exponent: $m > 0$ is mainly dictated by the the statistical property of the fading and device mobility pattern. Hence, after posing: $\alpha \triangleq 1/m$ and, then, introducing \eqref{eq:eqA.3} into \eqref{eq:eqA.2}, Eq. \eqref{eq:eqx.7} follows from quite direct (but somewhat tedious) algebraic manipulations of the expressions of $I_c(j)$ and $D_c(j)$ in Table \ref{tab:tabB}.

\section{Gradient formulae for the adaptive implementation of the SCBM}
\label{sec:appB}

In this appendix we detail the scalar gradients of the Lagrangian function in Eq. \eqref{eq:eq3.62_P}. 

Specifically, the gradient of the $\mathcal{L}$ function with respect to $\widetilde{R}_0$ assumes the following expression:
\begin{equation}
\nabla_{\widetilde{R}_0} \mathcal{L} = \left( \frac{\partial \mathcal{E}_{tot}}{\partial \widetilde{R}_0} \right) - \lambda_1 \frac{1}{\Delta_{TM}} T_{TM} - \lambda_2 \frac{1}{\Delta_{DT}} T_{DT} - \lambda_{30} \: \beta \: \overline{w} \: e^{-\widetilde{R}_0},
\label{eq:eqA.14_P}
\end{equation}
where:
\begin{equation}
\begin{split}
\frac{\partial \mathcal{E}_{tot}}{\partial \widetilde{R}_0} &= \left( \alpha - 1 \right) K_0 M_0 \:e^{\left(\alpha - 1 \right) \widetilde{R}_0} - K_0 M_0 \:e^{-\widetilde{R}_0}  \left\{ \left( \overline{w} \right)^{1 + I_{MAX}} \: e^{\left[ \left( \alpha - 1 \right) \widetilde{R}_{I_{MAX} +1} - S \widetilde{R}_1  - \left( 1- \delta(Q -1) \right) S \left( \sum\limits_{k = 1}^{Q - 1} \widetilde{R}_{kS+1} \right) \right]} \right. \\
  & \hphantom{=}  + \left. \left(1 - \delta(I_{MAX}) \right) \left\{ \sum\limits_{m = 0}^{Q - 1} \left\{ \sum\limits_{l = mS + 1}^{(m + 1)S} \left( \overline{w} \right)^{l} \left\{ \delta(m) \: e^{\left( \alpha - l \right) \widetilde{R}_1} + \left( 1 - \delta(m) \right) \: e^{\left[ \left(\alpha + mS - l\right) \widetilde{R}_{mS+1} - S \widetilde{R}_1 - \left( 1 - \delta(m - 1) \right) \left(\sum\limits_{p=1}^{m-1} \widetilde{R}_{pS+1}\right) \right]} \right\} \right\} \right\} \right\}.
\end{split}
\label{eq:eqA.15_P}
\end{equation} 
The gradient of Eq. \eqref{eq:eq3.62_P} with respect to $\widetilde{R}_1$ is:
\begin{equation}
\nabla_{\widetilde{R}_1} \mathcal{L} = \left( \frac{\partial \mathcal{E}_{tot}}{\partial \widetilde{R}_1} \right) + \frac{\lambda_1}{\Delta_{TM}} \left( \frac{\partial T_{TM}}{\partial \widetilde{R}_1} \right) + \frac{ \lambda_2}{\Delta_{DT}} \left( \frac{\partial T_{DT}}{\partial \widetilde{R}_1} \right) - \lambda_{31} \: \beta \: \overline{w} \: e^{-\widetilde{R}_1},
\label{eq:eqA.16_P}
\end{equation}
where:
\begin{equation}
\begin{split}
\frac{\partial T_{TM}}{\partial \widetilde{R}_1} &= - M_0 \: e^{-\widetilde{R}_0} \left\{ S \left( \overline{w} \right)^{1 + I_{MAX}} \: e^{\left[-\widetilde{R}_{I_{MAX} +1} - S \widetilde{R}_1 - \left( 1 - \delta(Q - 1) \right) S \left( \sum\limits_{k=1}^{Q-1} \widetilde{R}_{kS+1} \right) \right]}  \right. \\
  & \phantom{0} + \left. \left( 1 - \delta(I_{MAX}) \right) \left\{ \sum\limits_{k=0}^{Q-1} \left\{ \sum\limits_{l=kS+1}^{(k+1)S} \left( \overline{w} \right)^{l} \left\{ \delta(k)\: l\: e^{-l\widetilde{R}_1} + \left(1 - \delta(k) \right) S \times e^{\left[ (kS-l)\widetilde{R}_{kS+1} - S \widetilde{R}_1 - \left(1 - \delta(k -1) \right) S \left( \sum\limits_{p=1}^{k-1} \widetilde{R}_{pS+1} \right) \right]} \right\} \right\} \right\} \right\},
\end{split}
\label{eq:eqA.17_P}
\end{equation} 
\begin{equation}
\frac{\partial T_{DT}}{\partial \widetilde{R}_1} = - SM_0 e^{-\widetilde{R_0}} \left( \overline{w} \right)^{1 + I_{MAX}} \left( 1 - \delta(1 + I_{MAX}) \right) \: e^{\left[ -\widetilde{R}_{I_{MAX} + 1} - S \widetilde{R}_1 - \left( 1 - \delta(Q-1) \right) S \left( \sum\limits_{k=1}^{Q-1}\widetilde{R}_{kS+1} \right) \right]},
\label{eq:eqA.18_P}
\end{equation} 
and,
\begin{equation}
\begin{split}
\frac{\partial \mathcal{E}_{tot}}{\partial \widetilde{R}_1} &= - K_0 M_0 e^{-\widetilde{R_0}} \left\{ S \left( \overline{w} \right)^{1 + I_{MAX}} \: e^{\left[ \left( \alpha - 1 \right) \widetilde{R}_{I_{MAX}+1} - S \widetilde{R}_1 - \left(1 - \delta(Q - 1) \right) S \left( \sum\limits_{k=1}^{Q-1} \widetilde{R}_{kS+1} \right) \right]} \right. \\
  & \hphantom{=} + \left. \left(1 - \delta(I_{MAX}) \right) \left\{ \sum\limits_{k=0}^{Q-1} \left\{ \sum\limits_{l=mS+1}^{(m+1)S} \left(\overline{w} \right)^{l} \left\{ \left( l-\alpha \right) \delta(m) \: e^{ \left( \alpha - l \right) \widetilde{R}_1} + S \left( 1 - \delta(m) \right)  \:  e^{\left[ (\alpha+mS-l)\widetilde{R}_{mS+1} - S \widetilde{R}_1 - \left( 1 - \delta(m-1) \right) \left( \sum\limits_{p=1}^{m-1} \widetilde{R}_{pS+1} \right) \right]} \right\} \right\} \right\} \right\}.
\end{split}
\label{eq:eqA.19_P}
\end{equation} 
The expression of the gradient of Eq. \eqref{eq:eq3.62_P} with respect to $\widetilde{R}_{jS+1}$ is given by:
\begin{equation}
\nabla_{\widetilde{R}_{jS+1}} \mathcal{L} = \left( \frac{\partial \mathcal{E}_{tot}}{\partial \widetilde{R}_{jS+1}} \right) + \frac{\lambda_1}{\Delta_{TM}} \left( \frac{\partial T_{TM}}{\partial \widetilde{R}_{jS+1}} \right) + \frac{\lambda_2}{\Delta_{DT}} \left( \frac{\partial T_{DT}}{\partial \widetilde{R}_{jS+1}} \right) - \lambda_{3(jS+1)} \: \beta \: \overline{w} \: e^{-\widetilde{R}_{jS+1}}, \quad j = 1, 2, \ldots, (Q - 1),
\label{eq:eqA.20_P}
\end{equation} 
with the following three auxiliary positions:
\begin{equation}
\frac{\partial T_{DT}}{\partial \widetilde{R}_{jS+1}} = - S \left( 1 - \delta(Q-1) \right) M_0 \: e^{-\widetilde{R}_0} \left( \overline{w} \right)^{1 + I_{MAX}} \left( 1 - \delta(1 + I_{MAX}) \right) e^{\left[ -\widetilde{R}_{I_{MAX}+1} - S \widetilde{R_1} - \left( 1 - \delta(Q-1) \right) S \left( \sum\limits_{k=1}^{Q-1} \widetilde{R}_{kS+1} \right) \right]}, 
\label{eq:eqA.21_P}
\end{equation} 
\begin{equation}
\begin{split}
\frac{\partial T_{TM}}{\partial \widetilde{R}_{jS+1}} &= M_0 e^{\widetilde{R}_0} \left\{ -\left(\overline{w}\right)^{1 + I_{MAX}} S \left(1 - \delta(Q-1)\right) \: e^{\left[ -\widetilde{R}_{I_{MAX}+1} - S \widetilde{R_1} - \left( 1-\delta(Q-1) \right) S \left( \sum\limits_{k=1}^{Q-1} \widetilde{R}_{kS+1} \right) \right]} \right. \\
  & \hphantom{=} +  \left( 1-\delta(I_{MAX}) \right) \left\{ \sum\limits_{k=j}^{Q-1} \left\{ \sum\limits_{l=kS+1}^{(k+1)S} \left(\overline{w} \right)^{l} \left\{ \delta(k-j)(jS-l) \: e^{\left[ (jS-l)\widetilde{R}_{jS+1} - S \widetilde{R}_1 - \left( 1-\delta(k-1) \right) S \left( \sum\limits_{p=1}^{j-1} \widetilde{R}_{pS+1} \right) \right]} \right. \right. \right. \\
	& \hphantom{0} - \left. \left. \left. \left. \left( 1 - \delta(k-j) \right) S \left( 1 - \delta(k-1) \right) \: e^{\left[ (kS-l)\widetilde{R}_{kS+1} - S \widetilde{R}_1 - \left( 1 - \delta(k-1) \right) S \left( \sum\limits_{p=1}^{k-1} \widetilde{R}_{pS+1} \right)\right]} \right\} \right\} \right\} \right\}, \quad j = 1, 2, \dots, (Q-1),
\end{split}
\label{eq:eqA.22_P}
\end{equation} 
and,
\begin{equation}
\begin{split}
\frac{\partial \mathcal{E}_{tot}}{\partial \widetilde{R}_{jS+1}} &= K_0 M_0 \: e^{\widetilde{R}_0} \left\{ - \left(1 - \delta(Q-1) \right) S \: e^{\left[ \left(\alpha - 1 \right) \widetilde{R}_{I_{MAX}+1} - S \widetilde{R_1} - \left( 1 - \delta(Q-1) \right) S \left( \sum\limits_{k=1}^{Q-1} \widetilde{R}_{kS+1} \right) \right]} + \left( 1 - \delta(I_{MAX}) \right) \right. \\
 & \hphantom{=} \times \left\{ \sum\limits_{m=j}^{Q-1} \left\{ \sum\limits_{l=mS+1}^{(m+1)S} \left( \overline{w} \right)^{l} \left( 1 - \delta(m) \right) \left\{ \delta(m-j) \left( \alpha + jS-l \right) \: e^{\left[ \left(\alpha +jS-l \right) \widetilde{R}_{jS+1} - S \widetilde{R}_{1} - \left( 1 - \delta(m-1) \right) \left( \sum\limits_{p=1}^{j-1} \widetilde{R}_{pS+1} \right) \right]} \right. \right. \right. \\
 & \hphantom{=} - \left. \left. \left. \left. \left( 1 - \delta(m-j) \right) \left( 1 - \delta(m-1) \right) \: e^{\left[ \left( \alpha+mS-l \right) \widetilde{R}_{mS+1} - S \widetilde{R}_1 - \left( 1 - \delta(m-1) \right) S \left( \sum\limits_{p=1}^{m-1} \widetilde{R}_{pS+1} \right) \right]} \right\} \right\} \right\} \right\}, \quad  j = 1, 2, \dots, (Q-1).
\end{split}
\label{eq:eqA.23_P}
\end{equation} 
In a dual way, the expression of the gradient of Eq. \eqref{eq:eq3.62_P} with respect to $\widetilde{R}_{I_{MAX}+1}$ is:
\begin{equation}
\nabla_{\widetilde{R}_{I_{MAX}+1}} \mathcal{L} = \left( \frac{\partial \mathcal{E}_{tot}}{\partial \widetilde{R}_{I_{MAX}+1}} \right) + \frac{\lambda_1}{\Delta_{TM}} \left( \frac{\partial T_{TM}}{\partial \widetilde{R}_{I_{MAX}+1}} \right) + \frac{\lambda_2}{\Delta_{DT}} \left(\frac{\partial T_{DT}}{\partial \widetilde{R}_{I_{MAX}+1}} \right),
\label{eq:eqA.24_P}  
\end{equation} 
with the following accompanying expressions:
\begin{equation}
 \frac{\partial T_{TM}}{\partial \widetilde{R}_{I_{MAX}+1}} = - M_0 \left( \overline{w} \right)^{1 + I_{MAX}} \: e^{-\widetilde{R}_0} \: e^{\left[-\widetilde{R}_{I_{MAX}+1} - S \widetilde{R}_1 - \left( 1 - \delta(Q-1) \right) S \left( \sum\limits_{k=1}^{Q-1} \widetilde{R}_{kS+1} \right) \right]},
\label{eq:eqA.25_P}
 \end{equation}
\begin{equation}
\frac{\partial T_{DT}}{\partial \widetilde{R}_{I_{MAX}+1}} = - M_0 \: e^{-\widetilde{R}_0} \left( \overline{w} \right)^{1+I_{MAX}} \left( 1 - \delta(1 + I_{MAX}) \right) \: e^{\left[-\widetilde{R}_{I_{MAX}+1} - S \widetilde{R}_1 - \left( 1 - \delta(Q-1) \right) S \left( \sum\limits_{k=1}^{Q-1} \widetilde{R}_{kS+1} \right) \right]}, 
\label{eq:eqA.26_P}
\end{equation}  
and,
\begin{equation}
\begin{split}
\frac{\partial \mathcal{E}_{tot}}{\partial \widetilde{R}_{I_{MAX}+1}} = K_0 M_0 \left( \overline{w} \right)^{1+I_{MAX}} \: e^{-\widetilde{R}_0} \left( \alpha - 1 \right) \: e^{\left[ -\left( \alpha - 1 \right) \widetilde{R}_{I_{MAX}+1} - S \widetilde{R}_1 - \left( 1 - \delta(Q-1) \right) S \left( \sum\limits_{k=1}^{Q-1} \widetilde{R}_{kS+1} \right) \right]}.  
\end{split}
\label{eq:eqA.27_P}
\end{equation}   
Finally, the gradients of Eq. \eqref{eq:eq3.62_P} with respect to the Lagrange multipliers equate the corresponding constraints, that is:
\begin{equation}
\nabla_{\lambda_{1}} \mathcal{L} = \left( \frac{T_{TM}}{\Delta_{TM}} \right) - 1,
\label{eq:eqA.28_P}
\end{equation} 
\begin{equation}
\nabla_{\lambda_{2}} \mathcal{L} = \left( \frac{T_{DT}}{\Delta_{DT}} \right) - 1,
\label{eq:eqA.29_P}
\end{equation} 
\begin{equation}
\nabla_{\lambda_{30}} \mathcal{L} = \beta \overline{w} \: e^{-\widetilde{R}_0} - 1,
\label{eq:eqA.30_P} 
\end{equation} 
\begin{equation}
\nabla_{\lambda_{31}} \mathcal{L} = \beta \overline{w} \: e^{-\widetilde{R}_1} - 1,
\label{eq:eqA.31_P}
\end{equation}
and,
\begin{equation}
\nabla_{\lambda_{3(jS+1)}} \mathcal{L} = \beta \overline{w} \: e^{-\widetilde{R}_{3(jS+1)}} - 1,  \quad j = 1, 2, \ldots, (Q - 1).
\label{eq:eqA.32_P}	 	  
\end{equation}  
%

\bibliography{Bibliografia}%

\section*{Authors' Biographies}

\begin{biography}{\includegraphics[width=66pt,height=44pt]{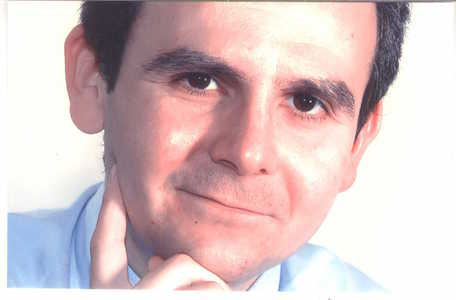}}{\textbf{Enzo Baccarelli.} He is Full Professor in Information and Communication Engineering at the DIET Department of Sapienza University of Rome. He received the Laurea degree in Electronic Engineering and the Ph.D. degree in Communication Theory and  Systems, both from Sapienza University of Rome. In 1995, he received the Post-Doctorate degree in Information Theory and Applications from the INFOCOM Department of Sapienza University of Rome. His current research focuses on data networks, distributed computing and networked data centers and Fog computing. From 2005 to 2010 he served as Associate Editor for IEEE Communications Letters. He was the national coordinator of several MIUR projects.}
\end{biography}


\begin{biography}{\includegraphics[width=66pt,height=82pt]{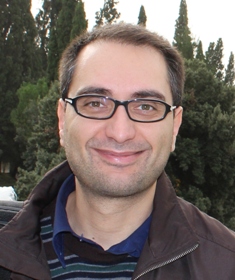}}{\textbf{Michele Scarpiniti.} He received the Laurea degree in Electrical Engineering with honors from the Sapienza University of Rome, Italy, in 2005 and the Ph.D. in Information and Communication Engineering in 2009. From March 2008 he is an Assistant Professor of Circuit Theory and Multimedia Signal Processing with the Department of Information Engineering, Electronics and Telecommunications, at Sapienza University of Rome. His present research interests include nonlinear adaptive filters, audio processing and neural networks for signal processing, ICA and blind signal processing. He is a member of the Intelligent Signal Processing and MultiMedia (ISPAMM) laboratory, an interdisciplinary research group working at the DIET department of Sapienza University of Rome. The ISPAMM research aims at the design and development of innovative methodologies for multimedia processing. He is a senior member of IEEE, a member of the ``Audio Engineering Society'' (AES) and a member of the ``Societ\`a Italiana Reti Neuroniche'' (SIREN). He is also in the board of the AES Italian Section.}
\end{biography}


\begin{biography}{\includegraphics[width=66pt,height=82pt]{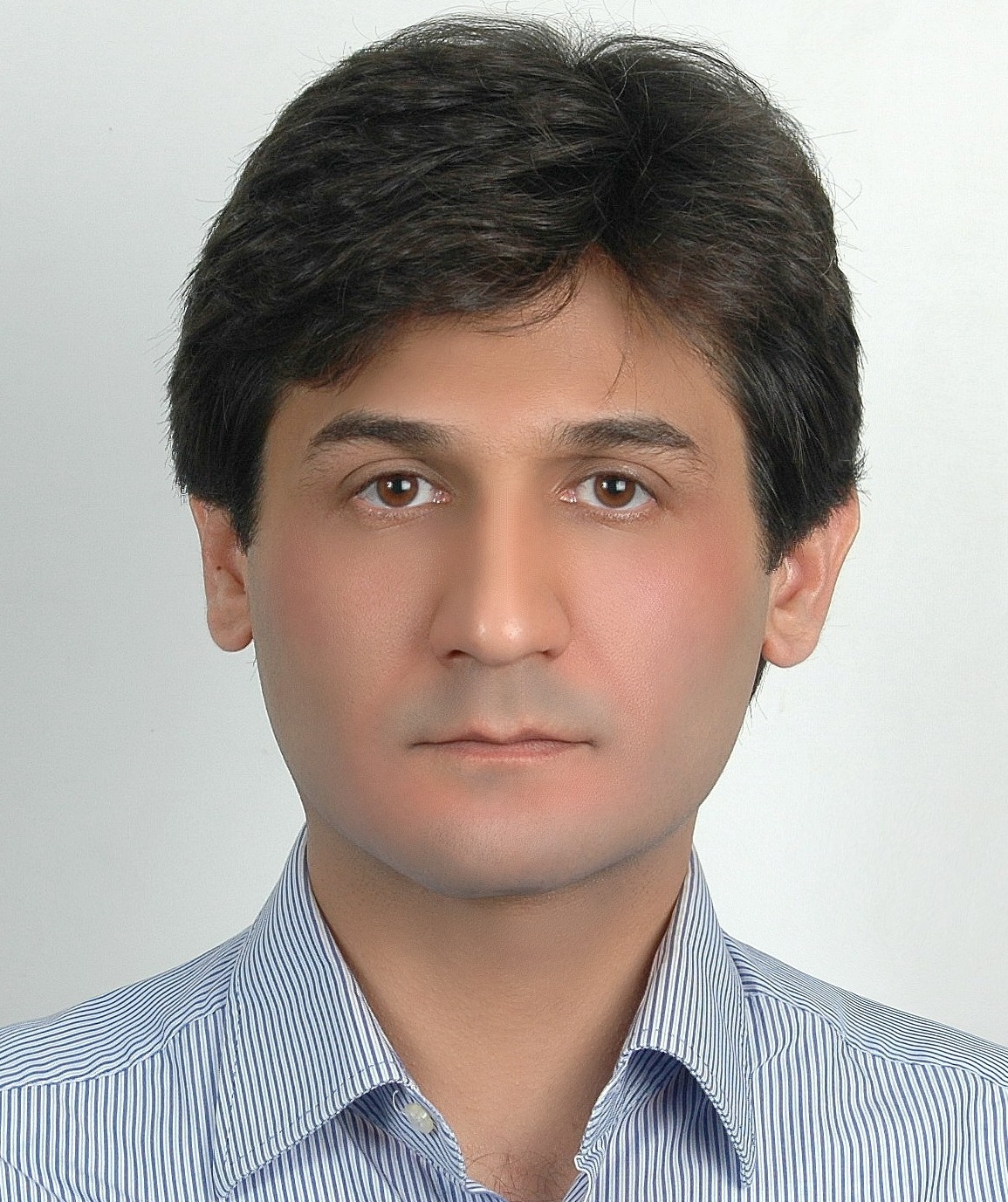}}{\textbf{Alireza Momenzadeh.} He is a researcher at the DIET Department of Sapienza University of Rome. He received a Bachelor of Civil Engineering from Estahban University in Iran and a Master degree in Structural Engineering from University Technology of Malaysia in 2015. His current research activity embraces nonlinear optimization through hybrid methods, Fog computing architectures and related sensor/actuator-based control applications.}
\end{biography}

\end{document}